\documentclass[twocolumn]{aastex62}
\usepackage[varg]{txfonts}
\usepackage{graphicx}
\usepackage{color}
\usepackage{eqnarray}
\begin{document}

\title{The JCMT BISTRO survey: alignment between outflows and magnetic fields in dense cores/clumps}

\author{Hsi-Wei Yen}
\affiliation{Academia Sinica Institute of Astronomy and Astrophysics, 11F of Astro-Math Bldg, 1, Sec. 4, Roosevelt Rd, Taipei 10617, Taiwan}

\author{Patrick M. Koch}
\affiliation{Academia Sinica Institute of Astronomy and Astrophysics, 11F of Astro-Math Bldg, 1, Sec. 4, Roosevelt Rd, Taipei 10617, Taiwan}

\author{Charles L. H. Hull}
\affiliation{National Astronomical Observatory of Japan, NAOJ Chile, Alonso de C\'ordova 3788, Office 61B, 7630422, Vitacura, Santiago, Chile}
\affiliation{Joint ALMA Observatory, Alonso de C\'ordova 3107, Vitacura, Santiago, Chile}
\affiliation{NAOJ Fellow}

\author{Derek Ward-Thompson}
\affiliation{Jeremiah Horrocks Institute, University of Central Lancashire, Preston PR1 2HE, UK}

\author{Pierre Bastien}
\affiliation{Centre de recherche en astrophysique du Qu\'{e}bec \& d\'{e}partement de physique, Universit\'{e} de Montr\'{e}al, C.P. 6128 Succ. Centre-ville, Montr\'{e}al, QC, H3C 3J7, Canada}

\author{Tetsuo Hasegawa}
\affiliation{National Astronomical Observatory of Japan, National Institutes of Natural Sciences, Osawa, Mitaka, Tokyo 181-8588, Japan}

\author{Woojin Kwon}
\affiliation{Department of Earth Science Education, Seoul National University, 1 Gwanak-ro, Gwanak-gu, Seoul 08826, Republic of Korea}
\affiliation{Korea Astronomy and Space Science Institute, 776 Daedeokdae-ro, Yuseong-gu, Daejeon 34055, Republic of Korea}

\author{Shih-Ping Lai}
\affiliation{Institute of Astronomy and Department of Physics, National Tsing Hua University, Hsinchu 30013, Taiwan}
\affiliation{Academia Sinica Institute of Astronomy and Astrophysics, 11F of Astro-Math Bldg, 1, Sec. 4, Roosevelt Rd, Taipei 10617, Taiwan}

\author{Keping Qiu}
\affiliation{School of Astronomy and Space Science, Nanjing University, 163 Xianlin Avenue, Nanjing 210023, People's Republic of China}
\affiliation{Key Laboratory of Modern Astronomy and Astrophysics (Nanjing University), Ministry of Education, Nanjing 210023, People's Republic of China}

\author{Tao-Chung Ching}
\affiliation{CAS Key Laboratory of FAST, National Astronomical Observatories, Chinese Academy of Sciences, People's Republic of China}
\affiliation{National Astronomical Observatories, Chinese Academy of Sciences, A20 Datun Road, Chaoyang District, Beijing 100012, People's Republic of China}

\author{Eun Jung Chung}
\affiliation{Department of Astronomy and Space Science, Chungnam National University, 99 Daehak-ro, Yuseong-gu, Daejeon 34134, Republic of Korea}

\author{Simon Coud\'e}
\affiliation{SOFIA Science Center, Universities Space Research Association, NASA Ames Research Center, Moffett Field, California 94035, USA}

\author{James Di Francesco}
\affiliation{NRC Herzberg Astronomy and Astrophysics, 5071 West Saanich Road, Victoria, BC V9E 2E7, Canada}
\affiliation{Department of Physics and Astronomy, University of Victoria, Victoria, BC V8W 2Y2, Canada}

\author{Pham Ngoc Diep}
\affiliation{Vietnam National Space Center, Vietnam Academy of Science and Technology, 18 Hoang Quoc Viet, Hanoi, Vietnam}

\author{Yasuo Doi}
\affiliation{Department of Earth Science and Astronomy, Graduate School of Arts and Sciences, The University of Tokyo, 3-8-1 Komaba, Meguro, Tokyo 153-8902, Japan}

\author{Chakali Eswaraiah}
\affiliation{CAS Key Laboratory of FAST, National Astronomical Observatories, Chinese Academy of Sciences, People's Republic of China}
\affiliation{National Astronomical Observatories, Chinese Academy of Sciences, A20 Datun Road, Chaoyang District, Beijing 100012, People's Republic of China}

\author{Sam Falle}
\affiliation{Department of Applied Mathematics, University of Leeds, Woodhouse Lane, Leeds LS2 9JT, UK}

\author{Gary Fuller}
\affiliation{Jodrell Bank Centre for Astrophysics, School of Physics and Astronomy, University of Manchester, Oxford Road, Manchester, M13 9PL, UK}

\author{Ray S. Furuya}
\affiliation{Tokushima University, Minami Jousanajima-machi 1-1, Tokushima 770-8502, Japan}
\affiliation{Institute of Liberal Arts and Sciences Tokushima University, Minami Jousanajima-machi 1-1, Tokushima 770-8502, Japan}

\author{Ilseung Han}
\affiliation{Korea Astronomy and Space Science Institute, 776 Daedeokdae-ro, Yuseong-gu, Daejeon 34055, Republic of Korea}
\affiliation{University of Science and Technology, Korea, 217 Gajeong-ro, Yuseong-gu, Daejeon 34113, Republic of Korea}

\author{Jennifer Hatchell}
\affiliation{Physics and Astronomy, University of Exeter, Stocker Road, Exeter EX4 4QL, UK}

\author{Martin Houde}
\affiliation{Department of Physics and Astronomy, The University of Western Ontario, 1151 Richmond Street, London N6A 3K7, Canada}

\author{Shu-ichiro Inutsuka}
\affiliation{Department of Physics, Graduate School of Science, Nagoya University, Furo-cho, Chikusa-ku, Nagoya 464-8602, Japan}

\author{Doug Johnstone}
\affiliation{NRC Herzberg Astronomy and Astrophysics, 5071 West Saanich Road, Victoria, BC V9E 2E7, Canada}
\affiliation{Department of Physics and Astronomy, University of Victoria, Victoria, BC V8W 2Y2, Canada}

\author{Ji-hyun Kang}
\affiliation{Korea Astronomy and Space Science Institute, 776 Daedeokdae-ro, Yuseong-gu, Daejeon 34055, Republic of Korea}

\author{Miju Kang}
\affiliation{Korea Astronomy and Space Science Institute, 776 Daedeokdae-ro, Yuseong-gu, Daejeon 34055, Republic of Korea}

\author{Kee-Tae Kim}
\affiliation{Korea Astronomy and Space Science Institute, 776 Daedeokdae-ro, Yuseong-gu, Daejeon 34055, Republic of Korea}
\affiliation{University of Science and Technology, Korea, 217 Gajeong-ro, Yuseong-gu, Daejeon 34113, Republic of Korea}

\author{Florian Kirchschlager}
\affiliation{Department of Physics and Astronomy, University College London, WC1E 6BT London, UK}

\author{Jungmi Kwon}
\affiliation{Department of Astronomy, Graduate School of Science, The University of Tokyo, 7-3-1 Hongo, Bunkyo-ku, Tokyo 113-0033, Japan}

\author{Chang Won Lee}
\affiliation{Korea Astronomy and Space Science Institute, 776 Daedeokdae-ro, Yuseong-gu, Daejeon 34055, Republic of Korea}
\affiliation{University of Science and Technology, Korea, 217 Gajeong-ro, Yuseong-gu, Daejeon 34113, Republic of Korea}

\author{Chin-Fei Lee}
\affiliation{Academia Sinica Institute of Astronomy and Astrophysics, 11F of Astro-Math Bldg, 1, Sec. 4, Roosevelt Rd, Taipei 10617, Taiwan}

\author{Hong-Li Liu}
\affiliation{Department of Astronomy, Yunnan University, Kunming, 650091, PR China}
\affiliation{Departamento de Astronom\'ia, Universidad de Concepci\'on, Av. Esteban Iturra s/n, Distrito Universitario, 160-C, Chile}
\affiliation{Shanghai Astronomical Observatory, Chinese Academy of Sciences, 80 Nandan Road, Shanghai 200030, People's Republic of China}

\author{Tie Liu}
\affiliation{Key Laboratory for Research in Galaxies and Cosmology, Shanghai Astronomical Observatory, Chinese Academy of Sciences, 80 Nandan Road, Shanghai 200030, People's Republic of China}

\author{A-Ran Lyo}
\affiliation{Korea Astronomy and Space Science Institute, 776 Daedeokdae-ro, Yuseong-gu, Daejeon 34055, Republic of Korea}

\author{Nagayoshi Ohashi}
\affiliation{Academia Sinica Institute of Astronomy and Astrophysics, 11F of Astro-Math Bldg, 1, Sec. 4, Roosevelt Rd, Taipei 10617, Taiwan}

\author{Takashi Onaka}
\affiliation{Department of Physics, Faculty of Science and Engineering, Meisei University, 2-1-1 Hodokubo, Hino, Tokyo 1191-8506, Japan}
\affiliation{Department of Astronomy, Graduate School of Science, The University of Tokyo, 7-3-1 Hongo, Bunkyo-ku, Tokyo 113-0033, Japan}

\author{Kate Pattle}
\affiliation{Centre for Astronomy, School of Physics, National University of Ireland Galway, University Road, Galway, Ireland}

\author{Sarah Sadavoy}
\affiliation{Department for Physics, Engineering Physics and Astrophysics, Queen's University, Kingston, ON, K7L 3N6, Canada}

\author{Hiro Saito}
\affiliation{Faculty of Pure and Applied Sciences, University of Tsukuba, 1-1-1 Tennodai, Tsukuba, Ibaraki 305-8577, Japan}

\author{Hiroko Shinnaga}
\affiliation{Department of Physics and Astronomy, Graduate School of Science and Engineering, Kagoshima University, 1-21-35 Korimoto, Kagoshima, Kagoshima 890-0065, Japan}

\author{Archana Soam}
\affiliation{SOFIA Science Center, Universities Space Research Association, NASA Ames Research Center, Moffett Field, California 94035, USA}

\author{Mehrnoosh Tahani}
\affiliation{Dominion Radio Astrophysical Observatory, Herzberg Astronomy and Astrophysics Research Centre, National Research Council Canada, P.O. Box 248, Penticton, BC V2A 6J9 Canada}

\author{Motohide Tamura}
\affiliation{Department of Astronomy, Graduate School of Science, The University of Tokyo, 7-3-1 Hongo, Bunkyo-ku, Tokyo 113-0033, Japan}
\affiliation{Astrobiology Center, 2-21-1 Osawa, Mitaka-shi, Tokyo 181-8588, Japan}
\affiliation{National Astronomical Observatory, 2-21-1 Osawa, Mitaka-shi, Tokyo 181-8588, Japan}

\author{Ya-Wen Tang}
\affiliation{Academia Sinica Institute of Astronomy and Astrophysics, 11F of Astro-Math Bldg, 1, Sec. 4, Roosevelt Rd, Taipei 10617, Taiwan}

\author{Xindi Tang}
\affiliation{Xinjiang Astronomical Observatory, Chinese Academy of Sciences, 830011 Urumqi, People's Republic of China}

\author{Chuan-Peng Zhang}
\affiliation{National Astronomical Observatories, Chinese Academy of Sciences, A20 Datun Road, Chaoyang District, Beijing 100012, People's Republic of China}
\affiliation{CAS Key Laboratory of FAST, National Astronomical Observatories, Chinese Academy of Sciences, People's Republic of China}

\correspondingauthor{Hsi-Wei Yen}
\email{hwyen@asiaa.sinica.edu.tw}

\begin{abstract}
We compare the directions of molecular outflows of 62 low-mass Class 0 and I protostars in nearby ($<$450 pc) star-forming regions with the mean orientations of the magnetic fields on 0.05--0.5 pc scales in the dense cores/clumps where they are embedded. The magnetic field orientations were measured using the JCMT POL-2 data taken by the BISTRO-1 survey and from the archive. The outflow directions were observed with interferometers in the literature. The observed distribution of the angles between the outflows and the magnetic fields peaks between 15$\arcdeg$ and 35$\arcdeg$. After considering projection effects, our results could suggest that the outflows tend to be misaligned with the magnetic fields by 50$\arcdeg$$\pm$15$\arcdeg$ in three-dimensional space and are less likely (but not ruled out) randomly oriented with respect to the magnetic fields. There is no correlation between the misalignment and the bolometric temperatures in our sample. In several sources, the small-scale (1000--3000 au) magnetic fields is more misaligned with the outflows than their large-scale magnetic fields, suggesting that the small-scale magnetic field has been twisted by the dynamics. In comparison with turbulent MHD simulations of core formation, our observational results are more consistent with models in which the energy densities in the magnetic field and the turbulence of the gas are comparable. Our results also suggest that the misalignment alone cannot sufficiently reduce the efficiency of magnetic braking to enable formation of the observed number of large Keplerian disks with sizes larger than 30--50 au. 
\end{abstract}

\keywords{Star formation (1569), Interstellar magnetic fields (845), Star forming regions (1565), Protostars (1302)}

\section{Introduction}
Star-forming regions are magnetized \citep[][]{Crutcher12,Planck16a}. 
The magnetic field is theoretically expected to be dynamically important during the star formation processes \citep[e.g.,][]{Galli93,Basu94,Allen03}.
On large scales, the relative importance between the magnetic field and the turbulence could affect structure formation in molecular clouds \citep{McKee07,Soler13} as well as the properties of dense cores formed in molecular clouds \citep[e.g.,][]{Burkert00, Gammie03}.
Inside dense cores, the magnetic field could influence the transfer of mass and angular momentum from large to small scales and subsequently the formation and evolution of circumstellar disks around protostars \citep[][]{Li14}.
Therefore, observational studies on magnetic field structures in comparison with other physical quantities in a large sample of molecular clouds and dense cores are essential to understand the role of the magnetic field in star formation \citep[e.g.,][]{Matthews09,Palmeirim13,Poidevin13,Koch14,Zhang14,Planck16b,Hull19,Pattle19b,Chen20}.

Alignment between the magnetic field and rotational axis in dense cores can be a diagnostic of the importance of the magnetic field in environments of core formation. 
Theoretically, if the magnetic field is dominant, dense cores are expected to have their rotational axes aligned with the magnetic field \citep{Mouschovias79}. 
The turbulent magnetohydrodynamic (MHD) simulations of core formation in converging flows also show that there are more dense cores with their rotational axes better aligned with the magnetic fields, when initially the magnetic field is dominating over the turbulence in converging flows \citep{Chen18}. 
In addition, the alignment between the magnetic field and rotational axis can be an important parameter during the collapse of dense cores. 
MHD simulations show that when the magnetic field is misaligned with the rotational axis in a collapsing dense core, 
the efficiency of magnetic braking decreases, 
and more angular momentum can be transferred to the vicinity of the central protostar, resulting in the formation of a larger rotationally supported disk \citep{Joos12,Li13,Hirano20}. 
Similar effects are also seen in MHD simulations with turbulence, where the local turbulence in a collapsing core can cause misalignment between the magnetic field and the rotational axis \citep[e.g.][]{Gray18, Lam19}.
Theoretically, an outflow is expected to launch along the rotational axis of a star-disk system \citep[e.g.,][]{Blandford82, Pudritz83}.
The MHD simulations also show that in a collapsing dense core, the direction of the bipolar outflow on a scale of a few thousand au is along the rotational axis of the dense core, even if the rotational axis of the dense core is initially misaligned with the magnetic field \citep{Ciardi10,Hirano20}.
Thus, observationally, the direction of a bipolar outflow in a protostellar source can be a proxy of the rotational axis of its natal dense core,
and statistical studies of the alignment between the magnetic fields and outflows in protostellar sources are crucial to better understand the formation of dense cores and circumstellar disks around protostars.

The alignment between the magnetic fields on the core scale of $\sim$0.1 pc and outflows in low-mass protostars has been studied with a sample of seven sources using the SHARP polarimeter at the Caltech Submillimeter Observatory \citep{Davidson11,Chapman13}. 
On the assumption that the observed magnetic field structures trace an hour-glass morphology, 
these single-dish studies suggest that the magnetic field tends to align with the outflow direction. 
On the contrary, 
the interferometric studies of $\sim$20--30 low-mass protostars with the Combined Array for Research in Millimeter-wave Astronomy (CARMA) at spatial resolutions of $\sim$1000 au show that the magnetic fields in the protostellar envelopes on a scale of a few thousand au are randomly oriented with respect to the outflow directions \citep{Hull13,Hull14,Hull19}. 
Similar studies of 12 sources with the Submillimeter Array (SMA) suggest a bimodal distribution, where the magnetic fields in protostellar envelopes are either aligned with or perpendicular to the outflow directions \citep{Galametz18}. 
The results obtained with these single-dish and interferometric observations seem to be inconsistent. 
However, the sample sizes of the previous single-dish studies are limited.  
On the other hand, in contrast with the single-dish observations, 
the magnetic field structures on the envelope scale observed with the interferometers are more likely to be affected by collapse and rotational motion in the protostellar envelopes, 
and possibly do not represent the initial configurations.
Therefore, single-dish observations toward a large sample of protostars to probe their large-scale magnetic fields, which may still preserve the initial configurations, are needed to investigate the alignment between the magnetic field and rotational axis in dense cores. 

The James Clerk Maxwell Telescope (JCMT) large program, B-fields In STar-forming Region Observations \citep[BISTRO;][]{Ward17}, provides excellent data sets to study the magnetic field in star-forming regions. 
The BISTRO program observes polarized thermal dust continuum emission at 850 $\mu$m and 450 $\mu$m using the polarimeter POL-2 \citep{Friberg16} for the bolometer SCUBA-2 \citep{Holland13}.
The first part of the survey, BISTRO-1, observed 16 fields in dense parts of the Gould Belt star-forming regions and covered a total area of $>$110 arcmin$^2$ with a total observing time of 224 hours. 
The BISTRO-1 observations have been completed.
There are two on-going follow-up BISTRO surveys, 
BISTRO-2 to continue the observations of the Gould Belt regions and to include other intermediate- and high-mass star-forming regions 
and BISTRO-3 to observe regions at different evolutionary stages and different spatial scales.
Each follow-up survey also has a total observing time of 224 hours and targets 16 fields.
A series of studies on the structures and strengths of the magnetic field and the properties of the dust grains in several star-forming regions have been conducted with the BISTRO program \citep{Pattle17,Pattle18,Pattle19,Kwon18,Soam18,Coude19,Liu19,WangJ19,Doi20}.
In the present paper, 
we study the alignment between the magnetic fields and outflows in 62 low-mass protostars using BISTRO and archival POL-2 data, 
and discuss the observed distribution of the angles between the magnetic fields and the outflows in the context of the formation of dense cores and circumstellar disks.

\section{Sample}\label{sample}

\begin{figure*}
\centering
\includegraphics[width=0.98\textwidth]{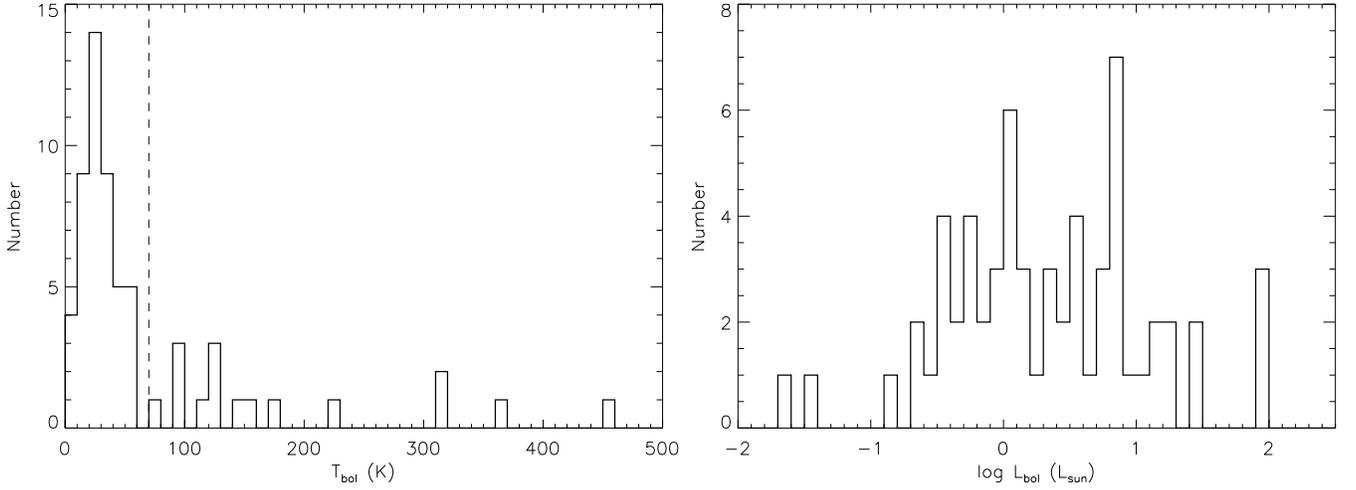}
\caption{Distributions of bolometric temperature ($T_{\rm bol}$; left) and luminosity ($L_{\rm bol}$; right) in our sample. A vertical dashed line denotes $T_{\rm bol}$ of 70 K. If $T_{\rm bol}$ and $L_{\rm bol}$ of a sample source only have upper limits, the upper limits are plotted. Details are given in Table~1.}\label{dist_sample}
\end{figure*} 

Our sample sources were selected from dense molecular clouds in nearby low-mass star-forming regions at distances less than 450 pc observed by the JCMT BISTRO survey (program ID: M16AL004 and M17BL011), including Taurus, Perseus, Ophiuchus, and Serpens Main. 
A part of the BISTRO results of these regions has been published by \citet{Kwon18}, \citet{Soam18}, \citet{Coude19}, \citet{Liu19}, \citet{Pattle19}, and \citet{Doi20}, 
and the remaining data of regions in Taurus and Serpens Main will be presented in the forthcoming papers by Eswaraiah et al.~(in prep), Kwon et al.~(in prep.), and Tang et al.~(in prep.) in detail. 
The Auriga star-forming region was also observed by the BISTRO survey (Ngoc et al. in prep.). 
Since no young protostars are associated with clear bipolar outflows in the Auriga BISTRO field, 
these data are not included here.
In addition, we searched the JCMT data archive for the POL-2 data of other nearby molecular clouds and Bok globules, and we retrieved the data taken with the regular projects M17AP067 (PI: P.~Koch), M17AP073 (PI: W.~Kwon), M17BP058 (PI: W.~Kwon) and M17BP070 (PI: A.~Soam).
Some of these data have been published by \citet{Soam19} and \citet{Yen19}.
Then, from these star-forming regions, 
we selected low-mass Class 0 and I protostars whose molecular outflows have been imaged with interferometric observations at spatial resolutions of a few hundred au and show more or less well defined axes or cavity walls, 
so that the directions of their outflows could be accurately determined\footnote{We note that this selection criteria may bias our sample and exclude pole-on outflows. Nevertheless, we have tested this effect in our simulations of distributions of misalignment angles (Section~\ref{numdist}) by excluding pairs of vectors with one of them aligned with the line of sight within a certain degree. We confirmed that the simulated distributions are not affected.}. 
These selection criteria led to a sample of 62 Class 0 and I protostars. 
Table~1 lists all the sample sources with their outflow directions obtained from the literature and their basic properties. 
This represents the complete sample of young protostars which have been observed to host clear bipolar outflows in the BISTRO-1 survey fields of the low-mass star-forming regions.
The distributions of the bolometric temperature ($T_{\rm bol}$) and luminosity ($L_{\rm bol}$) in our sample are presented in Fig.~\ref{dist_sample}.
We note that there are more than a factor of two more Class 0 and I protostars in these star-forming regions than those included here \citep{Dunham08, Enoch08, Evans09}, 
but there are no suitable measurements of their outflow directions with interferometric observations of these additional sources. 
Thus, they are not included in this work.


\startlongtable
\begin{deluxetable*}{cccccccccc}
\tablecaption{Sample list}
\centering
\tablehead{Name & Region & Distance & RA (J2000) & Dec (J2000) &  $T_{\rm bol}$ & $L_{\rm bol}$ & Outflow & Mean B field & Reference \\
 & & (pc) & & & (K) & ($L_\sun$) & orientation & orientation & }
\startdata
L1448 IRS 2 & L1448 & 288 & 03:25:22.41 & +30:45:13.3 &  43 &  3.6 & 118\arcdeg & 138\arcdeg$\pm$2\arcdeg & 1,2 \\
L1448 IRS 2E & L1448 & 288 & 03:25:25.66 & +30:44:56.7 &  15 &   0.05 & 165\arcdeg &   9\arcdeg$\pm$2\arcdeg & 1,2 \\
L1448 IRS 3Bc\tablenotemark{a} & L1448 & 288 & 03:25:35.67 & +30:45:34.1 &  57 &   $<$8.3 & 128\arcdeg & 169\arcdeg$\pm$1\arcdeg & 1,2 \\
L1448 IRS 3Ba\tablenotemark{a,b} & L1448 & 288 & 03:25:36.38 & +30:45:14.7 &  57 &   $<$8.3 & 122\arcdeg &  17\arcdeg$\pm$3\arcdeg & 1,2 \\
L1448 IRS 3Bb\tablenotemark{a,b} & L1448 & 288 & 03:25:36.50 & +30:45:21.9 &  57 &   $<$8.3 &  38\arcdeg &  17\arcdeg$\pm$3\arcdeg & 1,2 \\
L1448-mm\tablenotemark{b} & L1448 & 288 & 03:25:38.88 & +30:44:05.3 &  47 &   8.4 & 162\arcdeg &  21\arcdeg$\pm$30\arcdeg & 1,2 \\
L1448C-S\tablenotemark{b} & L1448 & 288 & 03:25:39.14 & +30:43:57.9 & 163 &   0.68 &  43\arcdeg &  21\arcdeg$\pm$30\arcdeg & 1,2 \\
Per-emb 17 & L1455 & 279 & 03:27:39.10 & +30:13:03.1 &  39 &   4.2 &  10\arcdeg &  65\arcdeg$\pm$4\arcdeg & 1,2 \\
L1455 IRS 4 & L1455 & 279 & 03:27:43.28 & +30:12:28.8 &  65 &   1.4 & 115\arcdeg &  80\arcdeg$\pm$3\arcdeg & 1,2 \\
Per-emb 3 & NGC1333 & 299 & 03:29:00.58 & +31:12:00.2 &  32 &   0.5 &  96\arcdeg &  53\arcdeg$\pm$6\arcdeg & 1,2 \\
NGC1333 IRAS 4A & NGC1333 & 299 & 03:29:10.54 & +31:13:30.9 &  29 &   7 &  35\arcdeg &  60\arcdeg$\pm$1\arcdeg & 1,2 \\
NGC1333 IRAS 4B1\tablenotemark{a,b} & NGC1333 & 299 & 03:29:12.02 & +31:13:08.0 &  28 &   $<$4 &   0\arcdeg &  71\arcdeg$\pm$1\arcdeg & 1,2 \\
NGC1333 IRAS 4B2\tablenotemark{a,b} & NGC1333 & 299 & 03:29:12.84 & +31:13:06.9 &  28 &   $<$4 &  90\arcdeg &  71\arcdeg$\pm$1\arcdeg & 1,2 \\
NGC1333 IRAS 7\tablenotemark{b} & NGC1333 & 299 & 03:29:11.26 & +31:18:31.1 &  59 &   2.8 & 150\arcdeg &  92\arcdeg$\pm$1\arcdeg & 1,2 \\
Per-emb 21\tablenotemark{b} & NGC1333 & 299 & 03:29:10.67 & +31:18:20.2 &  45 &   6.9 &  48\arcdeg &  92\arcdeg$\pm$\arcdeg & 1,2 \\
NGC1333 IRAS 2A1\tablenotemark{a,b} & NGC1333 & 299 & 03:28:55.57 & +31:14:37.0 &  69 &  $<$19 &  14\arcdeg &  79\arcdeg$\pm$2\arcdeg & 1,2 \\
NGC1333 IRAS 2A2\tablenotemark{a,b} & NGC1333 & 299 & 03:28:55.56 & +31:14:36.4 &  69 &  $<$19 & 104\arcdeg &  79\arcdeg$\pm$2\arcdeg & 1,2 \\
NGC1333 IRAS 1a\tablenotemark{a,b} & NGC1333 & 299 & 03:28:37.09 & +31:13:30.8 & 103 &   $<$9.1 & 123\arcdeg &  83\arcdeg$\pm$8\arcdeg & 1,2 \\
NGC1333 IRAS 1b\tablenotemark{a,b} & NGC1333 & 299 & 03:28:37.22 & +31:13:31.8 & 103 &   $<$9.1 & 169\arcdeg &  83\arcdeg$\pm$8\arcdeg & 1,2 \\
NGC1333 IRAS 2B & NGC1333 & 299 & 03:28:57.37 & +31:14:15.8 & 106 &   5.3 &  24\arcdeg &  55\arcdeg$\pm$4\arcdeg & 1,2 \\
SVS 13A\tablenotemark{b} & NGC1333 & 299 & 03:29:03.77 & +31:16:03.8 & 188 &  32.5 & 130\arcdeg & 164\arcdeg$\pm$1\arcdeg & 1,2 \\
RNO15-FIR\tablenotemark{b} & NGC1333 & 299 & 03:29:04.1 & +31:14:46.2 &  36 &   0.4 & 145\arcdeg & 164\arcdeg$\pm$1\arcdeg & 1,2 \\
Per-emb 37 & NGC1333 & 299 & 03:29:19.0 & +31:23:14.3 &  22 &   0.5 &  37\arcdeg & 131\arcdeg$\pm$3\arcdeg & 1,2 \\
Per-emb 49 & NGC1333 & 299 & 03:29:12.95 & +31:18:14.3 & 239 &   1.1 &  25\arcdeg &  92\arcdeg$\pm$1\arcdeg & 1,2 \\
Per-emb 50 & NGC1333 & 299 & 03:29:07.77 & +31:21:57.1 & 128 &  23.2 & 104\arcdeg & 175\arcdeg$\pm$1\arcdeg & 1,2 \\
Per-emb 58 & NGC1333 & 299 & 03:28:58.42 & +31:22:17.5 & 322 &   0.63 & 167\arcdeg & 119\arcdeg$\pm$2\arcdeg & 1,2 \\
SVS 13B\tablenotemark{b} & NGC1333 & 299 & 03:29:03.08 & +31:15:51.7 &  20 &   1 & 160\arcdeg & 164\arcdeg$\pm$1\arcdeg & 1,2 \\
SVS 13C\tablenotemark{b} & NGC1333 & 299 & 03:29:01.97 & +31:15:38.1 &  21 &   1.5 &   8\arcdeg & 164\arcdeg$\pm$1\arcdeg & 1,2 \\
Per-emb 6 & Perseus B1 & 301 & 03:33:14.4 & +31:07:10.9 &  52 &   0.3 &  60\arcdeg & 144\arcdeg$\pm$2\arcdeg & 1,2 \\
Per-emb 10 & Perseus B1 & 301 & 03:33:16.45 & +31:06:52.5 &  30 &   0.6 &  51\arcdeg & 156\arcdeg$\pm$1\arcdeg & 1,2 \\
B1-a & Perseus B1 & 301 & 03:33:16.66 & +31:07:55.2 & 132 &   1.5 & 101\arcdeg & 146\arcdeg$\pm$1\arcdeg & 1,2 \\
B1-c & Perseus B1 & 301 & 03:33:17.85 & +31:09:32 &  48 &   3.7 & 131\arcdeg &  95\arcdeg$\pm$1\arcdeg & 1,2 \\
B1-b\tablenotemark{b} & Perseus B1 & 301 & 03:33:20.96 & +31:07:23.8 & 157 &   0.17 &  30\arcdeg & 157\arcdeg$\pm$1\arcdeg & 1,2 \\
B1-bN\tablenotemark{b} & Perseus B1 & 301 & 03:33:21.21 & +31:07:43.7 &  14 &   0.32 &  90\arcdeg & 157\arcdeg$\pm$1\arcdeg & 1,2 \\
B1-bS\tablenotemark{b} & Perseus B1 & 301 & 03:33:21.36 & +31:07:26.4 &  17 &   0.70 & 112\arcdeg & 157\arcdeg$\pm$1\arcdeg & 1,2 \\
HH211-mms & IC348 & 295 & 03:43:56.81 & +32:00:50.2 &  27 &   1.8 & 116\arcdeg & 152\arcdeg$\pm$1\arcdeg & 1,2 \\
IC348 MMSa\tablenotemark{a,b} & IC348 & 295 & 03:43:57.07 & +32:03:04.8 &  30 &   $<$1.5 & 167\arcdeg & 153\arcdeg$\pm$2\arcdeg & 1,2 \\
IC348 MMSb\tablenotemark{a,b} & IC348 & 295 & 03:43:57.69 & +32:03:10.0 &  30 &   $<$1.5 &  36\arcdeg & 153\arcdeg$\pm$2\arcdeg & 1,2 \\
Per-emb 16\tablenotemark{b} & IC348 & 295 & 03:43:50.98 & +32:03:24.1 &  39 &   0.4 &  10\arcdeg & 113\arcdeg$\pm$3\arcdeg & 1,2 \\
Per-emb 28\tablenotemark{b} & IC348 & 295 & 03:43:51.01 & +32:03:08.0 &  45 &   0.7 & 112\arcdeg & 113\arcdeg$\pm$3\arcdeg & 1,2 \\
Per-emb 62 & IC348 & 295 & 03:44:12.98 & +32:01:35.4 & 378 &   1.8 &  24\arcdeg & 145\arcdeg$\pm$9\arcdeg & 1,2 \\
IRAS 04169+2702 & B211/B213 & 140 & 04:19:58.46 & +27:09:56.9 & 133 &   0.77 &  64\arcdeg & 102\arcdeg$\pm$6\arcdeg & 3,4 \\
IRAS 04166+2706 & B211/B213 & 140 & 04:19:42.50 & +27:13:36.0 & 139 &   0.3 &  30\arcdeg &  47\arcdeg$\pm$3\arcdeg & 5,6 \\
L1521F & L1521F & 140 & 04:28:38.9 & +26:51:35.0 &  20 &   0.03 &  70\arcdeg &  21\arcdeg$\pm$6\arcdeg & 7,8 \\
L1527 & L1527 & 140 & 04:39:53.88 & +26:03:09.7 &  44 &   1.9 &  92\arcdeg &  82\arcdeg$\pm$11\arcdeg & 9,10 \\
HH212 & HH212 & 414 & 05:43:51.41 & -01:02:53.1 &  41 &   9 &  23\arcdeg &  35\arcdeg$\pm$4\arcdeg & 11,12 \\
HH111 & HH111 & 414 & 05:51:46.25 & +02:48:29.7 &  69 &  20 &  97\arcdeg &  67\arcdeg$\pm$2\arcdeg & 11,13 \\
GSS 30 IRS 3 & Ophiuchus A & 138 & 16:26:21.72 & -24:22:50.9 &  86 &  33 &  20\arcdeg &  79\arcdeg$\pm$1\arcdeg & 14,15 \\
VLA 1623A & Ophiuchus A & 138 & 16:26:26.39 & -24:24:30.7 &  10 &   1.10 & 125\arcdeg &  75\arcdeg$\pm$1\arcdeg & 16,17 \\
Elias 32 & Ophiuchus B & 138 & 16:27:28.4 & -24:27:21.7 & 321 &   5 &  91\arcdeg & 151\arcdeg$\pm$3\arcdeg & 14,18 \\
Elias 33 & Ophiuchus B & 138 & 16:27:30.2 & -24:27:43.9 & 460 &  12 & 129\arcdeg & 153\arcdeg$\pm$2\arcdeg & 14,18 \\
S68N\tablenotemark{b} & Serpens Main & 436 & 18:29:48.09 & +01:16:43.3 &  30 &  14 & 131\arcdeg &  85\arcdeg$\pm$1\arcdeg & 19 \\
S68Nc1\tablenotemark{b} & Serpens Main & 436 & 18:29:48.72 & +01:16:55.6 &  $<$40 &   $<$2.1 & 109\arcdeg &  85\arcdeg$\pm$1\arcdeg & 19 \\
S68Nb1\tablenotemark{b} & Serpens Main & 436 & 18:29:49.51 & +01:17:10.9 &  $<$60 &   $<$0.9 &  68\arcdeg &  85\arcdeg$\pm$1\arcdeg & 19 \\
Serpens SMM1b\tablenotemark{a,b} & Serpens Main & 436 & 18:29:49.67 & +01:15:21.2 &  39 & $<$109 & 165\arcdeg &  98\arcdeg$\pm$1\arcdeg & 9,20 \\
Serpens SMM1a\tablenotemark{a,b} & Serpens Main & 436 & 18:29:49.8 & +01:15:20.3 &  39 & $<$109 & 135\arcdeg &  98\arcdeg$\pm$1\arcdeg & 9,20 \\
Serpens SMM1d\tablenotemark{a,b} & Serpens Main & 436 & 18:29:49.99 & +01:15:23.0 &  39 & $<$109 &  80\arcdeg &  98\arcdeg$\pm$1\arcdeg & 9,20 \\
Serpens SMM4B\tablenotemark{a,b} & Serpens Main & 436 & 18:29:56.53 & +01:13:11.5 &  $<$30 &   $<$2.6 &  76\arcdeg &  48\arcdeg$\pm$3\arcdeg & 19 \\
Serpens SMM4A\tablenotemark{a,b} & Serpens Main & 436 & 18:29:56.72 & +01:13:15.6 &  $<$30 &   $<$2.6 &  14\arcdeg &  48\arcdeg$\pm$3\arcdeg & 19 \\
Serpens SMM11\tablenotemark{b} & Serpens Main & 436 & 18:30:00.39 & +01:11:44.6 &  $<$29 &  $<$ 0.9 &  76\arcdeg &  94\arcdeg$\pm$1\arcdeg & 19 \\
B335 & B335 & 165 & 19:37:00.90 & +07:34:09.5 &  36 &   1.4 &  99\arcdeg & 111\arcdeg$\pm$2\arcdeg & 9,10 \\
L1157 & L1157 & 352 & 20:39:06.27 & +68:02:15.7 &  42 &   7 & 163\arcdeg & 159\arcdeg$\pm$1\arcdeg & 21,22 \\
\enddata
\tablenotetext{a}{This source is in a multiple system, where individual protostars are not resolved in infrared observations. Thus, the same $T_{\rm bol}$ and $L_{\rm bol}$ are assigned to it and its companions. The assigned $L_{\rm bol}$ is the total luminosity of the system and should be considered as an upper limit for the individual sources.}
\tablenotetext{b}{In this source, there are multiple sample protostars in the same dense core or clump.}
\tablecomments{The outflow orientations and the mean magnetic field orientations on 0.05--0.5 pc scales are presented as position angles increasing from north to east. Typically, the uncertainties in the outflow orientations are $\sim$10$\arcdeg$ \citep{Stephens17}. The uncertainties in the mean magnetic field orientations are calculated with the error propagation of the uncertainties of the individual polarization detections. If the blue- and redshifted lobes of a bipolar outflow are not symmetric, we adopt the position angle of the mean axis of the two lobes as the outflow orientation. For each protostar, the first reference is for the bolometric temperature ($T_{\rm bol}$) and luminosity ($L_{\rm bol}$), and the second reference is for the outflow orientation. For HH~212 and HH~111, $L_{\rm bol}$ is from the second reference. If only one reference is listed, both $T_{\rm bol}$ and outflow orientation are from the same reference. The distances to Perseus, Ophiuchus, and Serpens are adopted from \citet{Ortiz18a,Ortiz18b}, the distance to Orion from \citet{Menten07}, the distance to Taurus and Cepheus from \citet{Torres09} and \citet{Zucker19}, and the distance to B335 from \citet{Watson20}.} 
Reference. (1) \citet{Tobin16} and references therein; (2) \citet{Stephens17}; (3) \citet{Young03}; (4) \citet{Takakuwa18}; (5) \citet{Chen95}; (6) \citet{WangL19}; (7) \citet{Hsieh17}; (8) \citet{Takahashi13}; (9) \citet{Kristensen12}; (10) \citet{Hull14}; (11) \citet{Tafalla13}; (12) \citet{Lee17}; (13) \citet{Lee16}; (14) \citet{Kempen09}; (15) \citet{Friesen18}; (16) \citet{Murillo13}; (17) \citet{Santangelo15}; (18) \citet{Kamazaki19}; (19) \citet{Aso19}; (20) \citet{Tychoniec19}; (21) \citet{Motte01}; (22) \citet{Maury19}.
\end{deluxetable*}

\section{POL-2 data and analysis}
In this work, we used the POL-2 data at 850 $\mu$m.
The angular resolution of the JCMT POL-2 observations at 850 $\mu$m is 14\farcs6, corresponding to spatial scales from 2000 au to 6000 au at the distances to our sample protostars.
We reduced all the POL-2 data with the software {\it Starlink} \citep{Currie14} and the task {\it pol2map} of the version updated on November 6 in 2019. 
We followed the standard procedure of the data reduction as described in \citet{Pattle17,Pattle19}.
The instrumental polarization model of ``JAN2018'' was adopted\footnote{The default instrumental polarization model used by {\it pol2map} was changed from ``JAN2018'' to ``Aug2019'' in April 2020. The difference at 850~$\mu$m between the two models is expected to be lower than the noise (\url{https://www.eaobservatory.org/jcmt/2020/04/change-to-the-default-ip-model-used-by-pol2map/}).}.
The data were first reduced with the default pixel size of 4$\arcsec$.  
Then every $3 \times 3$ pixels in the final Stokes {\it IQU} maps were binned up to have a pixel size of 12$\arcsec$, which is comparable to the angular resolution, to extract polarization detections.
Our detection criteria of the polarized emission are signal-to-noise ratios of Stokes {\it I} and polarized intensities both higher than three and polarization percentage lower than 20\%, 
which is typically the maximum polarization percentage observed in star-forming regions by {\it Planck} \citep{Planck15}.
Thus, the uncertainty in the polarization orientation of each detection is always $\lesssim$9$\arcdeg$.
The detected polarization orientations were rotated by 90$\arcdeg$ to infer the orientations of the magnetic field.

To measure the mean orientations of the magnetic fields in the dense cores or clumps\footnote{Single-dish polarimetric observations of molecular clouds and Bok globules show that the polarized intensity generally increases with the increasing Stokes {\it I} intensity, even though the polarization percentage decreases with the increasing Stokes {\it I} intensity \citep[e.g.,][]{Wolf03, Coude19}. Thus, the polarized emission is expected to be proportional to density and is more sensitive to dense regions along the line of sight.}  associated with the protostars in our sample, 
we first applied the two-dimensional version of the core/clump identification algorithm {\it Clumpfind} \citep{Williams94} on the Stokes {\it I} maps with a pixel size of 4$\arcsec$ and separated the molecular clouds into individual dense cores or clumps. 
The results are shown in Section \ref{results} and Appendix \ref{clumps}.
Then, we calculated the mean Stokes {\it Q} and {\it U} of the polarization detections within the area of each dense core or clump identified by {\it Clumpfind}, 
and computed the mean magnetic field orientation from the mean Stokes {\it Q} and {\it U}.
For the isolated dense cores in our sample, L1521F, L1527, HH~212, HH~111, and B335, we simply included all the polarization detections in the observed fields to compute their mean Stokes {\it Q} and {\it U} and the resulting mean magnetic field orientations. 
Although L1157 is also an isolated dense core, 
there is extended emission along the northwest--southeast direction detected by the JCMT POL-2 observations, 
and this extended component is likely related to the powerful outflow in L1157 \citep{Bachiller01,Tafalla15} and is distinct from the central compact core (Fig.~\ref{subplt2}). 
Thus, for L1157, we only used the detections within the central core identified by {\it Clumpfind} to compute its mean magnetic field orientation. 
The measured mean orientations of the magnetic field on a $\sim$0.1 pc scale in all the protostars in our sample are listed in Table 1. 

In our calculations of the mean magnetic field orientations, each polarization detection was weighted equally.
We have also computed the mean orientations by weighting the individual detections with their polarized intensities or signal-to-noise ratios.
The differences between the mean orientations calculated with the different weightings are typically less than 5$\arcdeg$, 
and are less than 9$\arcdeg$ in all the sources, except for L1448~mm and L1448C-S, where the differences are 19$\arcdeg$.
The mean magnetic field orientations in L1448~mm and L1448C-S also have larger uncertainties of 30$\arcdeg$ compared to all the other sources.
The comparisons of the mean magnetic field orientations computed with different weightings are listed in Appendix \ref{meanb}.
The computed mean orientation is not sensitive to the exact area or boundary of a dense core identified by {\it Clumpfind}, as discussed below.
In addition, we note that in some dense cores or clumps, there could be multiple protostars, which could have different outflow directions.
In this case, we consider all the protostars in this dense core or clump have the same mean magnetic field orientation, 
but we report different degrees of misalignment between their outflows and the mean magnetic field. 
They are considered as independent measurements in the following analysis.

\section{Results}\label{results}

\subsection{Magnetic structures in example sources} 
\begin{figure}
\centering
\includegraphics[width=0.5\textwidth]{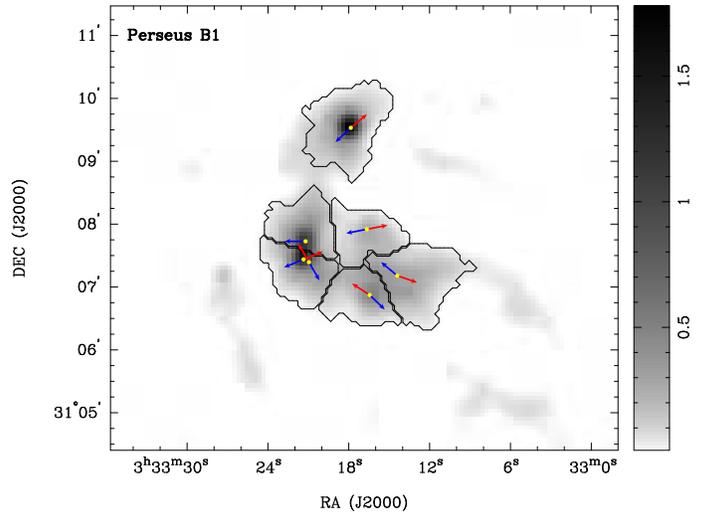}
\caption{Stokes {\it I} intensity map (gray scale; in units of Jy~Beam$^{-1}$) obtained with the JCMT POL-2 observations and the results of the core/clump identification using {\it Clumpfind} in the Perseus B1 region. Contours delineate the area of identified dense cores or clumps associated with our sample protostars. Yellow dots show the locations of our sample protostars, and blue and red arrows present the orientations of their blue- and redshifted outflow lobes.}\label{region1}
\end{figure} 

\begin{figure*}
\centering
\includegraphics[angle=90,width=0.9\textwidth]{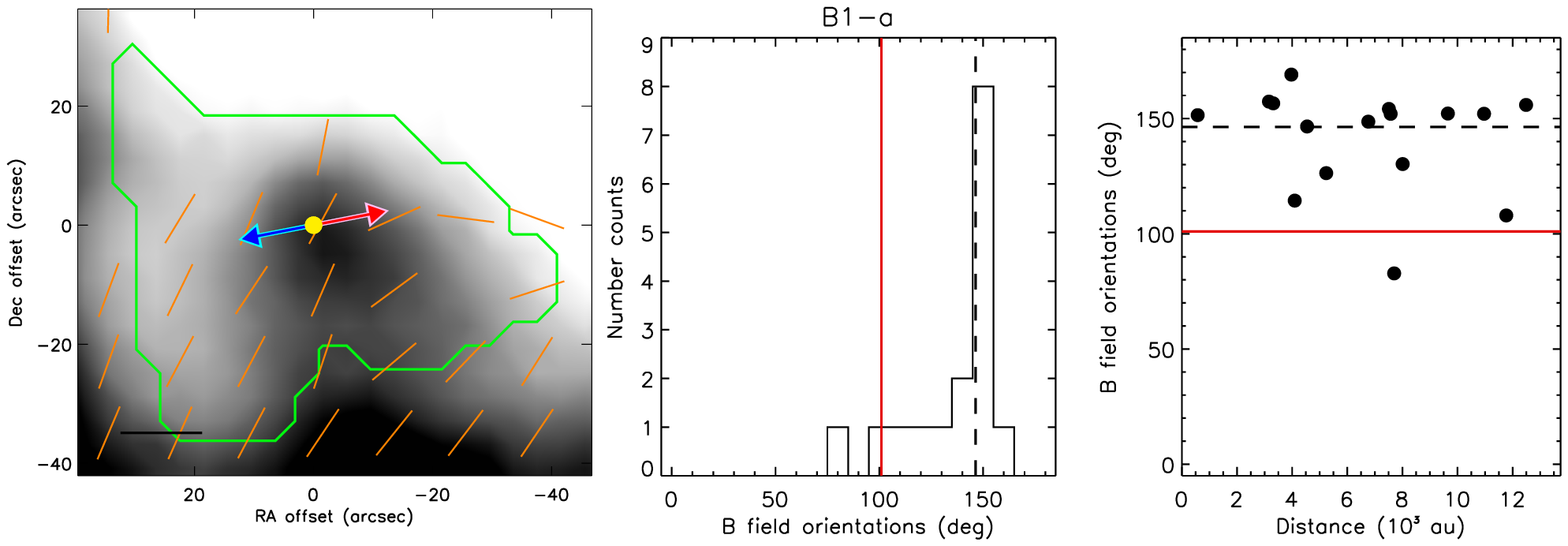}
\includegraphics[angle=90,width=0.9\textwidth]{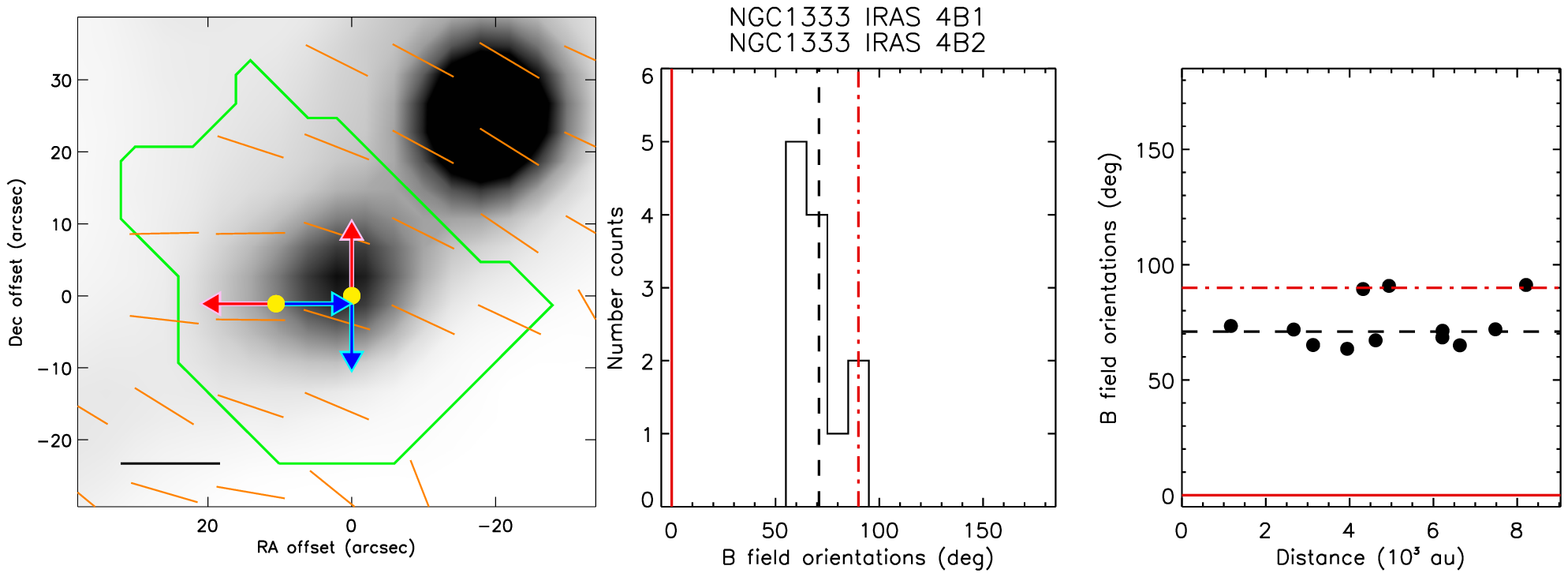}
\includegraphics[angle=90,width=0.9\textwidth]{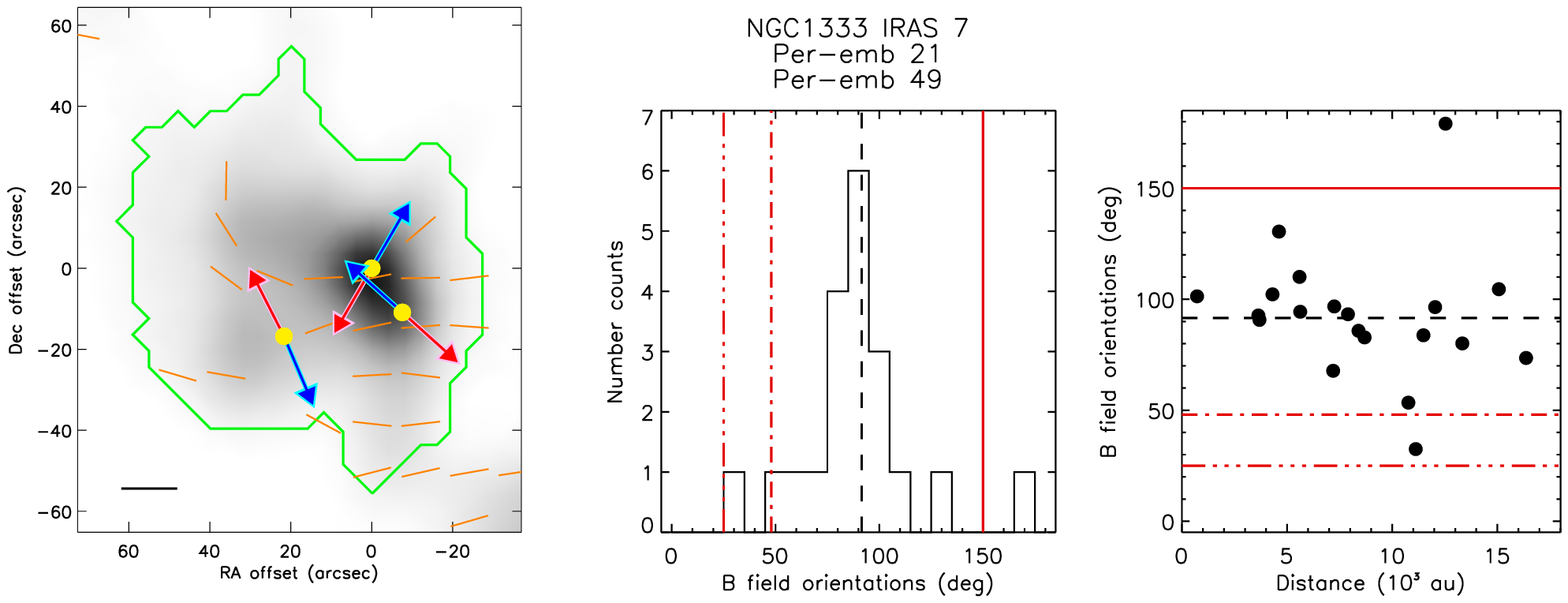}
\caption{Magnetic field orientations in B1-a (upper row), NGC~1333~IRAS~4B (middle row), and NGC~1333~IRAS~7 (lower row; including Per-emb~21 and 49) observed with the JCMT POL-2 observations. B1-a is located in the Perseus B1 region (Fig.~\ref{region1}).  NGC~1333~IRAS~4B and NGC~1333~IRAS~7 are located in the NGC~1333 region in Perseus (Fig.~\ref{regions}). Left panels show the Stokes {\it I} maps (gray scale) of the dense cores, where the sample protostars are embedded, and green contours delineate the area of the dense cores identified by {\it clumpfind}. Yellow dots show the locations of the protostars, and blue and red arrows present the orientations of their blue- and redshifted outflows. Black horizontal segments denote the spatial scale of 0.02 pc (or $\sim$4000 au). Middle panels present the distributions of the position angles of the individual magnetic field orientations detected with the JCMT POL-2 observations. Right panels present the distributions of the position angles of the magnetic field orientations as a function of distance from the protostars. If multiple protostars are present in one core, the distances between the magnetic field orientations and the protostar closest to the Stokes {\it I} intensity peak are calculated. In the middle and right panels, black dashed lines denote the mean magnetic field orientations computed from the averaged Stokes {\it Q} and {\it U} emission in the area of the dense cores, and red solid and dashed-dotted (if there are multiple outflows) lines show the position angles of the outflows. The names of the protostars are labeled in the order from western to eastern sources above the middle panel in each row. }\label{ex1}
\end{figure*}

\begin{figure*}
\centering
\includegraphics[angle=90,width=0.9\textwidth]{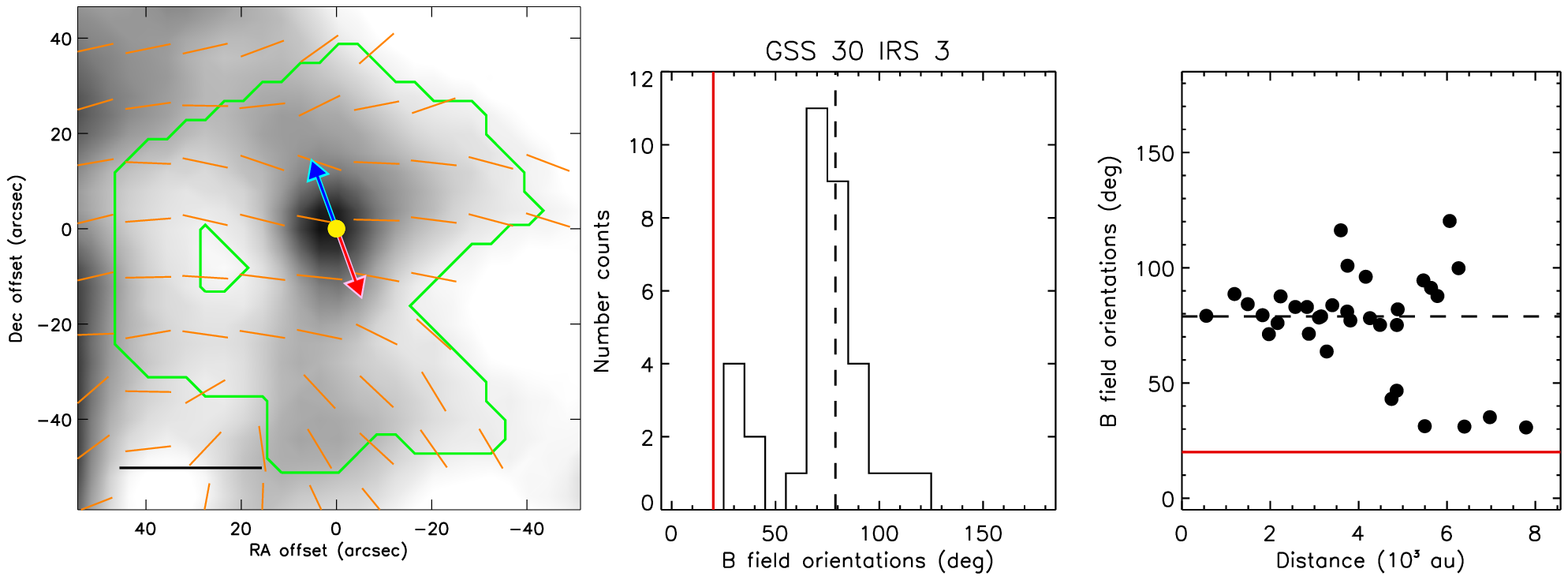}
\includegraphics[angle=90,width=0.9\textwidth]{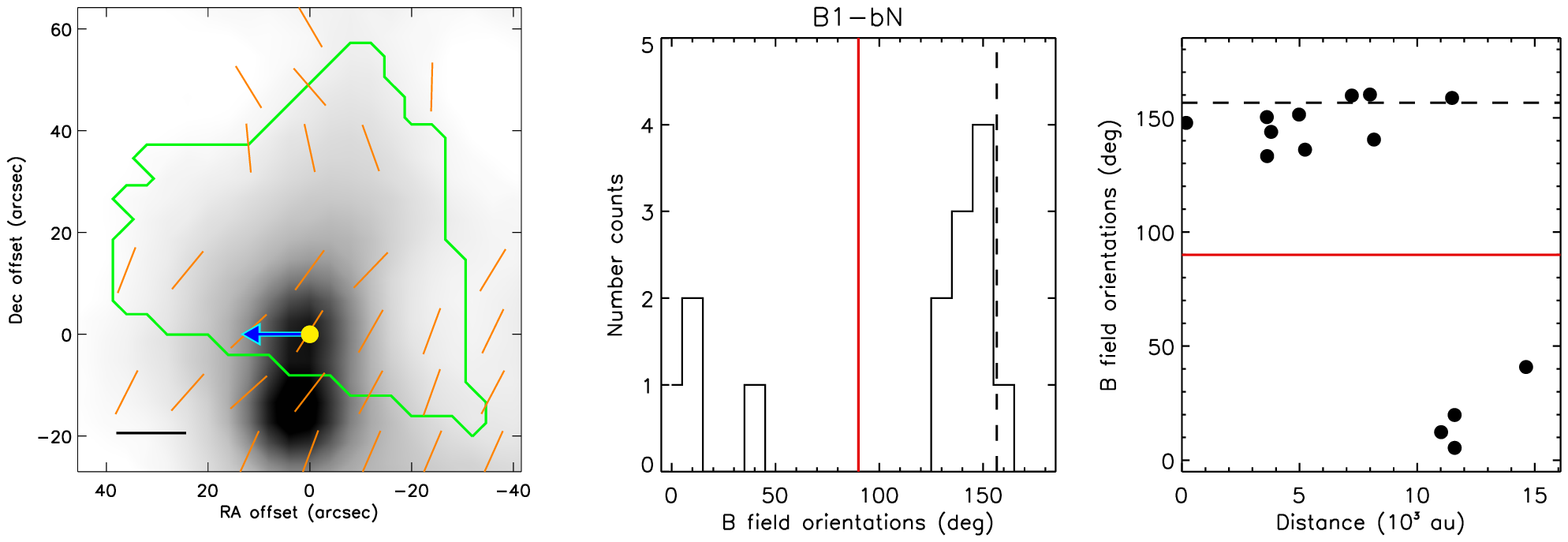}
\includegraphics[angle=90,width=0.9\textwidth]{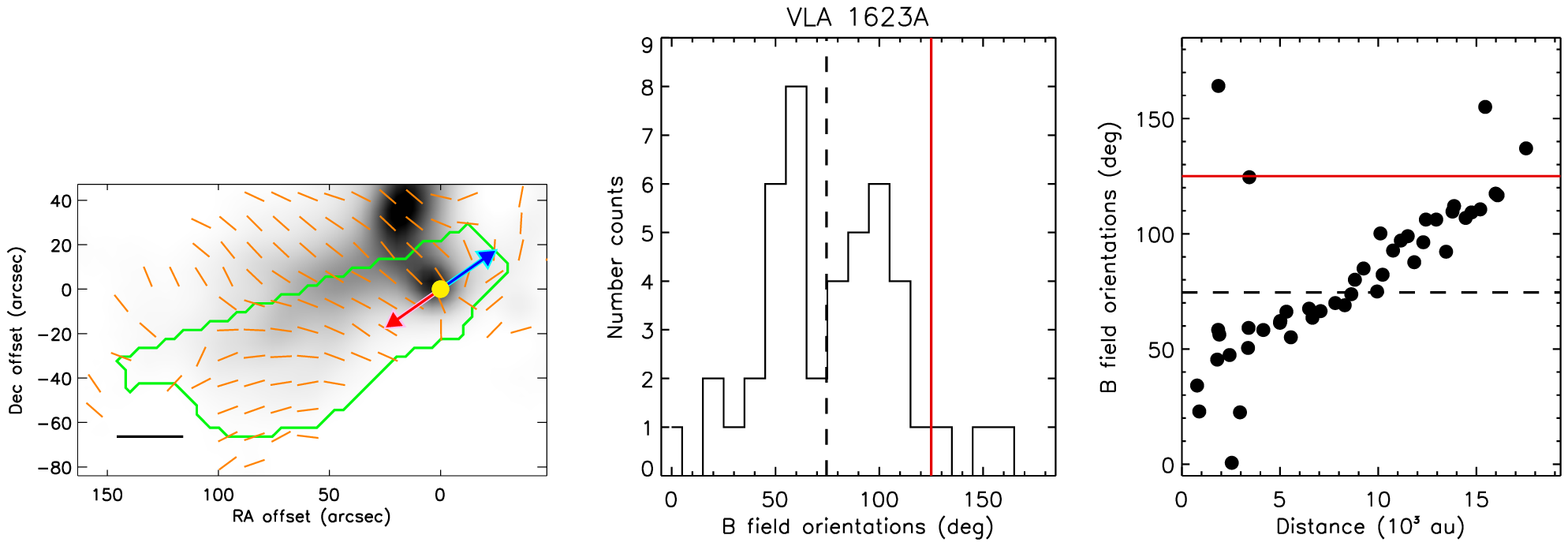}
\caption{Same as Figure \ref{ex1} but for  GSS~30~IRS~3 (top row), B1-bN (middle row), and VLA~1623A (bottom row).}\label{ex2}
\end{figure*}

Figure \ref{region1} presents the results of our core/clump identification in the Perseus B1 region as an example.   
The contours delineate the area of the dense cores or clumps associated with our sample protostars identified by {\it Clumpfind}.
The locations of the sample protostars and their outflow directions are also plotted.
The results of the core/clump identification for the remaining sample protostars are shown in Appendix \ref{clumps}.
The detected magnetic field orientations in individual dense cores and clumps associated with our sample protostars are shown in Fig.~\ref{ex1} and \ref{ex2} and Appendix \ref{allsample}.
In most of the dense cores and clumps on scales of 0.05 pc to 0.5 pc in our sample, 
the magnetic fields tend to show uniform structures, 
so representative mean orientations can be derived. 
Figure \ref{ex1} shows the maps of the detected magnetic field segments in B1-a, NGC~1333~IRAS~4B, and NGC~1333~IRAS~7 as examples.
The distributions of the magnetic field orientations inferred from the individual polarization detections in these protostellar sources are clustered and peak closely to the derived mean orientations of the magnetic fields in their dense cores (middle panels).
In addition, there is no significant dependence of the magnetic field orientations on the radial distances of the individual detections from the protostellar positions (right panels).
Thus, our results show that the magnetic field orientations do not change significantly as a function of position within these dense cores,  
although there are a few sources showing broader distributions of the magnetic field orientations in the dense cores, such as L1448~IRS~3Ba,b, L1448-mm (L1448C-S), HH211-mms, L1527, and Serpens~SMM4A (Appendix \ref{allsample}).

We note that in a few sources, the magnetic field orientations at larger distances of $>$6000--10\,000 au from the protostars are different from those closer to the protostars. 
In Figure~\ref{ex2}, we present the results of GSS~30~IRS~3, B1-bN, and VLA~1623A as examples.
In GSS~30~IRS~3, the magnetic field segments at distances larger than 6000~au show different orientations compared with those at smaller distances, 
but the number of these segments is relatively small. 
As a result, the mean magnetic field orientation in GSS~30~IRS~3 is not affected by those detections at larger distances.
The other source, B1-bN, shows a similar trend that the magnetic field orientations change by $\sim$40$\arcdeg$ at distances larger than 10\,000 au, 
and those polarization detections at the larger distances also do not affect the computed mean orientation significantly.
VLA~1623A is an extreme case in our sample.
The magnetic field orientations in VLA~1623A clearly change as a function of distance to the protostar. 
In VLA~1623A, the magnetic field close to the protostar within a radius of 6000~au is more misaligned with the outflow, 
and the magnetic field becomes more aligned with the outflow as the distance to the protostar increases. 
Nevertheless, the number of the sample sources showing clear distance dependence is small ($<$5) compared with the total sample size (Appendix \ref{allsample}). 
We have confirmed that only including the polarization detections at distances less than 6000 au or 10\,000~au from the protostars has little impact on the final statistics, and does not affect our discussions and conclusions. 

\subsection{Distribution of the misalignments}\label{numdist}
\begin{figure}
\centering
\includegraphics[width=0.48\textwidth]{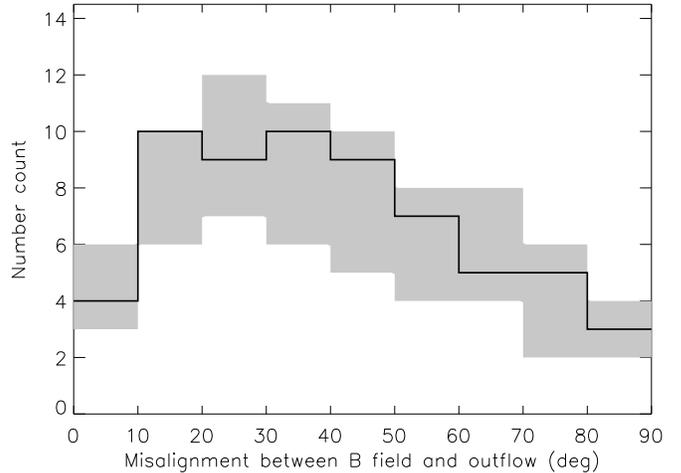}
\caption{Distribution of the misalignment angles between the outflow orientations and the mean magnetic field orientations in the dense cores and clumps  (thick step curve). Gray area presents the 1$\sigma$ uncertainty of the distribution, assuming that there is a 10$\arcdeg$ uncertainty in each measurement of the misalignment. We note that the total number count is conserved when interpreting the uncertainty of the distribution. If the number count in one bin decreases, the number counts in nearby bins increase.
}\label{counts}
\end{figure}

We computed the misalignment angle between the outflow orientation and the mean orientation of the magnetic field in the dense core or clump for each sample protostar. 
The misalignment angle is defined to be between 0$\arcdeg$ and 90$\arcdeg$.
Figure~\ref{counts} presents the distribution of the misalignment angles, where the bin size is 10$\arcdeg$.
The distribution peaks at misalignment angles between 15$\arcdeg$ and 35$\arcdeg$, 
and there are fewer protostars with their magnetic fields perfectly aligned with or perpendicular to the outflows.
The fraction of our sample sources having misalignment angles larger than 45$\arcdeg$ is 40\% (25/62) and larger than 70$\arcdeg$ is 13\% (8/62).
To investigate whether there is any potential bias in our results due to source properties and spatial resolutions, 
we compared the misalignment angles with the sizes, total fluxes, mean intensities, and mean polarization percentages of the dense cores as well as the distances to the sources, 
and there is no dependence of the misalignment angles on these parameters (Appendix \ref{bias}).

\begin{figure*}
\centering
\includegraphics[width=0.98\textwidth]{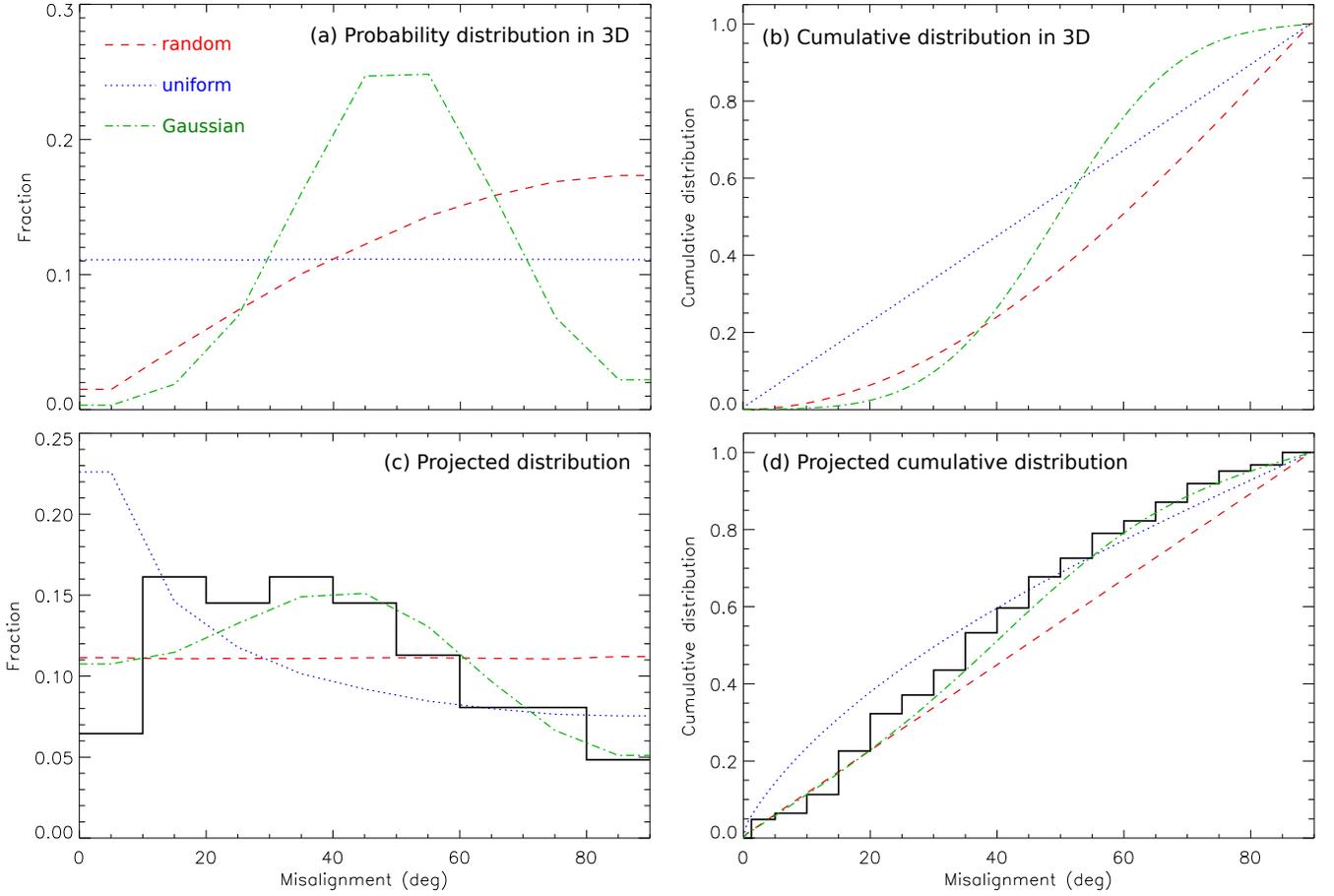}
\caption{(a) Distributions of the misalignment angles in the 3D space in our mock samples. (b) Cumulative distributions of the 3D misalignment angles in our mock samples. (c) Distributions of the misalignment angles projected onto the plane of the sky in our mock samples in comparison with the observed distribution. (d) Cumulative distributions of the projected misalignment angles in our mock samples in comparison with the observations. Blue dotted and green dashed-dotted curves show the cases where the probability distributions of the 3D misalignment is uniform and a Gaussian function, respectively. This Gaussian function has a center at 50$\arcdeg$ and a 1$\sigma$ dispersion of 15$\arcdeg$. Red dashed curves show the case where two vectors are randomly oriented with respect to each other in the 3D space. The observed distributions are shown in black solid step curves. The bin size in (a) and (c) is 10$\arcdeg$. The step size of the observed cumulative distribution in (d) is 5$\arcdeg$.
}\label{mock_counts}
\end{figure*} 

Due to projection effects, 
two misaligned vectors in three-dimensional (3D) space have a higher probability to appear more aligned on the plane of the sky. 
Thus, to study the intrinsic distribution of the misalignment in the 3D space, 
we generated mock samples of pairs of vectors assuming different probability distributions of the 3D misalignment. 
Then, we randomly defined a line-of-sight direction and projected these pairs of vectors on the assumed plane of the sky to compare with the observations.
We assumed three different probability distributions of the 3D misalignment, 
(1) two vectors are randomly orientated with respect to each other, 
(2) the misalignment between two vectors has uniform probability from 0$\arcdeg$ to 90$\arcdeg$, 
and (3) the probability distribution of the misalignment is a Gaussian function. 
The probability distributions of the 3D misalignment for these scenarios are shown in Fig.~\ref{mock_counts}a.
In the case of two randomly oriented vectors, there is a higher probability to have an orthogonal configuration in the 3D space.
Figure~\ref{mock_counts}b presents the cumulative distributions of the 3D misalignment. 
Then we projected these distributions onto the plane of the sky (Fig.~\ref{mock_counts}c).   

If two vectors are randomly oriented with respect to each other in the 3D space (red dashed curves in Fig.~\ref{mock_counts}), 
the distribution of the projected misalignment angles is flat (when projected onto the plane of the sky), which is different from the observed distribution.
We performed a Kolmogorov-Smirnov (K-S) test on these mock and observed samples. 
The probability that the two are drawn from the same distribution is 20\%.
For the mock sample with a uniform probability distribution of the 3D misalignment (blue dotted curves in Fig.~\ref{mock_counts}), 
after the projection, the distribution of the misalignment is not uniform,   
and there are more sources showing smaller misalignment angles.  
The K-S test on these mock and observed samples suggests that the probability that the two are drawn from the same distribution is $<$1\%.
For the mock samples with the Gaussian probability distribution of the 3D misalignment, 
we generated several different samples by varying the center $\theta_{\rm c}$ and dispersion $\delta_\theta$ of the Gaussian probability distributions. 
Then, we projected these distributions of the 3D misalignments and performed the K-S tests. 
We found that the probability of the mock and observed samples drawn from the same distribution is higher than 90\% when $\theta_{\rm c}$ is close to 50$\arcdeg$ and $\delta_\theta$ is close to 15$\arcdeg$ (green dashed-dotted curves in Fig.~\ref{mock_counts}).
Therefore, our results show that the 3D misalignment angles between the outflows and magnetic fields in our sample sources are not uniformly distributed. 
Our results could suggest that the outflows tend to be misaligned with the mean magnetic field orientations in the associated dense cores or clumps by 50$\arcdeg$$\pm$15$\arcdeg$ and are less likely randomly oriented with respect to the magnetic fields in the 3D space.
Nevertheless, our results do not rule out the possibility of the random orientations of the outflows and the magnetic fields.
A similar trend that outflows may have preferred orientations with respect to the magnetic fields has also been suggested in the W43-MM1 high-mass star-forming region  \citep{Arce20}, 
although the possibility of random orientations of outflows and magnetic fields is also not ruled out in that study.

\subsection{Dependence of the misalignments on T$_{\rm bol}$ and spatial scale}

\begin{figure}
\centering
\includegraphics[width=0.5\textwidth]{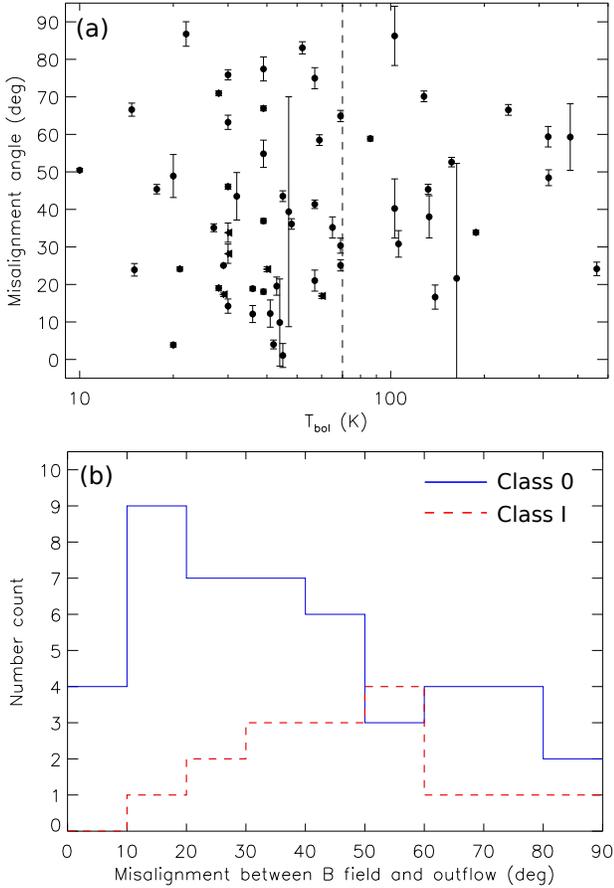}
\caption{(a) Angles between the outflow orientations and the mean orientations of the large-scale magnetic field in the dense cores and clumps as a function of bolometric temperature ($T_{\rm bol}$) of our sample protostars. A vertical dashed line denotes $T_{\rm bol}$ of 70 K, which has been adopted to classify Class 0 and I protostars \citep{Chen95}. Error bars present the uncertainties in the mean magnetic field orientations from the error propagation of the uncertainties of the individual polarization detections. For several sources, the error bars are smaller than the symbol size. There is an additional uncertainty in the misalignment angles due to the uncertainty in the outflow orientations, which is typically 10$\arcdeg$. If $T_{\rm bol}$ of a protostar in our sample only has an upper limit (Table 1), then that upper limit is plotted as a left-facing triangle. (b) Distributions of the misalignment angles of the Class 0 (blue step curve) and I (red dashed step curve) protostars in our sample.}\label{tbol}
\end{figure} 

\begin{figure}
\centering
\includegraphics[width=0.4\textwidth]{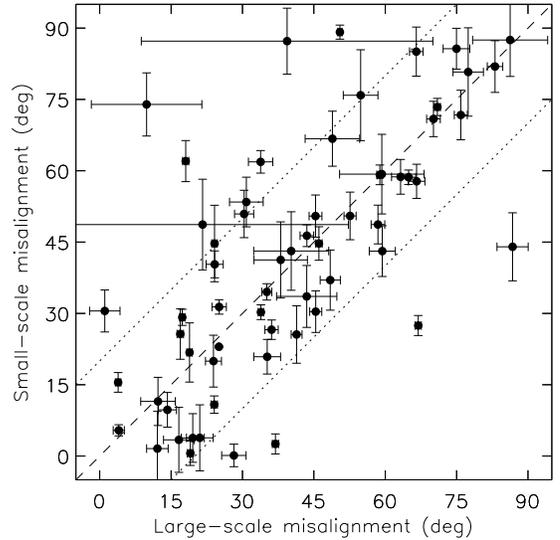}
\caption{Comparison of the misalignment between the outflows and the magnetic field orientations on two different scales. The vertical axis shows the angles between the outflows and the small-scale (1000--3000 au) magnetic fields, and the horizontal axis shows the angles between the outflows and the large-scale (0.05--0.5 pc or 10$^4$--10$^5$ au) magnetic fields. Two dotted lines denote angle difference of $\pm$20$\arcdeg$. Horizontal error bars present the uncertainties in the mean magnetic field orientations from the error propagation of the uncertainties of the individual polarization detections. For several sources, the horizontal error bars are smaller than the symbol size. Vertical error bars are the uncertainties in the polarization angles of the detections closest to the protostars.}\label{scale}
\end{figure} 

Figure \ref{tbol}a compares the misalignment angles with the bolometric temperatures ($T_{\rm bol}$) of our sample protostars. 
$T_{\rm bol}$ can be an evolutionary indicator \citep[e.g.,][]{Chen95}, 
although it also depends on the inclination and structures of the central disk and inner envelope in a protostellar source \citep{Young05, Crapsi08}.
Our results show that there is no clear dependence of the misalignment on the evolution as measured by $T_{\rm bol}$. 
For any given range of $T_{\rm bol}$, there is a wide range of misalignment angles. 
The correlation coefficient between $T_{\rm bol}$ and the misalignment angles is computed to be 0.1.
Nevertheless, we note that the Class 0 and I sources may have different distributions of the misalignments.
The misalignment angles in the Class I sources are clustered in a smaller range from 20$\arcdeg$ to 60$\arcdeg$ (Fig.~\ref{tbol}b).
However, in our sample, the number of the Class I sources is a factor of three fewer than the Class 0 sources.
A more uniform sample is needed to examine any possible difference in the distributions of the misalignments between the Class 0 and I sources.

We also compared the misalignments on small and large scales (Fig.~\ref{scale}).
For each protostar, we computed the small-scale misalignment from the single polarization detection closest to the protostellar position. 
Considering the resolution of the JCMT POL-2 observations of 14$\arcsec$, the magnetic field orientation inferred from the detection closest to the protostar could represent the averaged magnetic field structure in the protostellar envelopes on a scale of 1000--3000 au at the distances to our sample protostars.
This spatial scale is also comparable to those probed by interferometric polarimetric observations with CARMA and SMA \citep[e.g.,][]{Hull13,Hull14,Galametz18}.
As shown in previous studies which compare single-dish and interferometric results \citep[e.g.,][]{Hull14,Yen19,Doi20}, 
single-dish measurements at stellar positions can generally present the mean orientations of magnetic field on small scales observed by interferometers.
In 17 of our sample sources, the mean magnetic field orientations on a 1000 au scale were also measured with the CARMA, SMA, and/or ALMA observations \citep{Hull14,Galametz18,Sadavoy19}.
We compared our JCMT POL-2 and those interferometric results.
Our measured small-scale misalignments are indeed correlated with the interferometric measurements (Appendix \ref{int}).

In Figure \ref{scale}, most of the data points are scattered around the diagonal dashed line, suggesting that in most cases, the difference in the magnetic field orientations on the large- and small-scales is approximately 10$\arcdeg$--20$\arcdeg$ or less. 
The orientations of the large- and small-scale magnetic fields in 48\% (30/62) of the sample sources are consistent within 10$\arcdeg$ and 78\% (47/62) within 20$\arcdeg$.
The comparison between the large-scale misalignments observed by JCMT and the small-scale misalignments observed by the interferometers also shows the same trend as above that the difference between large- and small-scale magnetic fields is mostly 10$\arcdeg$--20$\arcdeg$ (Appendix \ref{int}).
In addition, in Figure~\ref{scale}, there are several data points distributed in the upper left half of the plot, but only a few in the lower-right half.
In 11 (18\%) of our sample protostars, the misalignment angles on the small scale increase by more than 20$\arcdeg$ compared to that on the large scale, 
and only four protostars (6\%) show the opposite. 
Thus, our results also suggest that the degrees of misalignment on the small scale tend to remain the same or become larger than those on the large scale in our sample. 

\section{Discussion}

\subsection{Comparison with interferometric results}\label{inter}
The CARMA observations of $\sim$20--30 low-mass protostars show that the magnetic fields in the protostellar envelopes on a scale of a few hundred to a few thousand au are randomly oriented with respect to the outflows \citep{Hull13,Hull14}.
The SMA observations of 12 protostars suggested a bimodal distribution of the misalignment angles, where the magnetic fields in protostellar envelopes are either aligned with or perpendicular to the outflows \citep{Galametz18}. 
Our JCMT POL-2 results show a number distribution of the misalignment angles different from the random or bimodal distributions. 
In these JCMT POL-2 data of 62 protostars, more than half of the sample sources have their magnetic fields misaligned with the outflows by 15$\arcdeg$--45$\arcdeg$ on the plane of the sky.
The different number distributions of misalignment found by the JCMT POL-2 and interferometric observations could be due to the different spatial scales probed by these observations. 
The large-scale magnetic fields observed with JCMT might have still preserved the initial morphologies.
On the other hand, 
as the infalling and rotational velocities increase with decreasing radii in the protostellar sources, 
the small-scale magnetic fields could be shaped by the gas motions, as suggested by the interferometric studies \citep[][]{Hull14, Galametz18}.
In addition, several interferometric observations at high angular resolutions, which well resolved the magnetic field structures in the protostellar envelopes on a scale of hundreds of au, revealed tangled, pinched and/or wrapped magnetic fields \citep{Girart06, Stephens13, Rao14, Hull17a, Hull17, Hull20, Cox18, Maury18, Sadavoy18, LeGouellec19, Lee19, Sadavoy19, Kwon19, Ko19, Yen19, Yen20}. 
Our JCMT POL-2 results also show that in 11 of our sample sources, the magnetic field orientations inferred from the detections closest to the protostars are more misaligned with the outflows than the magnetic field measured on larger scales, 
and there are only four sources showing the opposite trend. 
These results demonstrate the changes in the magnetic field structures from the large to small scales, which are possibly due to the dynamics of collapsing and rotating dense cores. 

\subsection{Implications for core formation}\label{sim_cores}
Given the assumption that the outflow directions trace the directions of the angular momenta of the dense cores in our sample sources\footnote{We note that in numerical simulations with turbulence, rotation in dense cores could be non-uniform \citep[e.g.,][]{Dib10,Zhang18,Verliat20}. In this case, the outflow direction may not represent the direction of the net angular momentum of an entire dense core but is related to the angular momentum of the material that has been accreted to form the central star-disk system.} \citep[e.g.][]{Blandford82, Pudritz83,Ciardi10,Hirano20}, 
our JCMT POL-2 results could suggest that the angular momenta of the dense cores tend to be misaligned with the magnetic fields by 50$\arcdeg$$\pm$15$\arcdeg$ if the distribution of the angles between the magnetic fields and the outflows in 3D space is a Gaussian function, 
although we do not rule out the possibility that the outflows are randomly oriented with respect to the magnetic fields (Section~\ref{numdist}).
In addition, several dense cores in our sample harbor multiple protostars with their outflows oriented in different directions, 
which could hint at non-uniform rotation in these dense cores.
In the classical picture of core formation without consideration of turbulence,
the angular momentum is expected to align with the magnetic field because the efficiency of magnetic braking is higher when the angular momentum and the magnetic field are perpendicular \citep[e.g.,][]{Mouschovias79}.
Our results do not support this classical picture. 

\begin{figure*}
\centering
\includegraphics[width=0.98\textwidth]{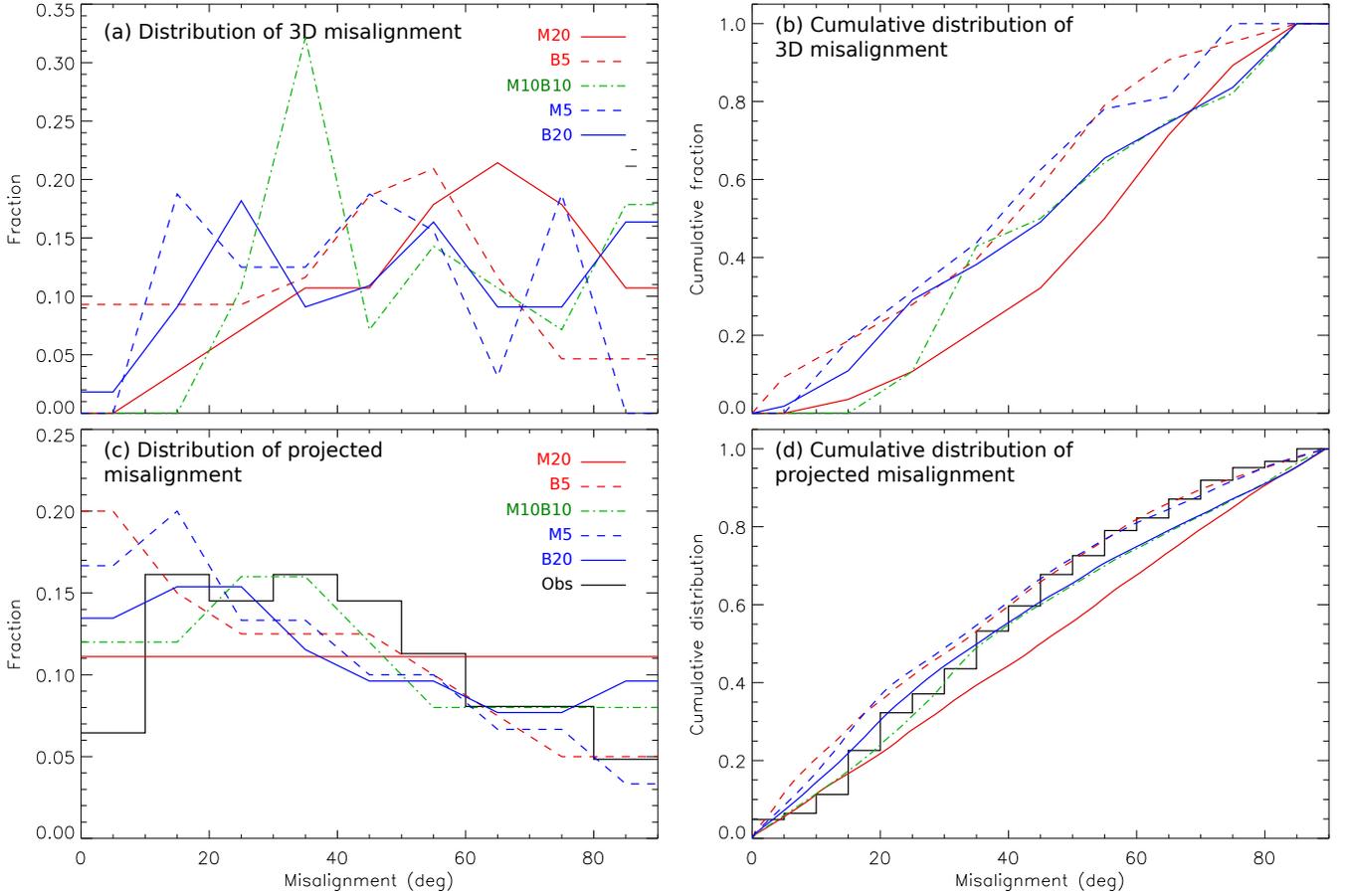}
\caption{(a) Distributions and (b) cumulative distributions of the misalignment angles between the magnetic fields and the angular momenta in the simulated dense cores in the 3D space \citep{Chen18}. Different lines with different colors show the distributions from different simulations (see Section \ref{sim_cores}). Red solid and dashed lines present the simulations with dominant turbulence, and blue solid and dashed lines the simulations with dominant magnetic field. The green dashed-dotted line presents the simulation with moderate magnetic field and turbulence. (c) and (d) are the same as (a) and (b) but after projecting the misalignment angles onto the plane of the sky. In (c) and (d), the black step curves show the observed distributions of the misalignment angles in our sample sources. The bin size in (a) and (c) is 10$\arcdeg$. The step size of the observed cumulative distribution in (d) is 5$\arcdeg$.}\label{sim_count}
\end{figure*}

Properties of the magnetic field and angular momentum of dense cores formed in magnetized and turbulent molecular clouds have been often studied with numerical simulations \citep[e.g.,][]{Burkert00, Gammie03, Li04, Dib10, Chen14, Chen15,LeeHull17,Kuznetsova20}. 
\citet{Chen18} compared the distributions of the misalignment between the angular momenta and the magnetic fields in the dense cores formed in their 3D turbulent MHD simulations of converging flows with different degrees of the magnetization and turbulence. 
Their simulations can be roughly classified into three groups, dominant magnetic field  (M5 and B20), dominant turbulence  (B5 and M20), and moderate magnetic field and turbulence (M10B10), where the labels in the parentheses are those adopted in \citet{Chen18} to represent different simulations. 
Figure \ref{sim_count}a presents the distributions of the 3D misalignment angles between the magnetic fields and the angular momenta in the dense cores in these simulations. 
In the simulations, the dense cores may not have uniform rotation and magnetic field structures. 
These misalignment angles are the angles between the mean magnetic field orientations and directions of the net angular momentum of the simulated dense cores.
In the simulations with dominant turbulence (B5 and M20), the distributions peak at large misalignment angles of 50$\arcdeg$--70$\arcdeg$.
In contrast, the simulations with the dominant magnetic field (M5 and B20) show flatter distributions of the misalignment angles without a dominant peak. 
The distribution of the misalignment angles in the simulation with the moderate magnetic field and turbulence (M10B10) is along this trend.
It has a peak at a misalignment angle of 30$\arcdeg$ on top of a flat distribution. 
As pointed out by \citet{Chen18}, 
the misalignment angle between the magnetic field and the angular momentum tends to increase as turbulence becomes more dominant. 
This trend is also seen in other simulations \citep[e.g.,][]{Joos13}.
Our observational results suggest that the distribution of the misalignment between the magnetic fields and the outflows on a 0.05--0.5 pc scale is not flat (Fig.~\ref{mock_counts}). 
This is different from the distributions in the simulations with dominant magnetic field (B20 and M5). 
In addition, our observational results could suggest a predominantly large number of protostars with their magnetic fields misaligned with the outflows by 50$\arcdeg$$\pm$15$\arcdeg$ on a 0.05--0.5 pc scale.
This observed distribution is more similar to those from the simulations of M10B10 and B5, which also peak at misalignment angles of 30$\arcdeg$--60$\arcdeg$, 
but differs from that of the simulation with the strongest turbulence (M20).
In the most turbulent simulation (M20), there is a larger fraction of cores ($\sim$45\%) with misalignment angles larger than 60$\arcdeg$ compared to the observations ($\sim$25\%).
Therefore, the comparison between our observational results and these numerical simulations suggests that the magnetic field is unlikely dominant during the core formation, 
and our results also hint at significant turbulence in the environment where these dense cores form. 

The distributions presented in Fig.~\ref{sim_count}a are the misalignment angles measured in the 3D space. 
To make a more direct comparison with the observational data, 
we projected these angles on an assumed plane of the sky, 
similar to the analysis in Section \ref{numdist}. 
We repeated this process 50\,000 times. 
Each time we randomly selected a direction of the line of sight, 
and computed the number distributions of the projected misalignment angles. 
All the misalignment angles of individual dense cores in the simulations were treated independently, 
and we did not consider any possible correlations between the directions of the magnetic fields or angular momenta of different cores, 
although they formed in the same molecular clouds.
Nevertheless, this simplification is appropriate because our sample protostars are located in several different molecular clouds. 
Finally, 
we computed number counts for different misalignment angles from these 50\,000 iterations. 
The distributions of the misalignment angles of these simulated dense cores projected onto the plane of the sky are plotted in Fig.~\ref{sim_count}c, 
and the cumulative distributions are plotted in Fig.~\ref{sim_count}d. 

The most turbulent simulation (M20), which has a distribution of the 3D misalignment skewed toward larger angles, shows a flat distribution after the projection, 
and it is different from the observed distribution (Fig.~\ref{sim_count}c). 
In contrast, 
the simulations with dominant magnetic field (M5 and B20), which have flatter distributions of the 3D misalignment angles, show distributions of the misalignment skewed toward smaller angles and peaking at $<$20$\arcdeg$ after the projection. 
The simulation B5, where turbulence dominates over the magnetic field, also shows a similar distribution after the projection, 
because there is a significant fraction of dense cores with well aligned magnetic fields and angular momenta in this simulation (Fig.~\ref{sim_count}a). 
Compared to the observations, these three simulations have more sources showing projected misalignment angles smaller than 20$\arcdeg$.
Among these five simulations, the distribution of the projected misalignment angles from M10B10 is the one most similar to that observed. 
The K-S test suggests a probability of 85\% that the observed distribution and the distribution in M10B10 are drawn from the same distribution, 
while for all the other models, the possibilities are below 30\% (27\% for B20, 3\% for M5, 2\% for B5 and 16\% for M20).
Nevertheless, we note that the numbers of dense cores in these simulations (55 in B20, 32 in M5, 43 in B5, and 28 in M20 and M10B10) may not be large enough to unveil underlying distributions of the misalignment in the simulated environments. 
Therefore, this comparison with the simulations only qualitatively demonstrates that to reproduce the observed distribution of the misalignment, there are likely only few protostellar sources having very small or large misalignment angles, and there is a dominant fraction of sources with intermediate misalignment angles, like $\sim$30$\arcdeg$ in the case of M10B10.
Overall, the observed distribution is more like the one in the simulation with the moderate magnetic field and turbulence (M10B10), 
and deviates more from those from the simulations with a dominant magnetic field or dominant turbulence (B20, M5, B5, and M20), 
but we do not rule out the scenarios of dominant magnetic field or turbulence in the environments of core formation from the comparison with these simulations.

\subsection{Implication for disk formation}
Ideal MHD simulations of a non-turbulent, rotating and collapsing dense core with its magnetic field aligned with the rotational axis show that magnetic braking is very efficient to transfer the angular momentum of collapsing material outward and suppress the formation of a Keplerian disk larger than 10 au around the central protostar \citep[e.g.,][]{Allen03,Mellon08}.
This is the so-called ``magnetic-braking catastrophe''.
Misalignment between the magnetic field and rotational axis in a dense core has been proposed to be a mechanism to reduce the efficiency of magnetic braking and to enable formation of a large Keplerian disk with a size of tens of au \citep[e.g.,][]{Joos12,Krumholz13}. 

Keplerian disks with sizes larger than tens of au have often been observed around protostars \citep{Williams11}. 
In the ALMA survey of the Lupus star-forming regions \citep{Ansdell18}, 
95 Class II sources were observed, 71 of them were detected in the 1.3 mm continuum, and 32 disks were resolved in the continuum. 
The resolutions of these observations are 0\farcs2--0\farcs3, corresponding to $\sim$40 au, 
so these observations are able to resolve Keplerian disks with sizes of tens of au. 
The line emission in the disks in a subset of the Lupus sample is bright, 
and the Keplerian rotation is indeed detected \citep[e.g.,][]{Ansdell18,Yen18}. 
Thus, the fraction of resolved Keplerian disks in the Lupus star-forming regions is 34\% (32/95) of the full sample and 45\% (32/71) of the detected sources. 
Similar surveys have also been conducted toward the $\rho$ Ophiuchus star-forming regions at similar spatial resolutions of $\sim$30 au \citep{Cieza19, Williams19}. 
\citet{Cieza19} presented observational results of 21 Class I protostars, 41 protostars with flat spectra, and 120 Class II protostars. 
The disks around 12, 16, and 31 of these Class I, flat spectrum, and Class II protostars were resolved in the 1.3 mm continuum, 
corresponding to fractions of resolved Keplerian disks of 57\%, 39\%, and 26\%, respectively.   
In the Orion star-forming regions, 
an ALMA survey found that 36\% (153/421) of the Class 0, I, and flat spectrum protostars have disk radii larger than 50 au in the continuum emission \citep[46\% for Class 0, 38\% for Class I, and 27\% for flat spectrum;][]{Tobin20}.
Therefore, these ALMA surveys suggest that the fractions of large Keplerian disks with sizes larger than 30--50 au around Class 0--II protostars range from approximately 30\% to 60\%. 

These fractions should be considered as lower limits. 
This is because these are shallow surveys with integration times of one or a few minutes per source, 
and hence faint large disks and large disks with steep intensity profiles may not be detected or resolved. 
In addition, the disk radii tend to be larger in the line emission than in the continuum \citep{Ansdell18,Yen18}, 
but these shallow surveys are not sensitive enough to detect line emission in most of the targeted disks. 
We note that here we only consider large disks with sizes of tens of au, 
and the fraction of protostars surrounded by a disk is much higher \citep[e.g.,][]{Haisch01,Williams11}. 
Whether the mechanism of the misalignment is able to explain the observed numbers of large Keplerian disks has been discussed in the literature assuming random alignment between the magnetic field and the rotational axis in the natal dense cores (see, e.g., \citet{Krumholz13} and \citet{Li13}). 
Here we revisit this question with our new constraint on the distribution of the misalignment angles and the observed number fractions of large Keplerian disks from the recent ALMA surveys.

The effects of the misalignment between the magnetic field and the rotational axis on disk formation have been studied with numerical simulations with ideal MHD conditions \citep{Joos12,Li13}, which exclude other possible mechanisms to reduce the efficiency of magnetic braking. 
These simulations show that when the median observed strength of the magnetic field of a mass-to-flux ratio\footnote{Mass-to-flux ratio is defined as $\frac{\Phi}{2\pi\sqrt{G}}$, where $\Phi$ is the magnetic flux and $G$ is the gravitational constant \citep{Nakano78}.} of two to three is adopted \citep{Crutcher12}, 
large Keplerian disks with sizes of tens of au only form if the initial magnetic field is almost perpendicular to the rotational axis. 
Our observational results show that fewer than 5\% of protostars have projected misalignment angles larger than 80$\arcdeg$. 
Even if we consider projection effects, 
based on our inferred Gaussian distributions of the 3D misalignment angles or the distribution in the numerical simulations M10B10 (the one most similar to our observational results; Section~\ref{sim_cores}), the fractions of protostars having misalignment angles larger than 80$\arcdeg$ are estimated to be lower than 10\%--20\% (Fig.~\ref{sim_count}). 
These numbers are approximately two to three times smaller than the lower limits of the observed fractions of large Keplerian disks with sizes larger than 30--50 au. 
We note that these MHD simulations may not be able to resolve formation of small Keplerian disks with sizes of $\lesssim$10 au. 
In our discussion of the observed fractions of the Keplerian disks, we only considered large Keplerian disks with sizes of tens of au.
Therefore, with the typical mass-to-flux ratio of two to three, the misalignment between the magnetic field and the rotational axis of a dense core cannot be the primary mechanism to enable the formation of a large Keplerian disk.

On the other hand, 
\citet{Crutcher10} suggest that the probability distribution of the mass-to-flux ratios in dense cores could be uniform with a minimum ratio of one. 
\citet{Joos12} show that if the initial mass-to-flux ratios are larger than five, large Keplerian disks form when the misalignment angle are larger than 70$\arcdeg$, 
and if the initial mass-to-flux ratios are larger than 17, large Keplerian disks always form regardless of the misalignment angles. 
Simulations with different initial density and velocity distributions by \citet{Li13} show that large Keplerian disks form when the initial mass-to-flux ratios are larger than ten and the misalignment angles are larger than 45$\arcdeg$. 
Following discussions in \citet{Krumholz13}, 
we estimate the expected number fraction of large Keplerian disks ($f_{\rm disk}$) as, 
\begin{eqnarray}
f_{\rm disk} & = & f_\lambda(5\mbox{--}10) \times f_{\rm ang}(>\!70\arcdeg) \nonumber \\ 
& + &  f_\lambda(10\mbox{--}17) \times f_{\rm ang}(>\!45\arcdeg) \nonumber \\
& + &  f_\lambda(>\!17), 
\end{eqnarray}
where $f_\lambda(5\mbox{--}10)$, $f_\lambda(10\mbox{--}17)$, and $f_\lambda(>\!17)$ are the fractions of sources having mass-to-flux ratios of 5--10, 10--17, and $>$17, respectively, and $f_{\rm ang}(>\!70\arcdeg)$ and $f_{\rm ang}(>\!45\arcdeg)$ are the fractions of sources having misalignment angles larger than 70\arcdeg and 45\arcdeg, respectively.
With a uniform distribution of the mass-to-flux ratio and our observed distributions of misalignment angles, the expected number fraction of large Keplerian disks is estimated to be $\sim$10\%, 
which is more than a factor of three lower than the observed number fraction.
If we also consider the parameters of mass-to-flux ratios of 3--5 and misalignment angles of 20\arcdeg--45\arcdeg that form sub-Keplerian disks with flat rotational profiles, which are supported by both rotation and magnetic pressure, in the MHD simulations by \citet{Joos12}, the expected number fraction increases to $\sim$20\%, 
but this fraction is still lower than the observed value. 
These estimated expected number fractions are similar to the minimal possibility estimated by \citet{Krumholz13}, 
but are lower than their maximum possibility because \citet{Krumholz13} include the parameters with which the simulations form sub-Keplerian disks or no disks in their estimation. 
On the other hand, 
observationally, large disks around protostars are often found to be Keplerian \citep[e.g.,][]{Pietu07, Simon17}. 
Thus, those cases of sub-Keplerian disks in the simulations may not represent observed disks. 
In addition, our observed distribution of the misalignment suggests a lower fraction of sources with large misalignments compared to the random distribution.

In summary, 
the numerical simulations suggest that in the limit of ideal MHD, the mass-to-flux ratio needs to be larger than five to ten together with a misalignment angle between the magnetic field and the rotational axis larger than 45$\arcdeg$ to form a Keplerian disk with a size of tens of au \citep{Joos12, Li13}. 
Observationally, the fraction of protostars having misalignment angles larger than 45$\arcdeg$ is likely $\sim$50\% (Section \ref{results}), 
and the probability of a dense core with a mass-to-flux ratio larger than five to ten is 10\%--20\% \citep{Crutcher10}. 
Therefore, with the misalignment alone, the expected fraction of large Keplerian disks with sizes of tens of au would be $<$10\%--20\%, a factor of two to three lower than the lower limits estimated from the observations. 
The combination of the observations and the MHD simulations suggests that the misalignment is unlikely a dominant mechanism to reduce the efficiency of magnetic braking and to enable formation of the large number of the observed Keplerian disks with sizes larger than 30--50 au. 
Nevertheless, our results do not exclude the possibility that the misalignment could still reduce the efficiency of magnetic braking and prompt disk formation in individual sources, such as the case in HH~211 \citep{Lee19}, 
but the number of sources with sufficiently large misalignment angles to have significant effects is small in our sample.
On the contrary, protostellar sources with well aligned magnetic fields and outflows can also exhibit significant rotational motion on a 1000 au scale, 
such as L1448~IRS~2 \citep{Yen15, Kwon19, Gaudel20}.
Thus, overall, the misalignment is unlikely a primary mechanism, 
and other mechanisms, such as non-ideal MHD effects, could play a more important role in disk formation and growth \citep{Tsukamoto17,Zhao16,Zhao18,Masson16,Matsumoto17,Lam19,Wurster19a,Wurster19b}.

\section{Summary}\label{summary}
We measure the mean orientations of the magnetic fields in the dense cores on scales of 0.05--0.5 pc associated with 62 Class 0 and I protostars using the JCMT BISTRO-1 and archival POL-2 data. 
We compared the mean magnetic field orientations with the orientations of the outflows launched from those protostars. 
The main results are summarized below.

\begin{enumerate}
\item{The observed distribution of the misalignment between the magnetic field and the outflow on the plane of the sky is not flat, but rather peaks at 15$\arcdeg$--35$\arcdeg$. There are 23\% (14/62) of the sample sources having the misalignment on the plane of the sky smaller than 20$\arcdeg$ and only 13\% (8/62) larger than 70$\arcdeg$. We have compared the measured misalignment angles with the sizes, total fluxes, mean intensities, and mean polarization percentages of the sample dense cores and clumps as well as the distances to the sample sources. The measured misalignment angles do not depend on these source properties and the spatial resolutions. Thus, we expect that there is no bias in the observed distribution of the misalignment introduced by the source properties or the non-uniform spatial resolutions in the sample.}
\item{After considering projection effects, the K-S test suggests that the observed distribution of the misalignment is different from a uniform distribution, and is also less likely (20\% probability) a random distribution of outflow and magnetic field orientations. If the distribution of the misalignment in 3D space is assumed to be a Gaussian function, the K-S tests suggest that the probability of most (68\%) of the sample sources having misalignment angles of 50$\arcdeg$$\pm$15$\arcdeg$ between the magnetic field and the outflow in the 3D space is higher than 90\%.}
\item{There is no significant correlation between the misalignment angles and bolometric temperatures of the protostars. For any given range of the bolometric temperatures, there is a wide range of the misalignment between the magnetic fields and the outflows in our sample. Thus, there is no clear sign of time evolution of the misalignment angles.}
\item{The observed distribution of the misalignment is different from the results obtained with the CARMA and SMA observations, 
which suggest random or bimodal distributions of the misalignment. 
The difference is most likely due to different spatial scales probed by the JCMT POL-2 and interferometric observations. 
The magnetic field on a 0.05--0.5 pc scale observed with the JCMT might have preserved the initial morphology, 
while the magnetic field in the protostellar envelopes on scales from hundreds to thousands of au observed with the interferometers can be already shaped by the infalling and rotational motions in the envelopes. 
As a matter of fact, the JCMT POL-2 observations, when limited to the vicinity of the protostar, show that the magnetic field orientations become more misaligned with the outflows than the large-scale magnetic field in several sample sources.
Thus, these results suggest changes in the magnetic field structures from the dense cores to the inner protostellar envelopes.}
\item{Given the assumption that the directions of the outflows trace the directions of the angular momenta of the dense cores in our sample sources, 
we compared the observed distribution of the misalignment with the turbulent MHD simulations of core formation in converging flows. 
The observed distribution is more similar to that in the simulations with moderate turbulence and magnetic field, where there are more dense cores with misalignment angles of 30\arcdeg--40$\arcdeg$ between the magnetic field and the rotational axis.
The simulations with a dominant magnetic field show flat distributions of the misalignment angles between the magnetic fields and rotational axes of dense cores, while the most turbulent simulation has a high fraction ($>$45\%) of cores having misalignment angles larger than 60$\arcdeg$.
The distributions of the misalignments in these simulations are different from our observational results. 
Therefore, our results could suggest that in the environment of the core formation, the energy density in the magnetic field is comparable to that in the turbulent velocity field of the gas, but the comparisons with these simulations do not have sufficient statistical significance to rule out the scenarios of dominant magnetic field or turbulence.}
\item{Misalignment between the magnetic field and rotational axis in a dense core has been proposed to be a mechanism to reduce the efficiency of magnetic braking and enable the formation of a large Keplerian disk with a size of tens of au around a protostar. MHD simulations show that when the mass-to-flux ratio is larger than 5--10 and the misalignment angle between the magnetic field and rotational axis is larger than 45$\arcdeg$ in a dense core, a large Keplerian disk with a size of tens of au can form even in the ideal MHD limit. Based on our observed distribution of the misalignment between the magnetic field and rotational axis in our sample and the assumption of a uniform probability distribution of the mass-to-flux ratios, the expected number fraction of protostars surrounded by a large Keplerian disk with a size of tens of au is $<$10\%--20\%. This is a factor of two to three lower than the number fractions ($>$30\%--60\%) of large disks with sizes larger than 30--50 au. around protostars in nearby star-forming regions observed with the recent ALMA surveys. Consequently, our results suggest that the misalignment is not the primary mechanism to reduce the efficiency of magnetic braking and to enable formation of the observed number of large Keplerian disks with sizes larger than 30--50 au.}
\end{enumerate}

\begin{acknowledgements} 
We thank Che-Yu Chen for fruitful discussions and providing their simulation results for us to compare with the observations. 
The James Clerk Maxwell Telescope is operated by the East Asian Observatory on behalf of The National Astronomical Observatory of Japan; Academia Sinica Institute of Astronomy and Astrophysics in Taiwan; the Korea Astronomy and Space Science Institute; Center for Astronomical Mega-Science (as well as the National Key R\&D Program of China with No.~2017YFA0402700). Additional funding support is provided by the Science and Technology Facilities Council of the United Kingdom and participating universities in the United Kingdom, Canada and Ireland.
Additional funds for the construction of SCUBA-2 were provided by the Canada Foundation for Innovation. 
H.-W.Y.\ acknowledges support from Ministry of Science and Technology (MOST) grant MOST 108-2112-M-001-003-MY2 in Taiwan.
P.M.K.\ acknowledges support from MOST 108-2112-M-001-012 and MOST 109-2112-M-001-022 in Taiwan, and from an Academia Sinica Career Development Award.
C.L.H.H.\ acknowledges the support of the NAOJ Fellowship and JSPS KAKENHI grants 18K13586 and 20K14527.
W.K.\ was supported by the New Faculty Startup Fund from Seoul National University and by the Basic Science Research Program through the National Research Foundation of Korea (NRF-2016R1C1B2013642).
E.J.C.\ was supported by the National Research Foundation of Korea(NRF) grant funded by the Korea government(MSIT) (No. NRF-2019R1I1A1A01042480).
D.J.\ is supported by the National Research Council of Canada and by a Natural Sciences and Engineering Research Council of Canada (NSERC) Discovery Grant. 
C.W.L.\ is supported by Basic Science Research Program through the National Research Foundation of Korea (NRF) funded by the Ministry of Education, Science and Technology (NRF-2019R1A2C1010851).
A.S.\ acknowledges support from the NSF through grant AST-1715876.
M.T.\ is supported by JSPS KAKENHI grant Nos.~18H05442, 15H02063, and 22000005.
This research is partially supported by Grants-in-Aid for Scientific Researches from the Japan Society for Promotion of Science (KAKENHI 19H0193810).
\end{acknowledgements} 


\begin{appendix}
\section{Core and clumps identification}\label{clumps}
\begin{figure*}
\centering
\includegraphics[width=0.45\textwidth]{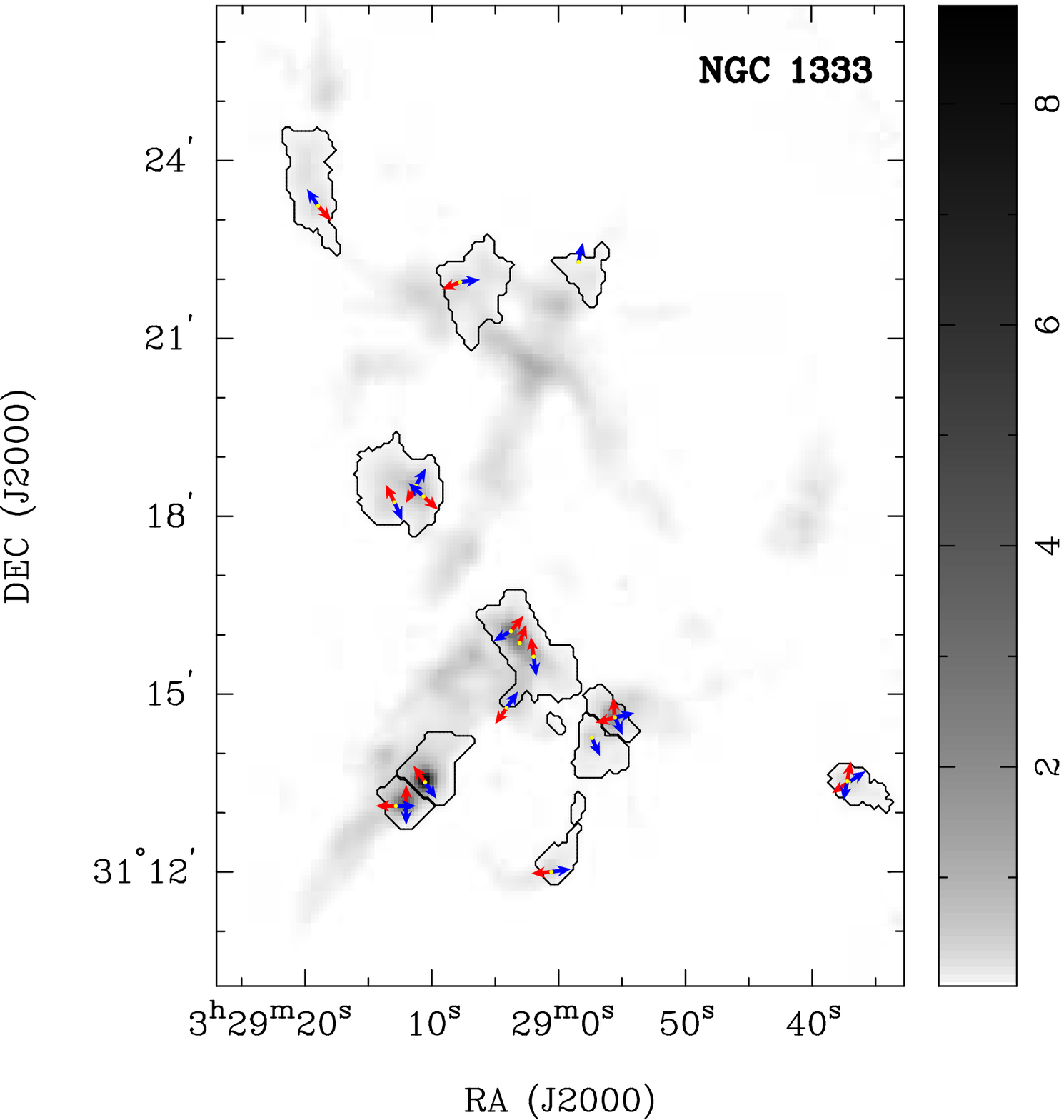}
\includegraphics[width=0.49\textwidth]{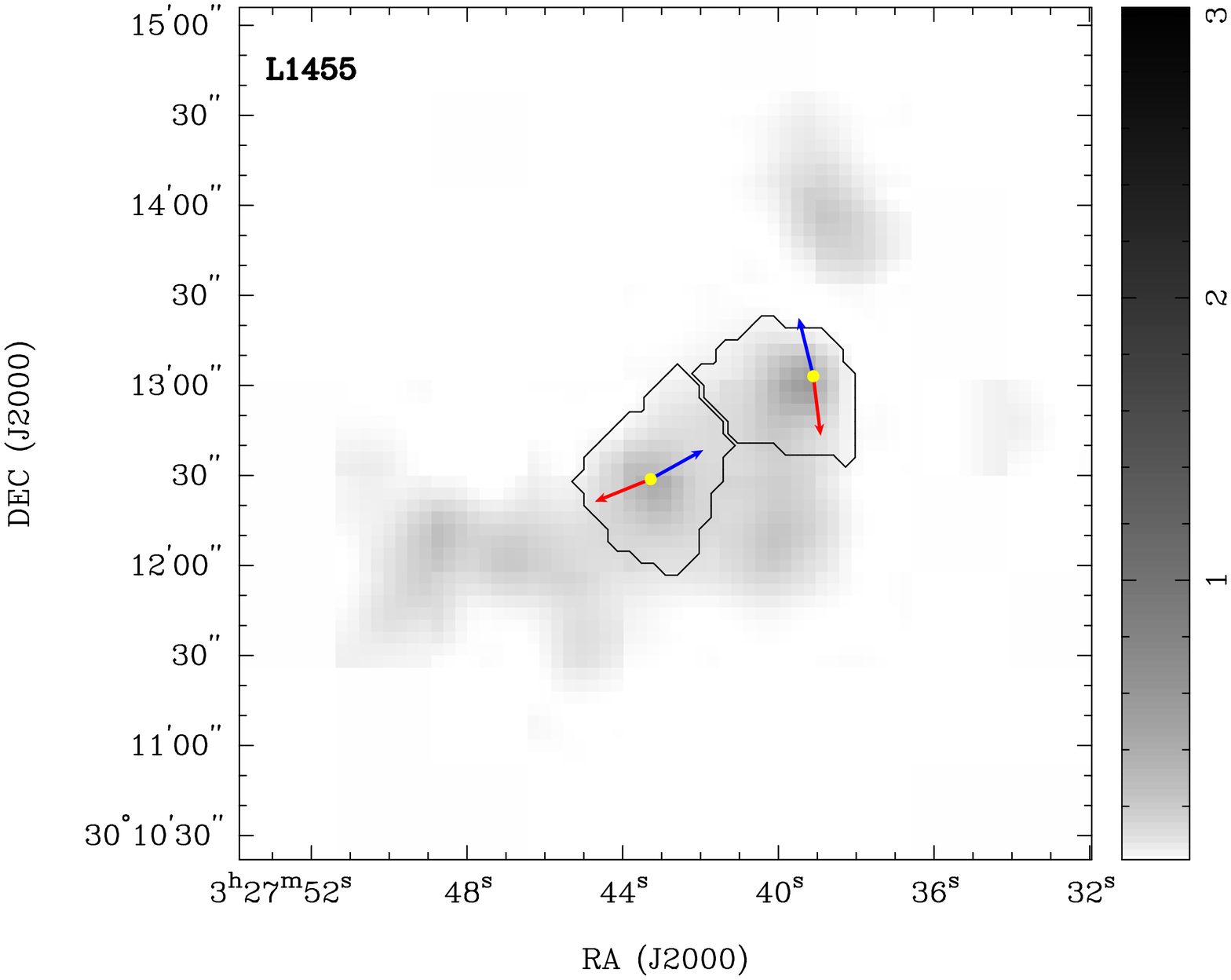}
\includegraphics[width=0.49\textwidth]{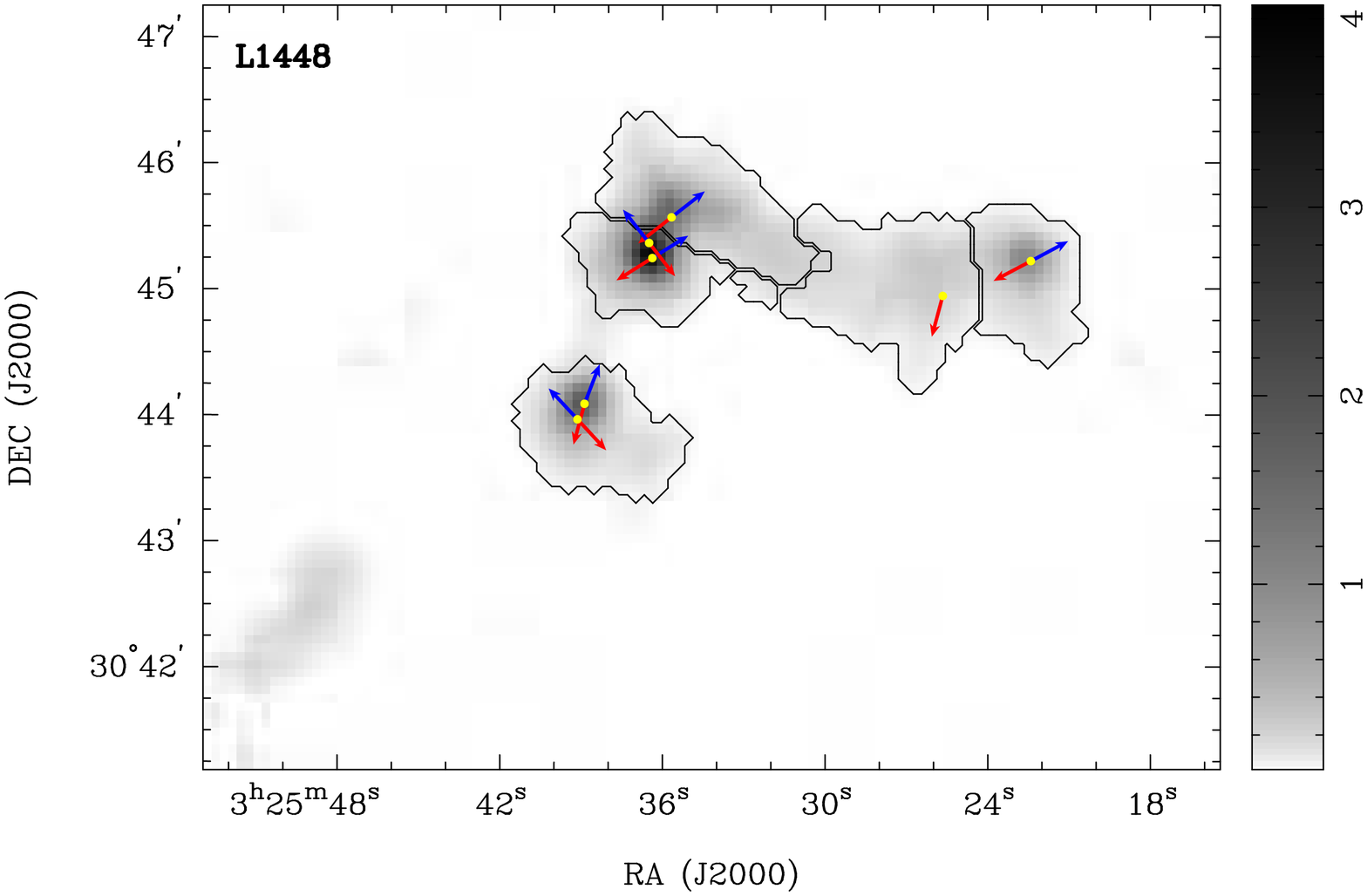}
\includegraphics[width=0.49\textwidth]{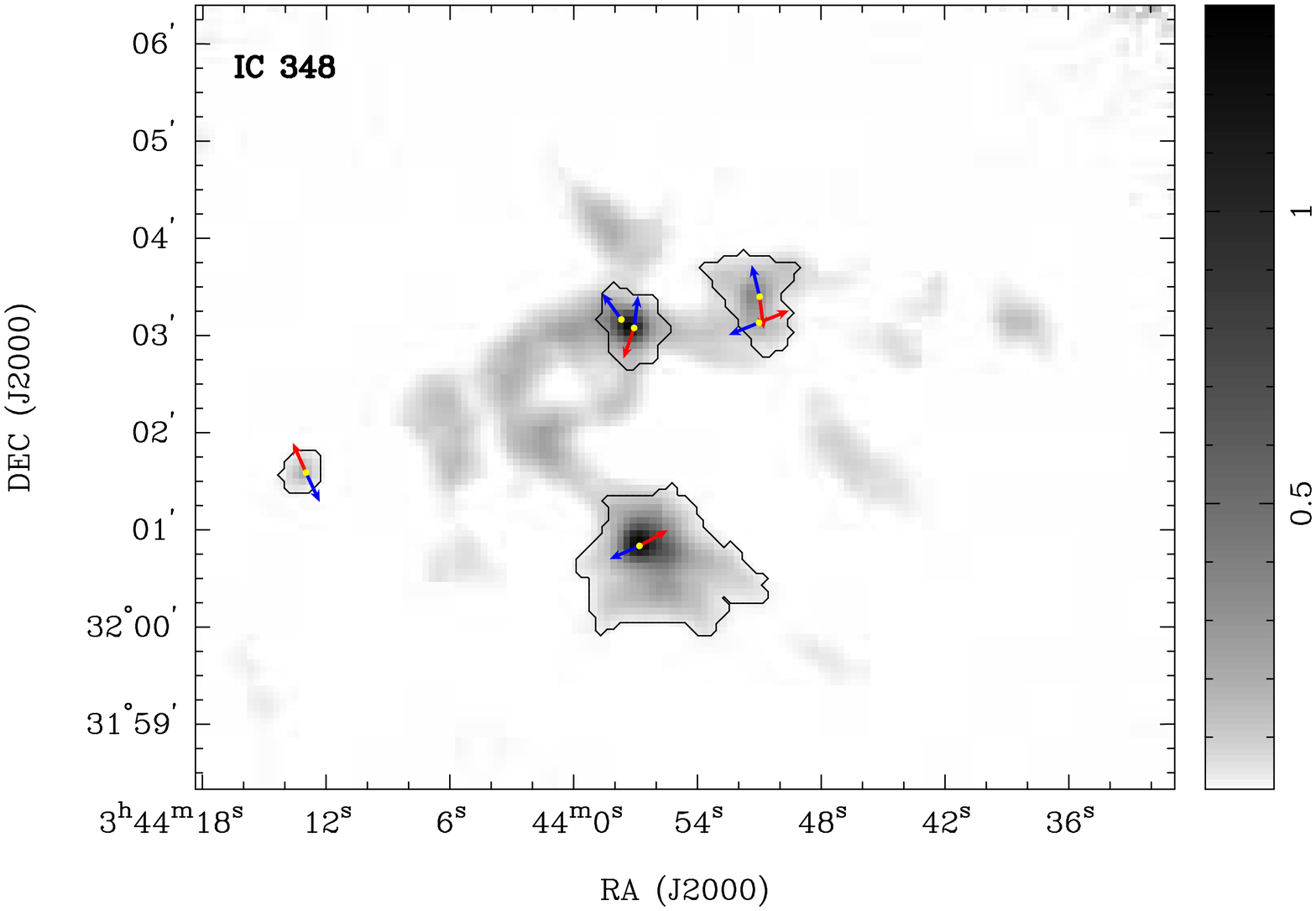}
\caption{Same as Figure~\ref{region1} but for the remaining regions in our sample. The name of the region is labelled in the upper right or left corner in each panel.}\label{regions}
\end{figure*}

Figure \ref{regions} presents the Stokes {\it I} maps and the area of individual cores and clumps identified using {\it Clumpfind}.

\begin{figure*}
\centering
\includegraphics[width=0.49\textwidth]{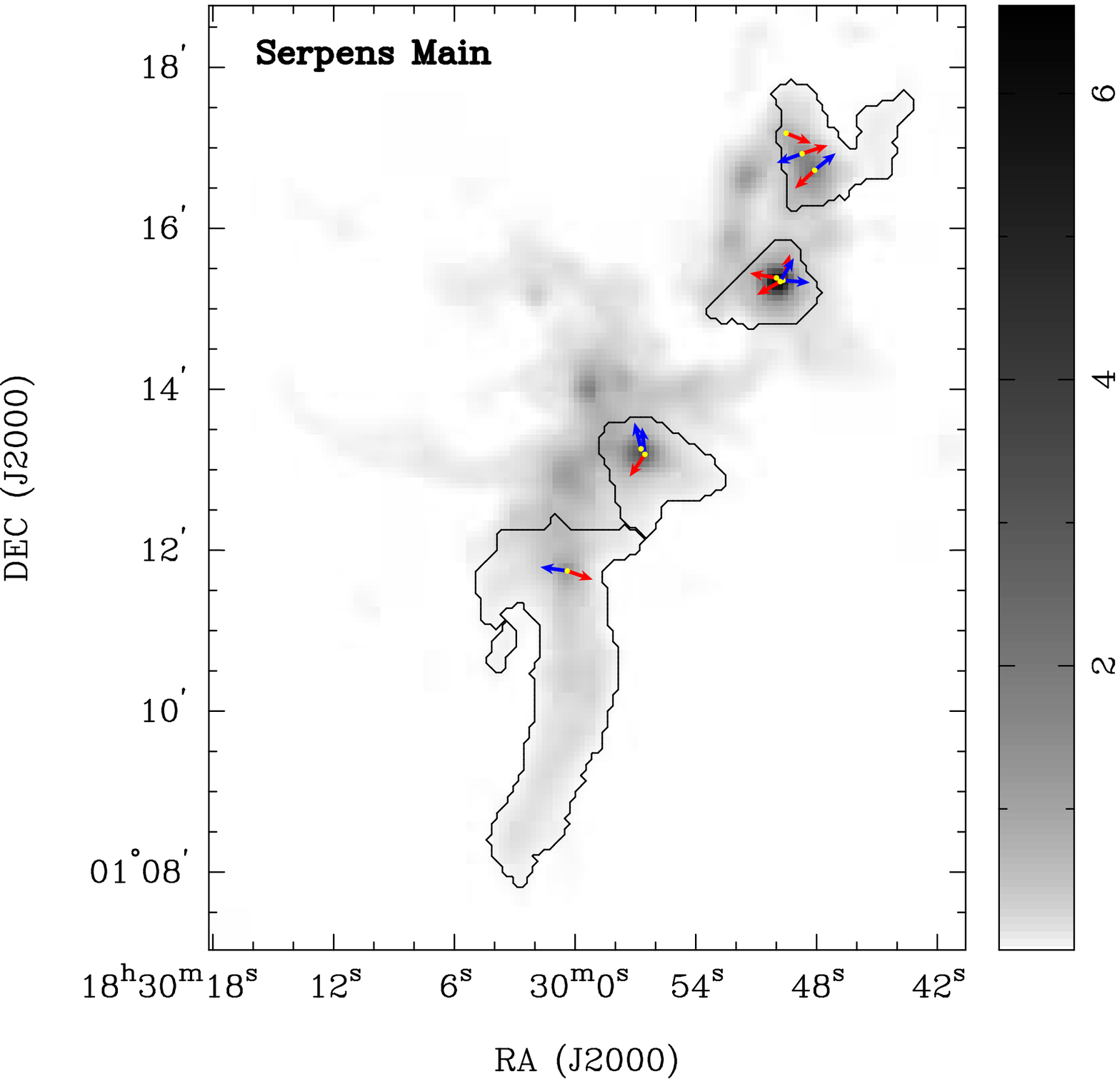}
\includegraphics[width=0.49\textwidth]{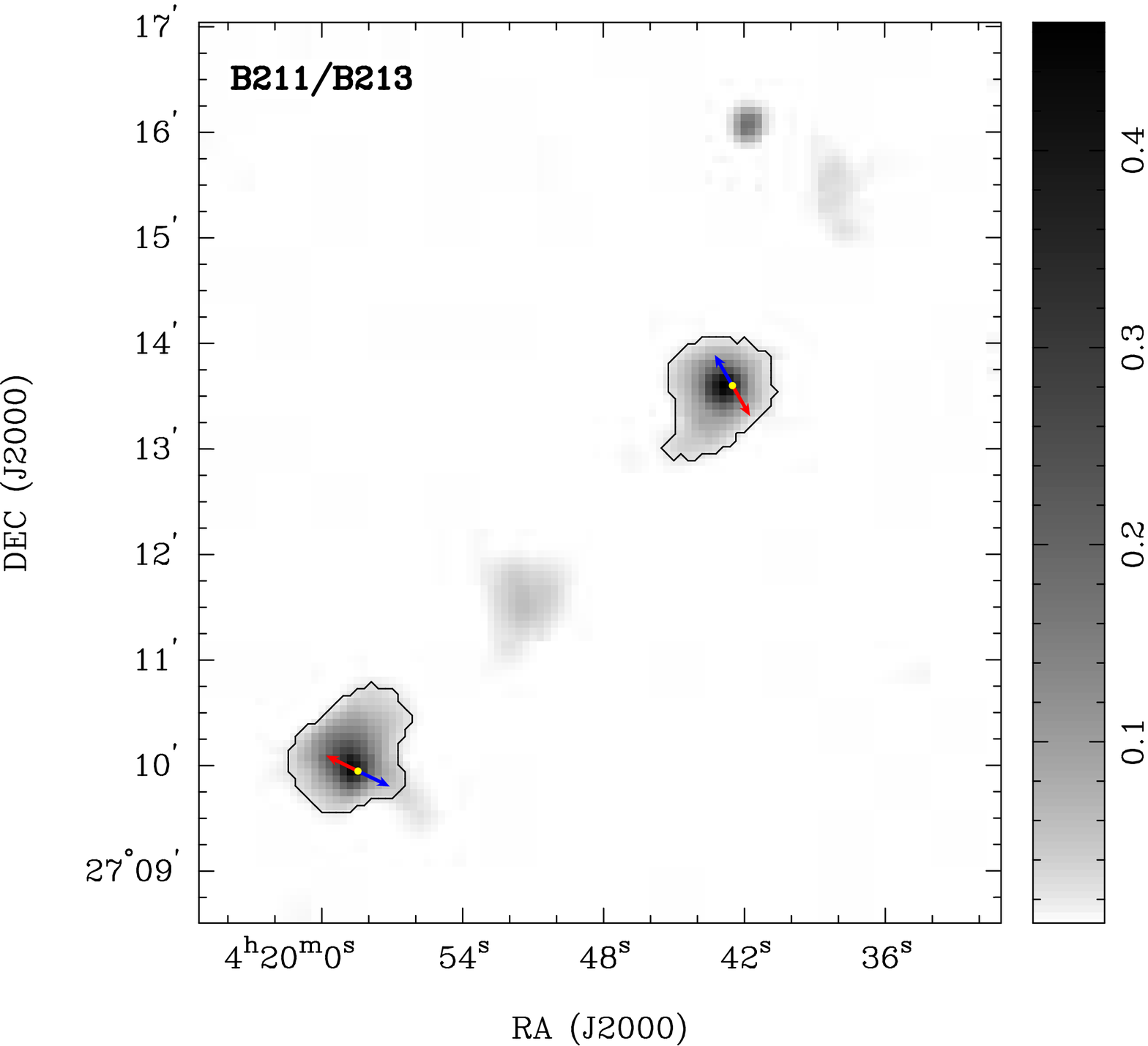}
\includegraphics[width=0.49\textwidth]{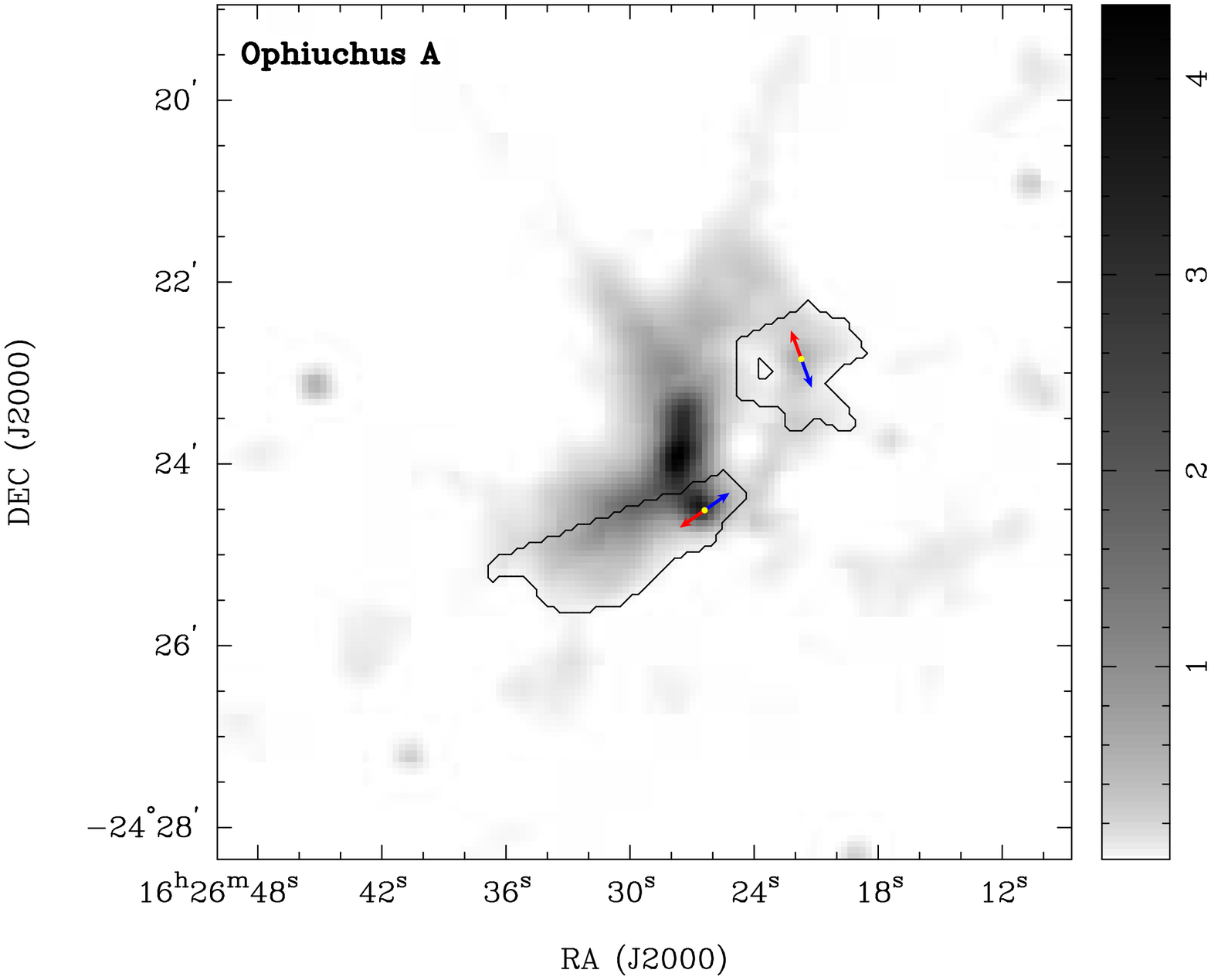}
\includegraphics[width=0.49\textwidth]{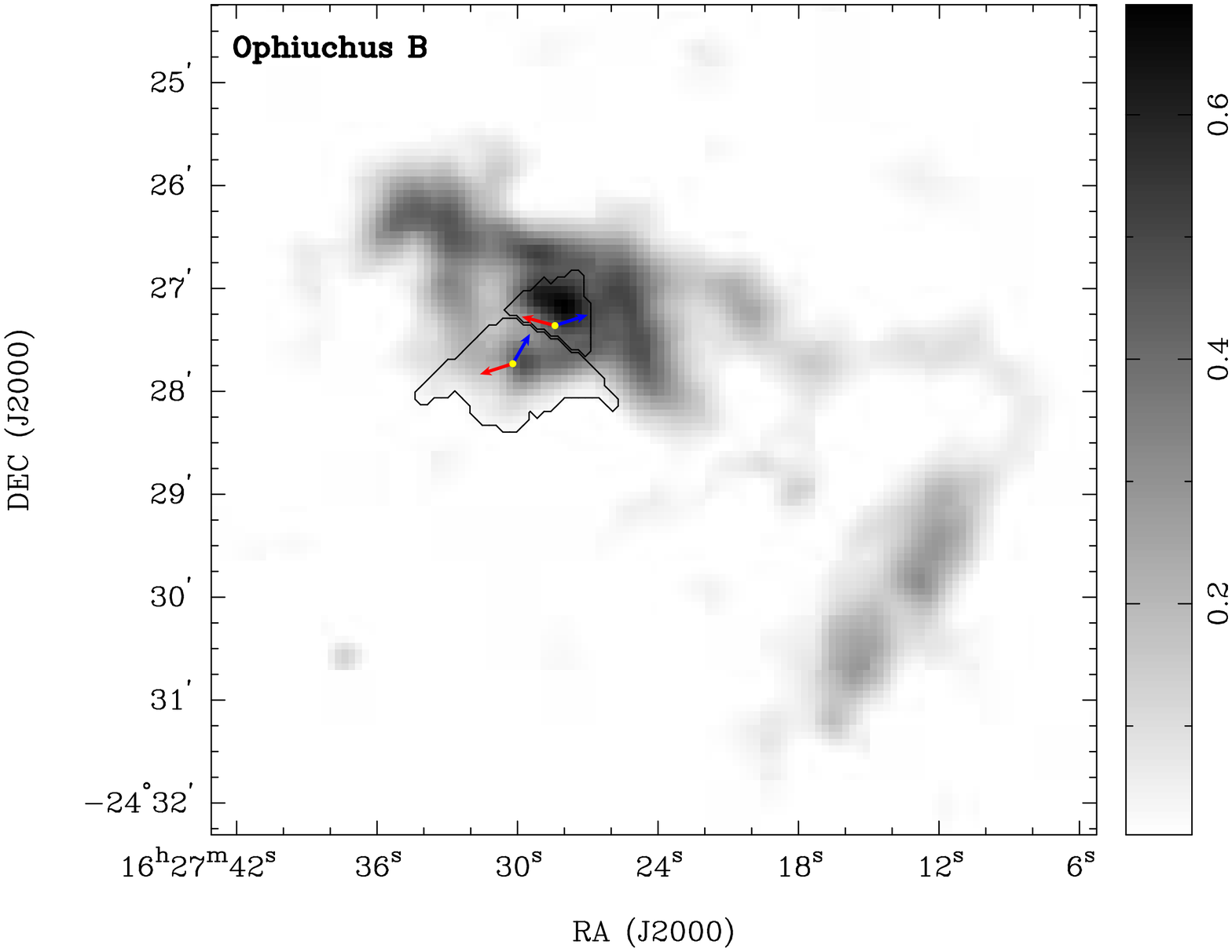}
\par
Figure \ref{regions} --- continued.
\end{figure*}

\section{Mean magnetic field orientations}\label{meanb}
Table 2 shows the mean magnetic field orientations computed without weighting and by weighting the individual polarization detections with their polarized intensities, signal-to-noise ratios, and Stokes {\it I} intensities. 
The differences between the mean orientations computed without weighting and with weighting by the polarized intensity and the signal-to-noise ratio are typically less than 5$\arcdeg$, 
and are less than 9$\arcdeg$ in all the sources, except for L1448~mm and L1448C-S. 
The mean orientations computed with weighting by the Stokes {\it I} intensity tend to show larger differences. 
This is because the dense cores typically exhibit high intensity contrast between the center and outer regions,  
and the resultant mean orientations are more biased by the central detections when the weighting by the Stokes {\it I} intensity is applied. 

\startlongtable
\begin{deluxetable}{ccccc}
\tablecaption{Sample list}
\centering
\tablehead{Name & $\bar{B}_{\theta, {\rm nw}}$ & $\bar{B}_{\theta,PI}$ & $\bar{B}_{\theta,{\rm SN}}$ & $\bar{B}_{\theta, I}$ }
\startdata
L1448IRS2 & 138\arcdeg$\pm$ 2\arcdeg & 136\arcdeg$\pm$ 2\arcdeg & 137\arcdeg$\pm$ 2\arcdeg & 129\arcdeg$\pm$ 2\arcdeg   \\
L1448IRS2E &   9\arcdeg$\pm$ 2\arcdeg &  10\arcdeg$\pm$ 2\arcdeg &   8\arcdeg$\pm$ 2\arcdeg &   8\arcdeg$\pm$ 2\arcdeg  \\
L1448IRS3Bc & 169\arcdeg$\pm$ 1\arcdeg & 168\arcdeg$\pm$ 1\arcdeg & 168\arcdeg$\pm$ 1\arcdeg & 167\arcdeg$\pm$ 2\arcdeg \\
L1448IRS3Ba &  17\arcdeg$\pm$ 3\arcdeg &  26\arcdeg$\pm$ 3\arcdeg &  19\arcdeg$\pm$ 3\arcdeg &  37\arcdeg$\pm$ 3\arcdeg \\
L1448IRS3Bb &  17\arcdeg$\pm$ 3\arcdeg &  26\arcdeg$\pm$ 3\arcdeg &  19\arcdeg$\pm$ 3\arcdeg &  37\arcdeg$\pm$ 3\arcdeg \\
L1448-mm &  21\arcdeg$\pm$31\arcdeg &  40\arcdeg$\pm$17\arcdeg &  30\arcdeg$\pm$30\arcdeg &  47\arcdeg$\pm$16\arcdeg \\
L1448C-S &  21\arcdeg$\pm$31\arcdeg &  40\arcdeg$\pm$17\arcdeg &  30\arcdeg$\pm$30\arcdeg &  47\arcdeg$\pm$16\arcdeg \\
Per-emb17 &  65\arcdeg$\pm$ 4\arcdeg &  64\arcdeg$\pm$ 4\arcdeg &  64\arcdeg$\pm$ 4\arcdeg &  70\arcdeg$\pm$ 5\arcdeg \\
L1455IRS4 &  80\arcdeg$\pm$ 3\arcdeg &  81\arcdeg$\pm$ 3\arcdeg &  75\arcdeg$\pm$ 3\arcdeg &  84\arcdeg$\pm$ 3\arcdeg \\
Per-emb3 &  53\arcdeg$\pm$ 6\arcdeg &  55\arcdeg$\pm$ 6\arcdeg &  55\arcdeg$\pm$ 6\arcdeg &  53\arcdeg$\pm$ 6\arcdeg \\
NGC1333IRAS4A &  60\arcdeg$\pm$ 1\arcdeg &  60\arcdeg$\pm$ 1\arcdeg &  60\arcdeg$\pm$ 1\arcdeg &  60\arcdeg$\pm$ 1\arcdeg \\
NGC1333IRAS4B1 &  71\arcdeg$\pm$ 1\arcdeg &  68\arcdeg$\pm$ 1\arcdeg &  67\arcdeg$\pm$ 1\arcdeg &  70\arcdeg$\pm$ 1\arcdeg \\
NGC1333IRAS4B2 &  71\arcdeg$\pm$ 1\arcdeg &  68\arcdeg$\pm$ 1\arcdeg &  67\arcdeg$\pm$ 1\arcdeg &  70\arcdeg$\pm$ 1\arcdeg \\
NGC1333IRAS7 &  92\arcdeg$\pm$ 1\arcdeg &  93\arcdeg$\pm$ 1\arcdeg &  93\arcdeg$\pm$ 1\arcdeg &  95\arcdeg$\pm$ 1\arcdeg \\
Per-emb21 &  92\arcdeg$\pm$ 1\arcdeg &  93\arcdeg$\pm$ 1\arcdeg &  93\arcdeg$\pm$ 1\arcdeg &  95\arcdeg$\pm$ 1\arcdeg \\
NGC1333IRAS2A1 &  79\arcdeg$\pm$ 2\arcdeg &  74\arcdeg$\pm$ 1\arcdeg &  74\arcdeg$\pm$ 1\arcdeg &  74\arcdeg$\pm$ 1\arcdeg \\
NGC1333IRAS2A2 &  79\arcdeg$\pm$ 2\arcdeg &  74\arcdeg$\pm$ 1\arcdeg &  74\arcdeg$\pm$ 1\arcdeg &  74\arcdeg$\pm$ 1\arcdeg \\
NGC1333IRAS1a &  83\arcdeg$\pm$ 8\arcdeg &  79\arcdeg$\pm$ 8\arcdeg &  81\arcdeg$\pm$ 8\arcdeg &  84\arcdeg$\pm$ 5\arcdeg \\
NGC1333IRAS1b &  83\arcdeg$\pm$ 8\arcdeg &  79\arcdeg$\pm$ 8\arcdeg &  81\arcdeg$\pm$ 8\arcdeg &  84\arcdeg$\pm$ 5\arcdeg \\
NGC1333IRAS2B &  55\arcdeg$\pm$ 4\arcdeg &  57\arcdeg$\pm$ 4\arcdeg &  58\arcdeg$\pm$ 4\arcdeg &  69\arcdeg$\pm$ 3\arcdeg \\
SVS13A & 164\arcdeg$\pm$ 1\arcdeg & 163\arcdeg$\pm$ 1\arcdeg & 163\arcdeg$\pm$ 1\arcdeg & 163\arcdeg$\pm$ 1\arcdeg \\
RNO15-FIR & 164\arcdeg$\pm$ 1\arcdeg & 163\arcdeg$\pm$ 1\arcdeg & 163\arcdeg$\pm$ 1\arcdeg & 163\arcdeg$\pm$ 1\arcdeg \\
Per-emb37 & 131\arcdeg$\pm$ 3\arcdeg & 132\arcdeg$\pm$ 3\arcdeg & 131\arcdeg$\pm$ 3\arcdeg & 127\arcdeg$\pm$ 4\arcdeg \\
Per-emb49 &  92\arcdeg$\pm$ 1\arcdeg &  93\arcdeg$\pm$ 1\arcdeg &  93\arcdeg$\pm$ 1\arcdeg &  95\arcdeg$\pm$ 1\arcdeg \\
Per-emb50 & 175\arcdeg$\pm$ 1\arcdeg & 174\arcdeg$\pm$ 1\arcdeg & 174\arcdeg$\pm$ 1\arcdeg & 171\arcdeg$\pm$ 1\arcdeg \\
Per-emb58 & 119\arcdeg$\pm$ 2\arcdeg & 117\arcdeg$\pm$ 2\arcdeg & 117\arcdeg$\pm$ 2\arcdeg & 120\arcdeg$\pm$ 2\arcdeg \\
SVS13B & 164\arcdeg$\pm$ 1\arcdeg & 163\arcdeg$\pm$ 1\arcdeg & 163\arcdeg$\pm$ 1\arcdeg & 163\arcdeg$\pm$ 1\arcdeg \\
SVS13C & 164\arcdeg$\pm$ 1\arcdeg & 163\arcdeg$\pm$ 1\arcdeg & 163\arcdeg$\pm$ 1\arcdeg & 163\arcdeg$\pm$ 1\arcdeg \\
Per-emb6 & 144\arcdeg$\pm$ 2\arcdeg & 143\arcdeg$\pm$ 2\arcdeg & 143\arcdeg$\pm$ 2\arcdeg & 147\arcdeg$\pm$ 2\arcdeg \\
Per-emb10 & 156\arcdeg$\pm$ 1\arcdeg & 153\arcdeg$\pm$ 1\arcdeg & 153\arcdeg$\pm$ 1\arcdeg & 153\arcdeg$\pm$ 1\arcdeg \\
B1-a & 146\arcdeg$\pm$ 1\arcdeg & 148\arcdeg$\pm$ 1\arcdeg & 147\arcdeg$\pm$ 1\arcdeg & 146\arcdeg$\pm$ 1\arcdeg \\
B1-c &  95\arcdeg$\pm$ 1\arcdeg & 100\arcdeg$\pm$ 1\arcdeg & 101\arcdeg$\pm$ 1\arcdeg & 100\arcdeg$\pm$ 1\arcdeg \\
B1-b & 157\arcdeg$\pm$ 1\arcdeg & 157\arcdeg$\pm$ 1\arcdeg & 156\arcdeg$\pm$ 1\arcdeg & 153\arcdeg$\pm$ 2\arcdeg \\
B1-bN & 157\arcdeg$\pm$ 2\arcdeg & 155\arcdeg$\pm$ 1\arcdeg & 155\arcdeg$\pm$ 1\arcdeg & 149\arcdeg$\pm$ 2\arcdeg \\
B1-bS & 157\arcdeg$\pm$ 1\arcdeg & 157\arcdeg$\pm$ 1\arcdeg & 156\arcdeg$\pm$ 1\arcdeg & 153\arcdeg$\pm$ 2\arcdeg \\
HH211-mms & 152\arcdeg$\pm$ 1\arcdeg & 155\arcdeg$\pm$ 1\arcdeg & 156\arcdeg$\pm$ 1\arcdeg & 154\arcdeg$\pm$ 1\arcdeg \\
IC348MMSa & 153\arcdeg$\pm$ 2\arcdeg & 153\arcdeg$\pm$ 2\arcdeg & 152\arcdeg$\pm$ 2\arcdeg & 154\arcdeg$\pm$ 2\arcdeg \\
IC348MMSb & 153\arcdeg$\pm$ 2\arcdeg & 153\arcdeg$\pm$ 2\arcdeg & 152\arcdeg$\pm$ 2\arcdeg & 154\arcdeg$\pm$ 2\arcdeg \\
Per-emb16 & 113\arcdeg$\pm$ 3\arcdeg & 113\arcdeg$\pm$ 3\arcdeg & 113\arcdeg$\pm$ 3\arcdeg & 113\arcdeg$\pm$ 4\arcdeg \\
Per-emb28 & 113\arcdeg$\pm$ 3\arcdeg & 113\arcdeg$\pm$ 3\arcdeg & 113\arcdeg$\pm$ 3\arcdeg & 113\arcdeg$\pm$ 4\arcdeg \\
Per-emb62 & 145\arcdeg$\pm$ 9\arcdeg & 145\arcdeg$\pm$ 8\arcdeg & 145\arcdeg$\pm$ 9\arcdeg & 145\arcdeg$\pm$ 9\arcdeg \\
IRAS04169+2702 & 102\arcdeg$\pm$ 6\arcdeg & 104\arcdeg$\pm$ 4\arcdeg & 104\arcdeg$\pm$ 5\arcdeg & 104\arcdeg$\pm$ 5\arcdeg \\
IRAS04166+2706 &  47\arcdeg$\pm$ 3\arcdeg &  47\arcdeg$\pm$ 3\arcdeg &  47\arcdeg$\pm$ 3\arcdeg &  47\arcdeg$\pm$ 4\arcdeg \\
L1521F &  21\arcdeg$\pm$ 6\arcdeg &  19\arcdeg$\pm$ 5\arcdeg &  18\arcdeg$\pm$ 5\arcdeg &  26\arcdeg$\pm$ 5\arcdeg \\
L1527 &  82\arcdeg$\pm$12\arcdeg &  81\arcdeg$\pm$10\arcdeg &  83\arcdeg$\pm$10\arcdeg &  40\arcdeg$\pm$ 6\arcdeg \\
HH212 &  35\arcdeg$\pm$ 4\arcdeg &  34\arcdeg$\pm$ 3\arcdeg &  32\arcdeg$\pm$ 4\arcdeg &  35\arcdeg$\pm$ 3\arcdeg \\
HH111 &  67\arcdeg$\pm$ 2\arcdeg &  66\arcdeg$\pm$ 2\arcdeg &  65\arcdeg$\pm$ 2\arcdeg &  62\arcdeg$\pm$ 2\arcdeg \\
GSS30IRS3 &  79\arcdeg$\pm$ 1\arcdeg &  80\arcdeg$\pm$ 1\arcdeg &  80\arcdeg$\pm$ 1\arcdeg &  80\arcdeg$\pm$ 1\arcdeg \\
VLA1623A &  75\arcdeg$\pm$ 1\arcdeg &  70\arcdeg$\pm$ 1\arcdeg &  71\arcdeg$\pm$ 1\arcdeg &  60\arcdeg$\pm$ 1\arcdeg \\
Elias32 & 151\arcdeg$\pm$ 3\arcdeg & 145\arcdeg$\pm$ 2\arcdeg & 142\arcdeg$\pm$ 2\arcdeg & 154\arcdeg$\pm$ 3\arcdeg \\
Elias33 & 153\arcdeg$\pm$ 2\arcdeg & 152\arcdeg$\pm$ 2\arcdeg & 151\arcdeg$\pm$ 2\arcdeg & 151\arcdeg$\pm$ 1\arcdeg \\
S68N &  85\arcdeg$\pm$ 1\arcdeg &  83\arcdeg$\pm$ 1\arcdeg &  83\arcdeg$\pm$ 1\arcdeg &  83\arcdeg$\pm$ 1\arcdeg \\
S68Nc1 &  85\arcdeg$\pm$ 1\arcdeg &  83\arcdeg$\pm$ 1\arcdeg &  83\arcdeg$\pm$ 1\arcdeg &  83\arcdeg$\pm$ 1\arcdeg \\
S68Nb1 &  85\arcdeg$\pm$ 1\arcdeg &  83\arcdeg$\pm$ 1\arcdeg &  83\arcdeg$\pm$ 1\arcdeg &  83\arcdeg$\pm$ 1\arcdeg \\
SerpensSMM1b &  98\arcdeg$\pm$ 1\arcdeg & 100\arcdeg$\pm$ 1\arcdeg &  98\arcdeg$\pm$ 1\arcdeg & 114\arcdeg$\pm$ 1\arcdeg \\
SerpensSMM1a &  98\arcdeg$\pm$ 1\arcdeg & 100\arcdeg$\pm$ 1\arcdeg &  98\arcdeg$\pm$ 1\arcdeg & 114\arcdeg$\pm$ 1\arcdeg \\
SerpensSMM1d &  98\arcdeg$\pm$ 1\arcdeg & 100\arcdeg$\pm$ 1\arcdeg &  98\arcdeg$\pm$ 1\arcdeg & 114\arcdeg$\pm$ 1\arcdeg \\
SerpensSMM4B &  48\arcdeg$\pm$ 3\arcdeg &  41\arcdeg$\pm$ 2\arcdeg &  39\arcdeg$\pm$ 2\arcdeg &  51\arcdeg$\pm$ 3\arcdeg \\
SerpensSMM4A &  48\arcdeg$\pm$ 3\arcdeg &  41\arcdeg$\pm$ 2\arcdeg &  39\arcdeg$\pm$ 2\arcdeg &  51\arcdeg$\pm$ 3\arcdeg \\
SerpensSMM11 &  94\arcdeg$\pm$ 1\arcdeg &  94\arcdeg$\pm$ 1\arcdeg &  93\arcdeg$\pm$ 1\arcdeg &  94\arcdeg$\pm$ 1\arcdeg \\
B335 & 111\arcdeg$\pm$ 2\arcdeg & 110\arcdeg$\pm$ 2\arcdeg & 111\arcdeg$\pm$ 2\arcdeg & 104\arcdeg$\pm$ 3\arcdeg \\
L1157 & 159\arcdeg$\pm$ 1\arcdeg & 157\arcdeg$\pm$ 1\arcdeg & 156\arcdeg$\pm$ 1\arcdeg & 157\arcdeg$\pm$ 1\arcdeg \\ 
\enddata
\tablecomments{$\bar{B}_{\theta, {\rm nw}}$, $\bar{B}_{\theta,PI}$, $\bar{B}_{\theta,{\rm SN}}$, and $\bar{B}_{\theta, I}$ are the mean magnetic field orientations of the dense cores or clumps computed from the mean Stokes {\it Q} and {\it U} without any weighting and with weighting by the polarized intensity, signal-to-noise ratio, and Stokes {\it I} intensity of the individual polarization detections, respectively. The uncertainties in the mean magnetic field orientations are calculated with the error propagation of the uncertainties of the individual polarization detections.}
\end{deluxetable}

\section{Number and radial distributions of magnetic field orientations}\label{allsample}
\begin{figure*}
\centering
\includegraphics[angle=90,width=0.9\textwidth]{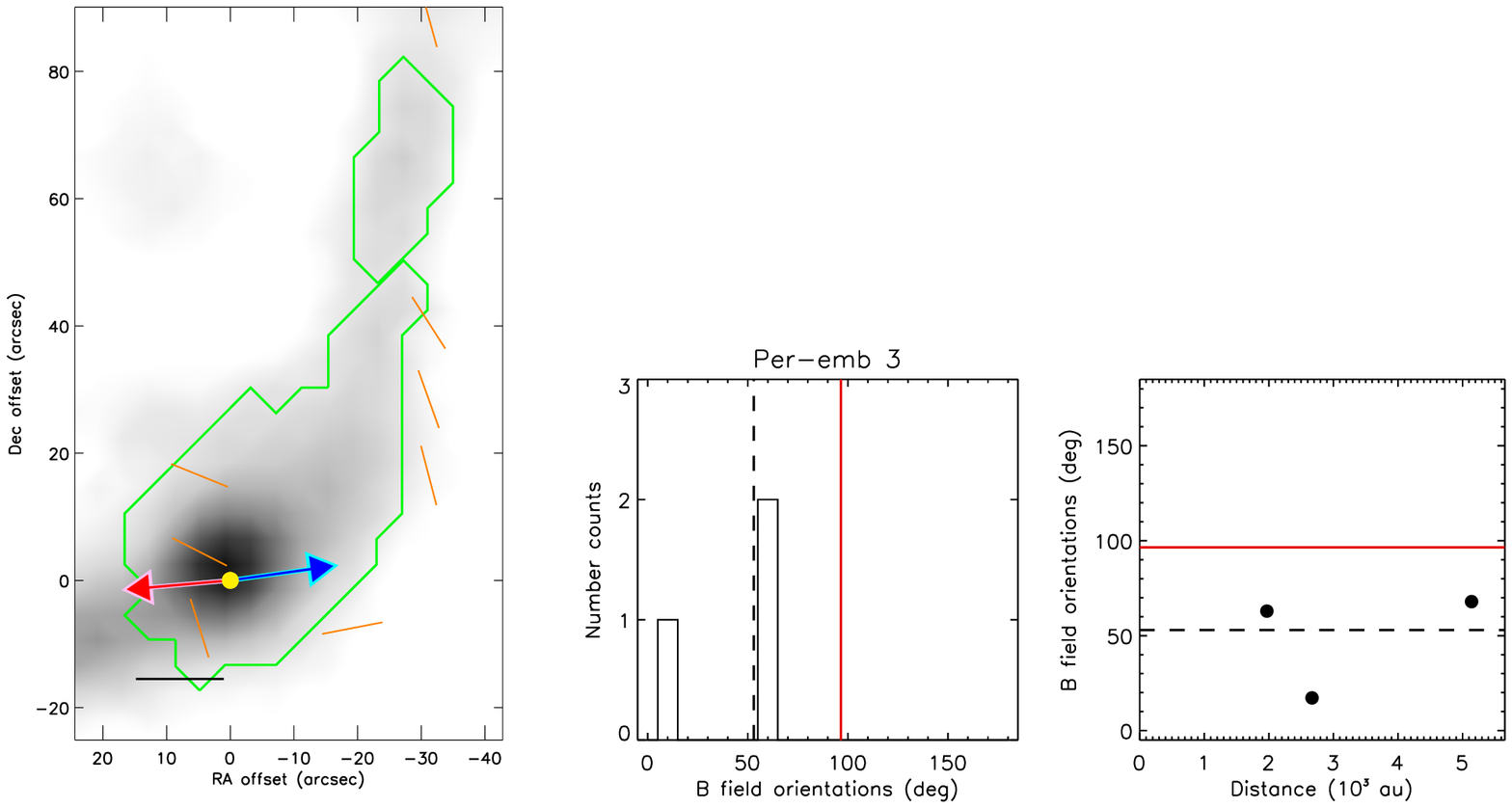}
\includegraphics[angle=90,width=0.9\textwidth]{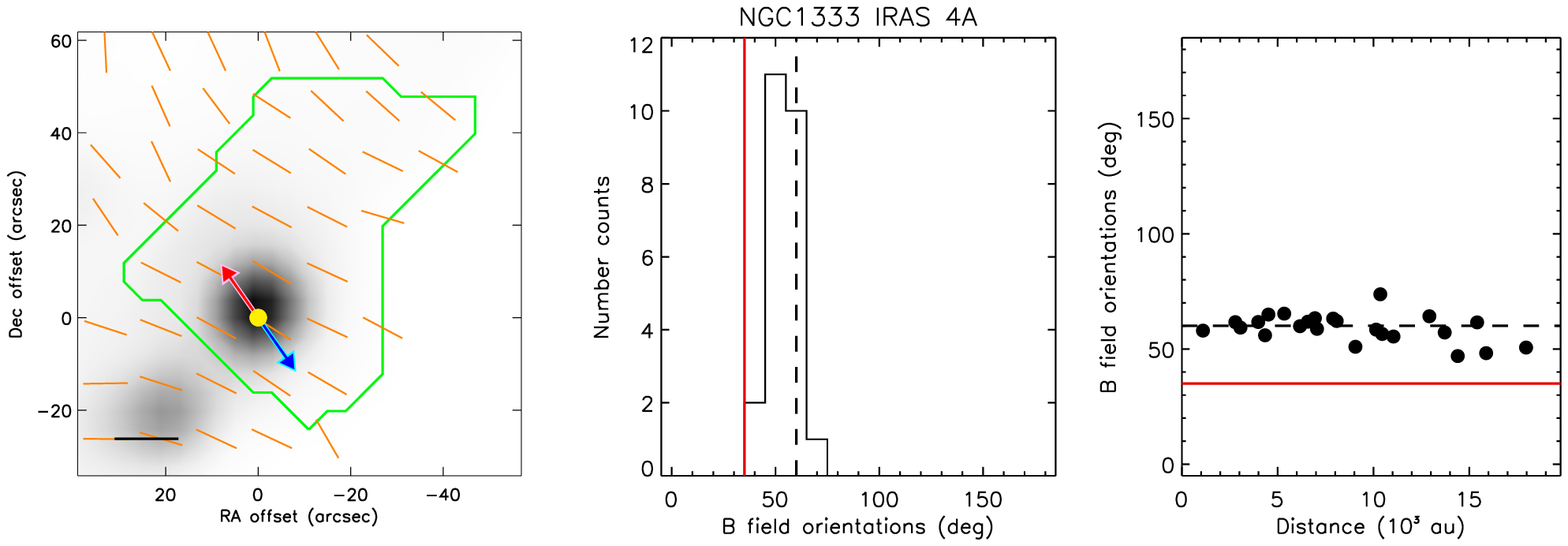}
\includegraphics[angle=90,width=0.9\textwidth]{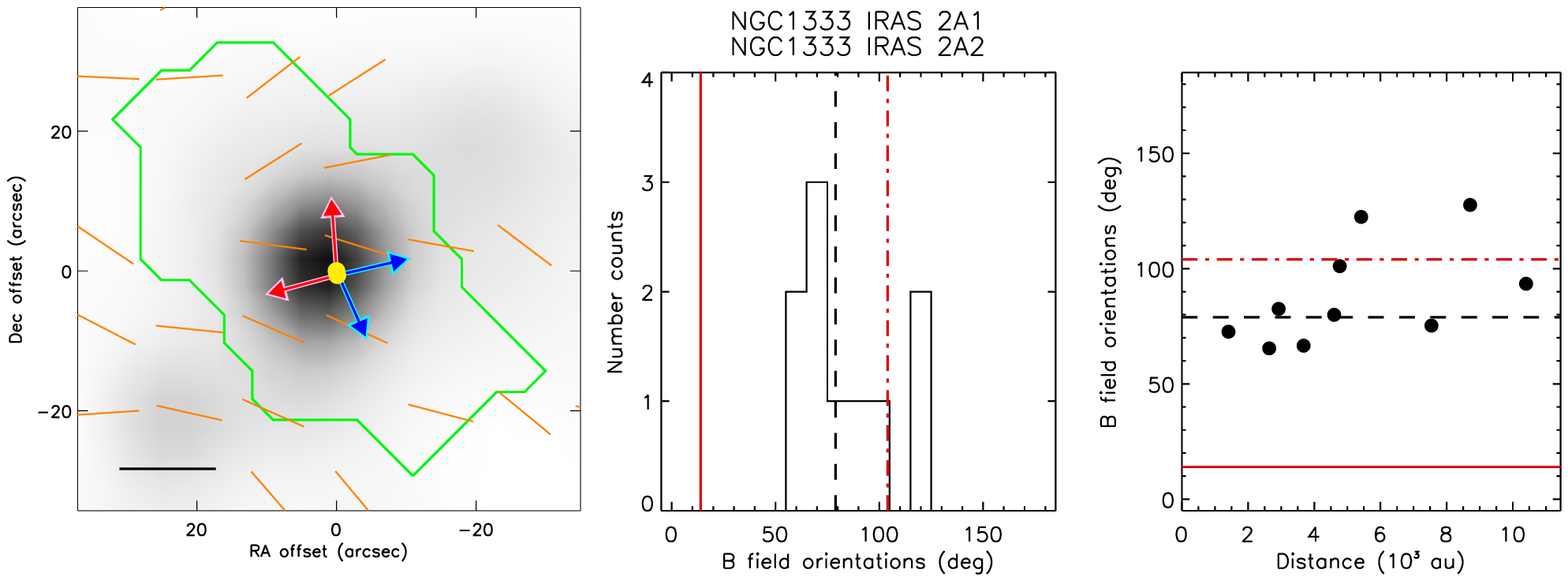}
\caption{Same as Figure~\ref{ex1} and \ref{ex2} but for the remaining sample sources. The source name is labelled above the middle panel of each row. Black horizontal segments denote the spatial scale of 0.02 pc (or $\sim$4000 au).}\label{subplt}
\end{figure*}

\begin{figure*}
\centering
\includegraphics[angle=90,width=0.9\textwidth]{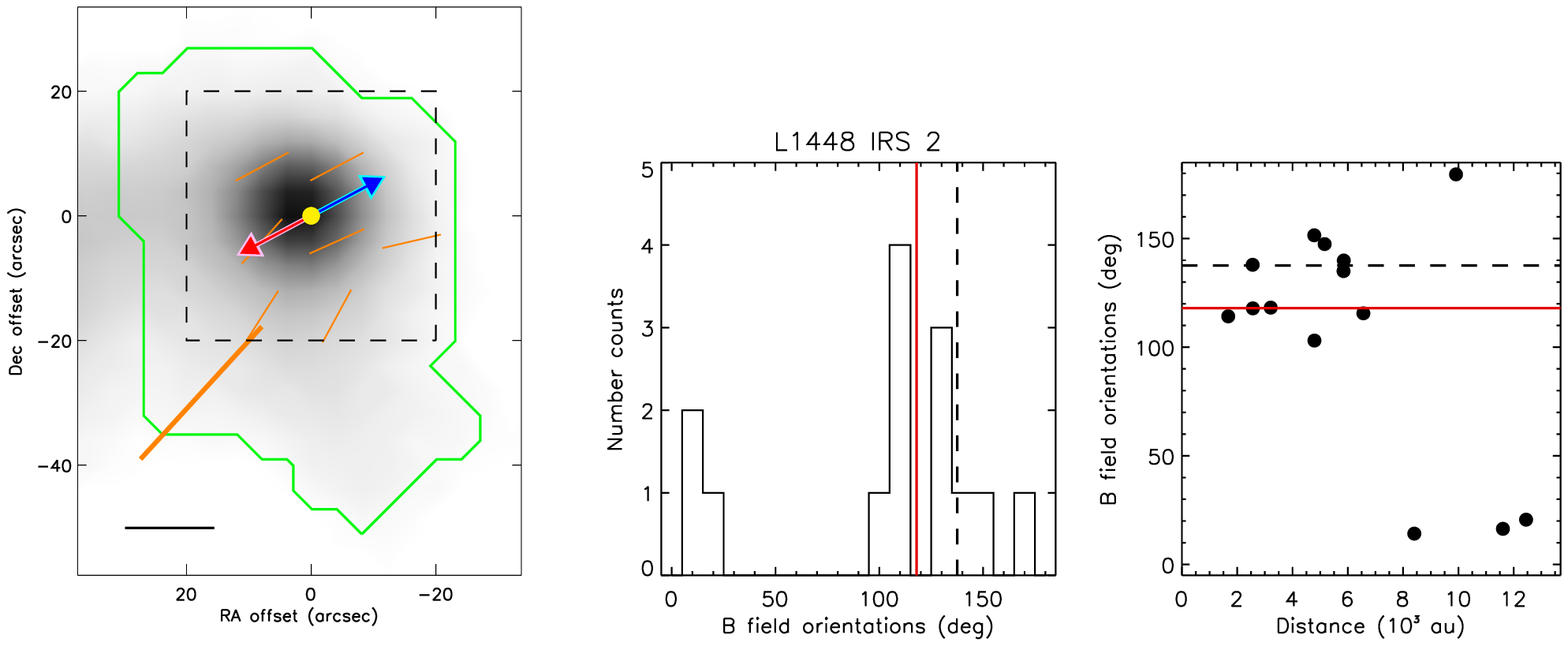}
\includegraphics[angle=90,width=0.9\textwidth]{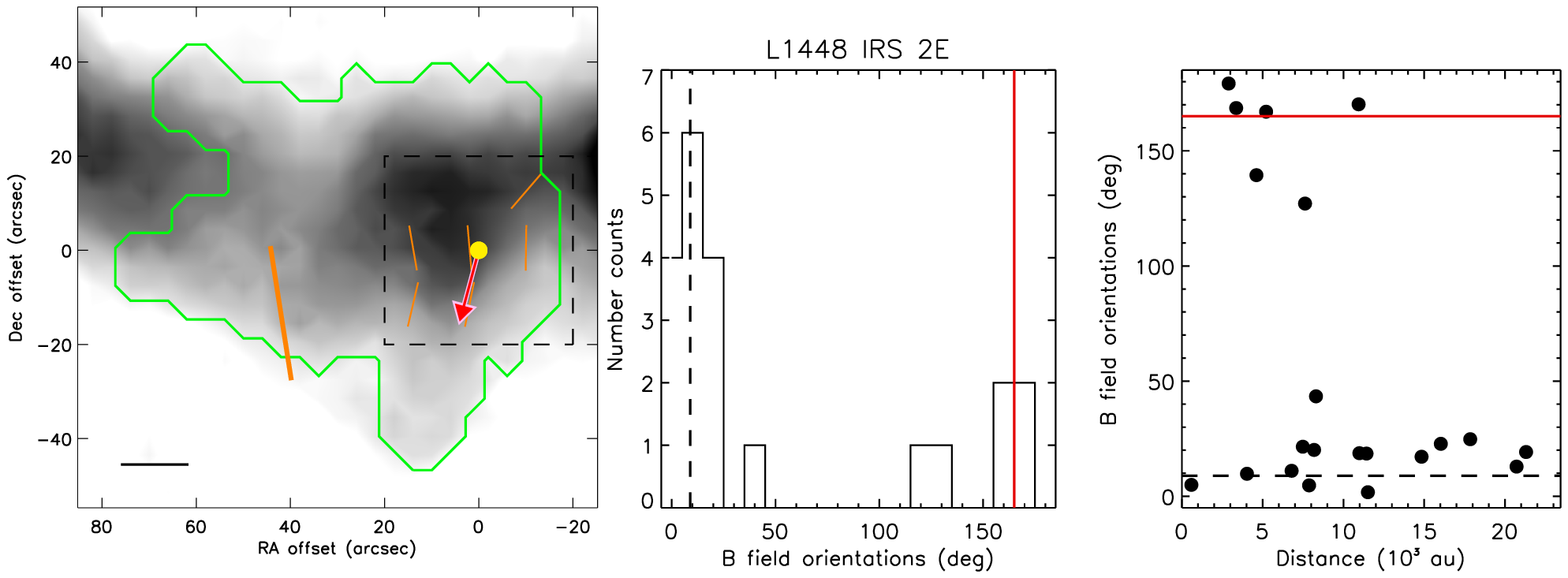}
\includegraphics[angle=90,width=0.9\textwidth]{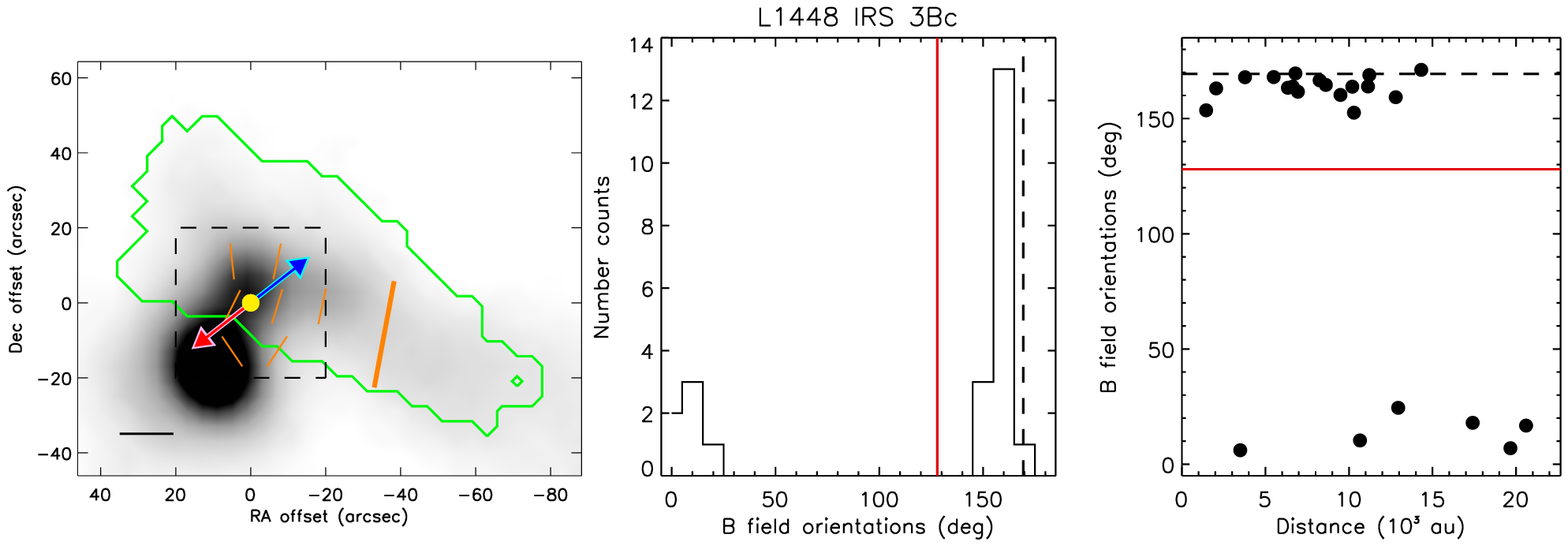}
\caption{Same as Figure~\ref{ex1} and \ref{ex2} but left panels only present the magnetic orientations (small orange segments) detected in the $40\arcsec \times 40\arcsec$ regions (dashed open squares) around the sample protostars as well as the mean orientations of the magnetic field (long thick orange segments) in the individual dense cores/clumps. The complete magnetic field structures in these sources will be presented separately and discussed in detail in the forthcoming papers by the BISTRO team. Middle and right panels still present the number distributions of all the magnetic field orientations detected in the dense cores and clumps.}\label{subplt2}
\end{figure*}

Figure \ref{subplt} and \ref{subplt2} present close-ups of the Stokes {\it I} maps, number distributions of the detected magnetic field orientations, and their radial dependences of all the sample sources expect for those presented in Figure \ref{ex1} and \ref{ex2}.

\begin{figure*}
\centering
\includegraphics[angle=90,width=0.9\textwidth]{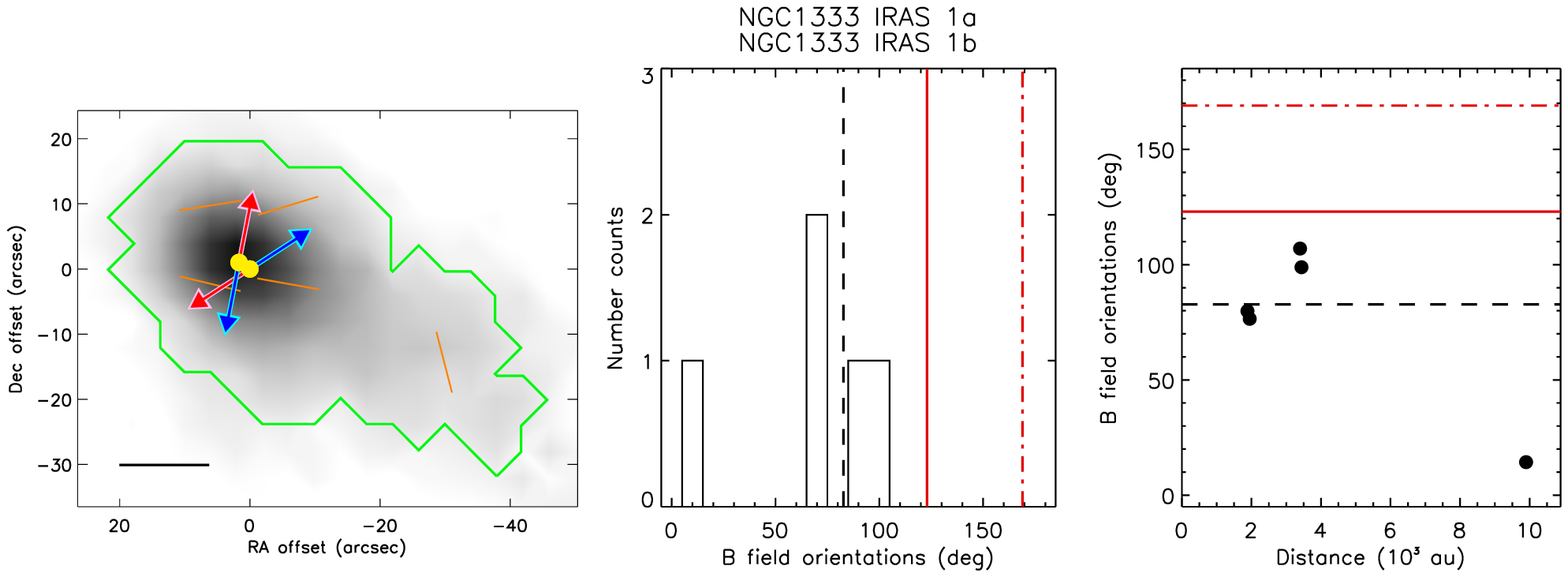}
\includegraphics[angle=90,width=0.9\textwidth]{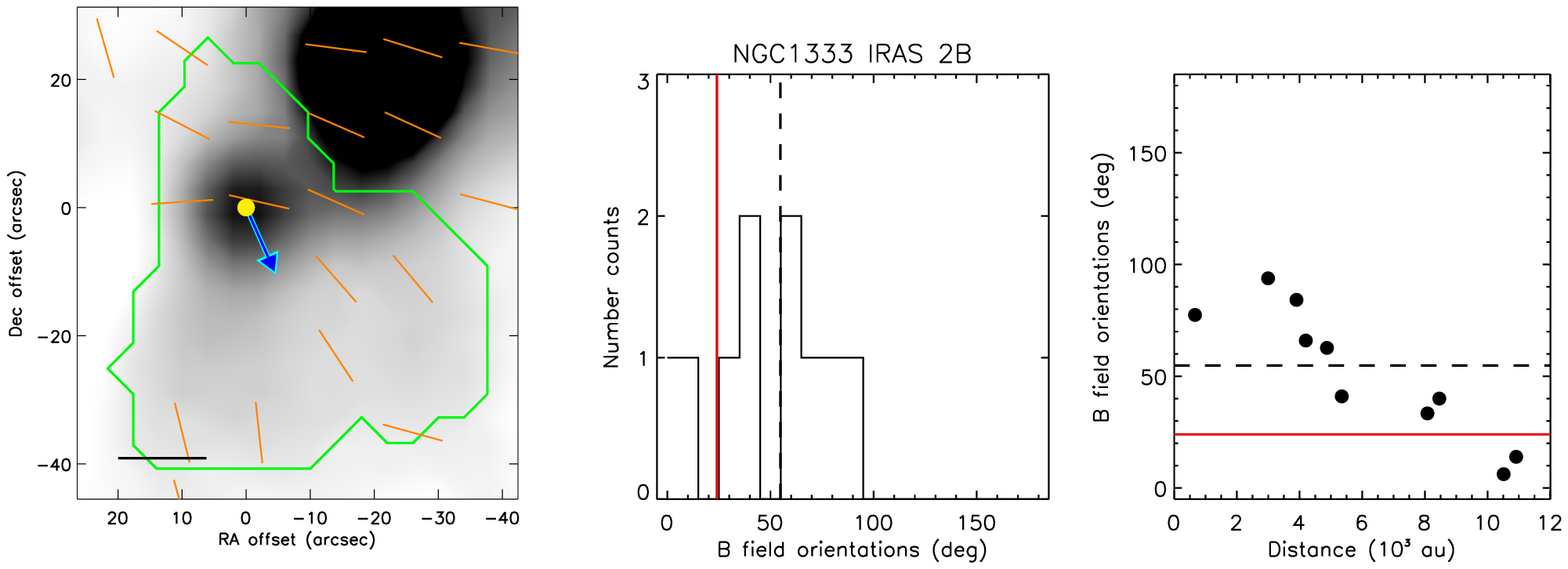}
\includegraphics[angle=90,width=0.9\textwidth]{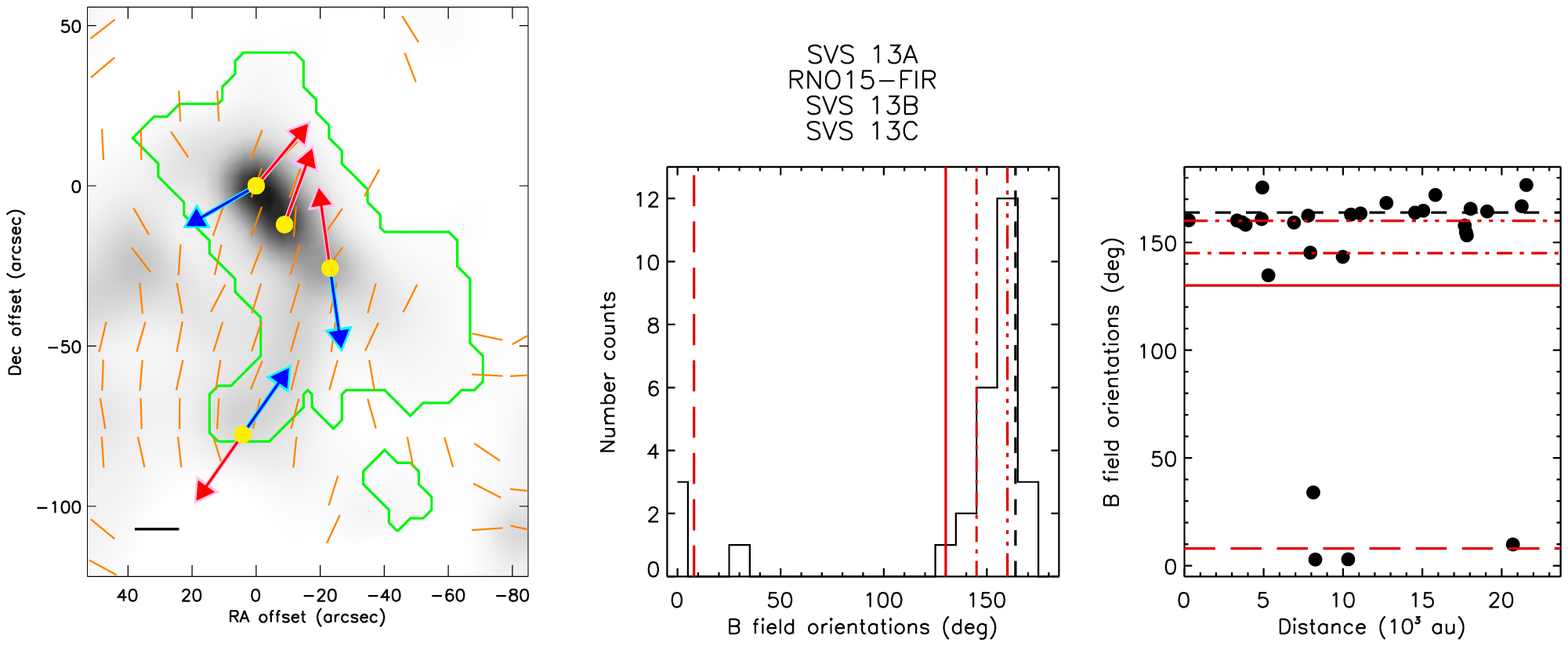}
\par
Figure \ref{subplt} --- continued.
\end{figure*}

\begin{figure*}
\centering
\includegraphics[angle=90,width=0.83\textwidth]{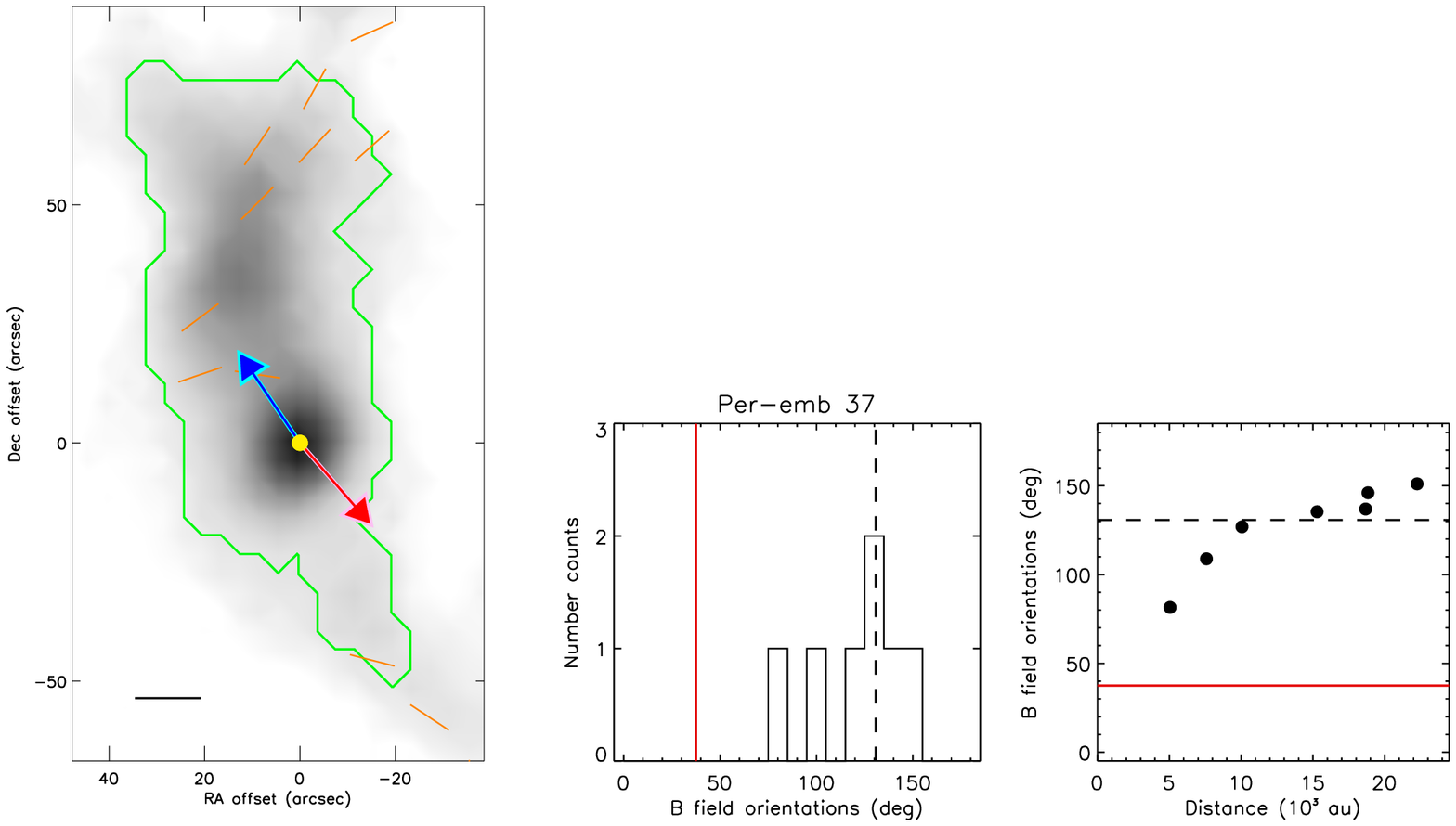}
\includegraphics[angle=90,width=0.9\textwidth]{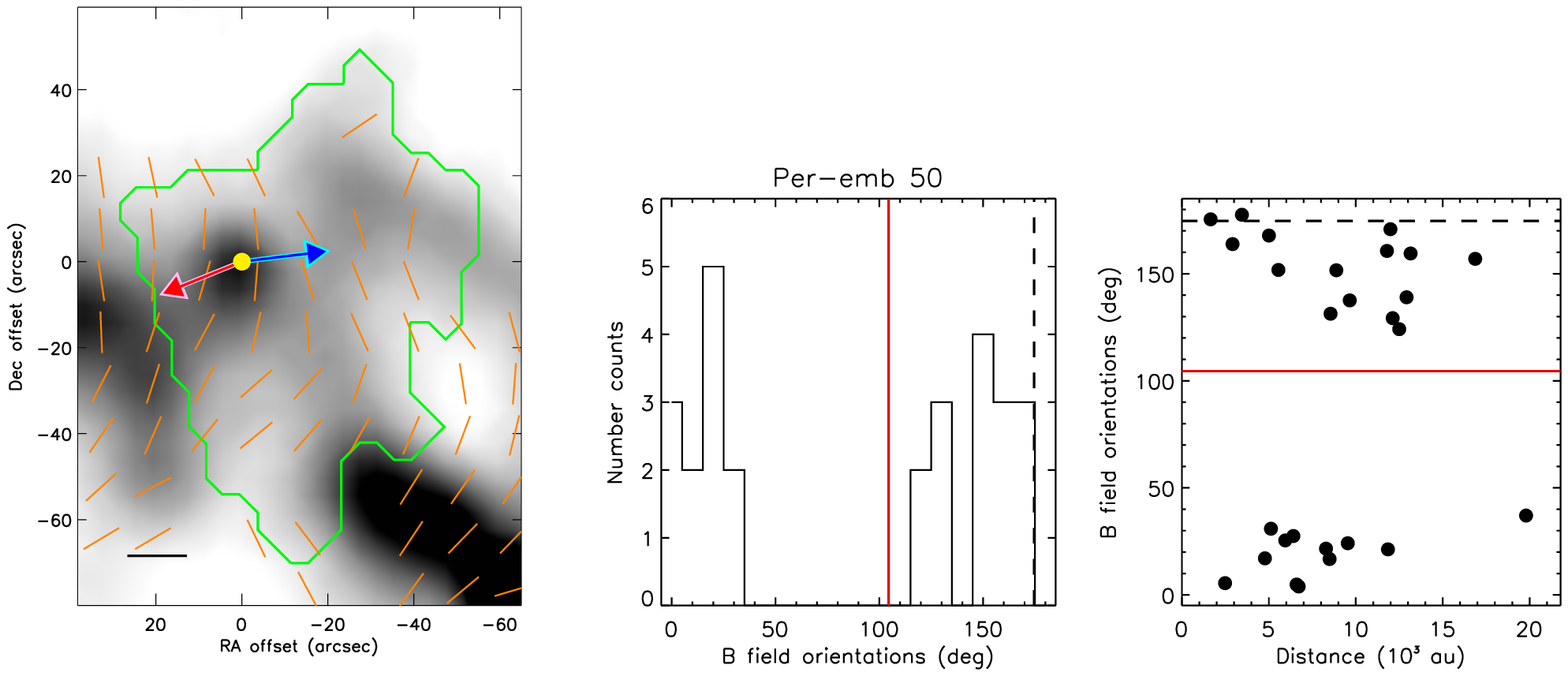}
\par
Figure \ref{subplt} --- continued.
\end{figure*}

\begin{figure*}
\centering
\includegraphics[angle=90,width=0.9\textwidth]{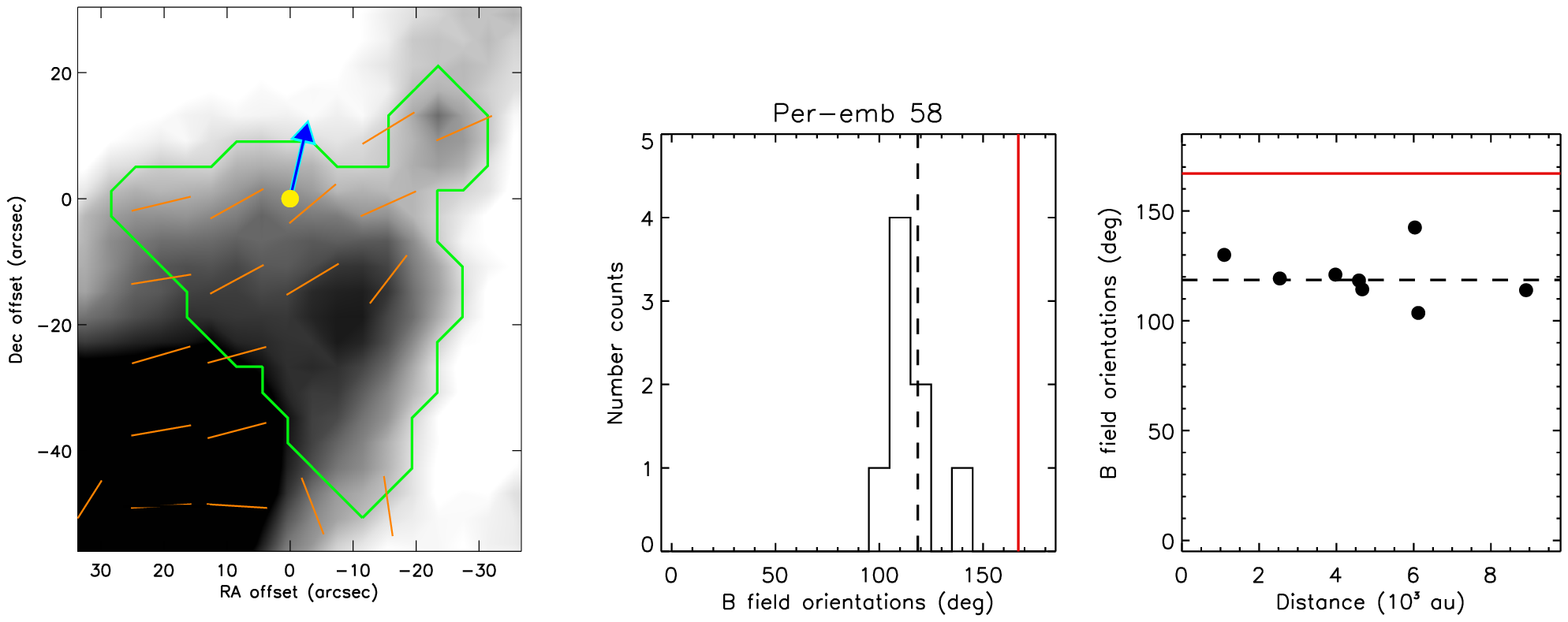}
\includegraphics[angle=90,width=0.9\textwidth]{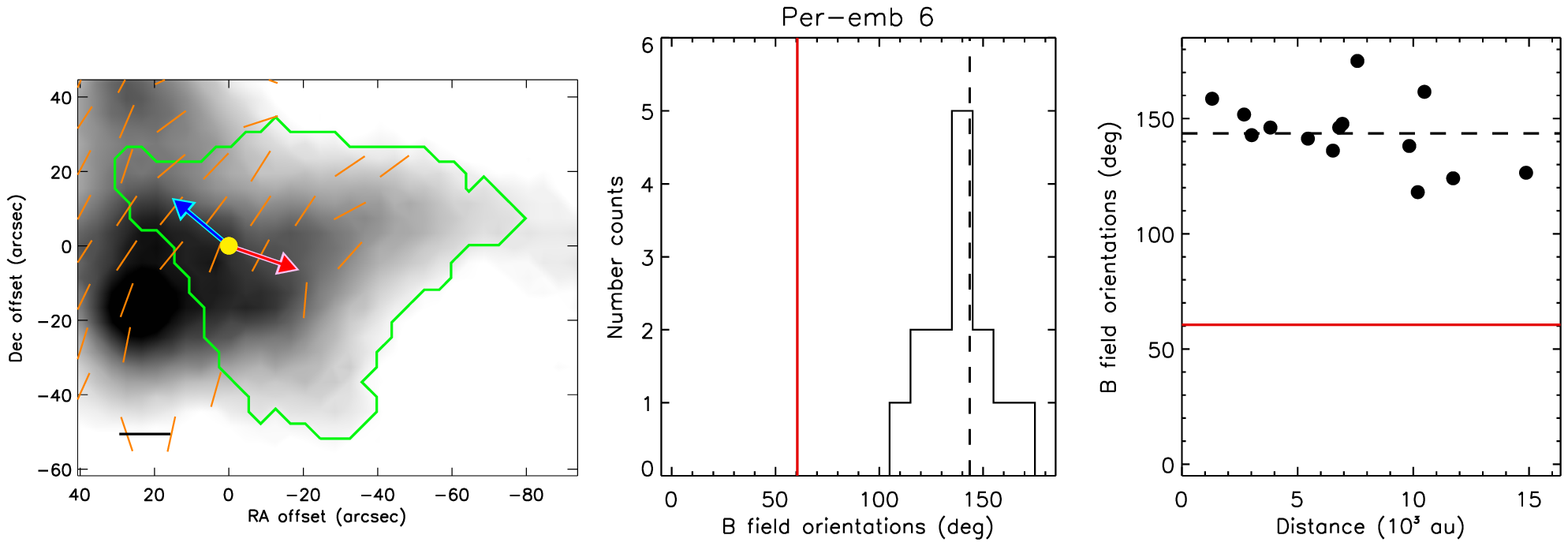}
\includegraphics[angle=90,width=0.9\textwidth]{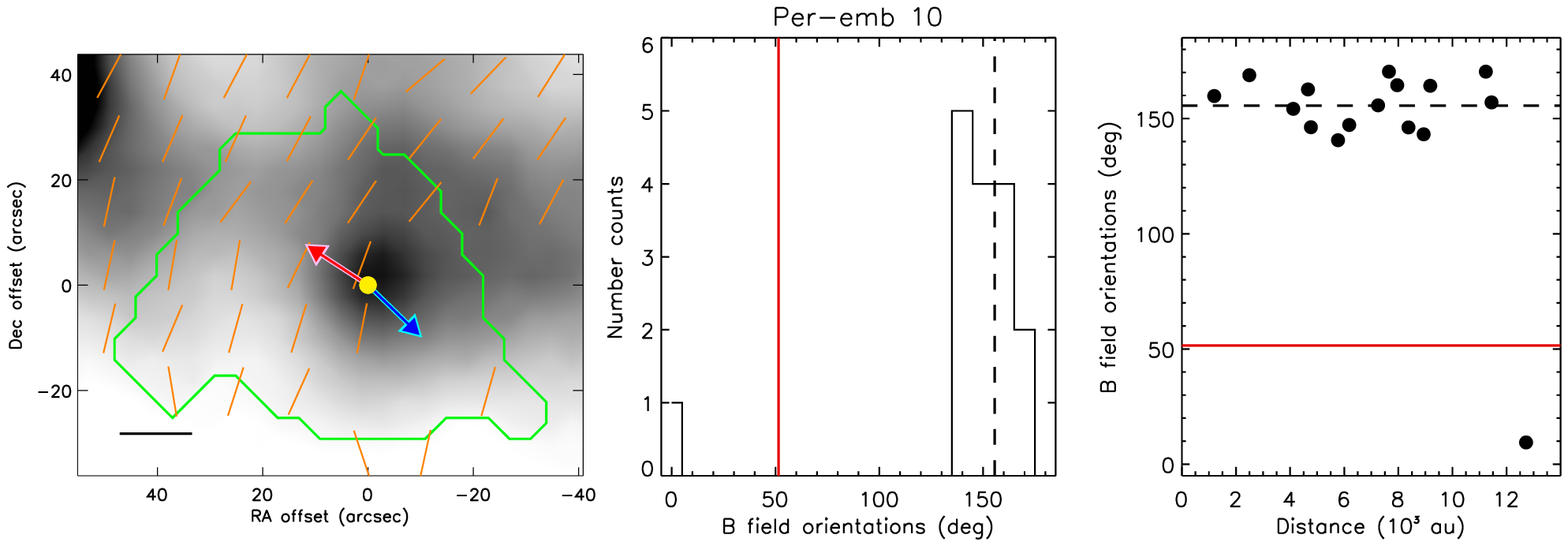}
\par
Figure \ref{subplt} --- continued.
\end{figure*}

\begin{figure*}
\centering
\includegraphics[angle=90,width=0.9\textwidth]{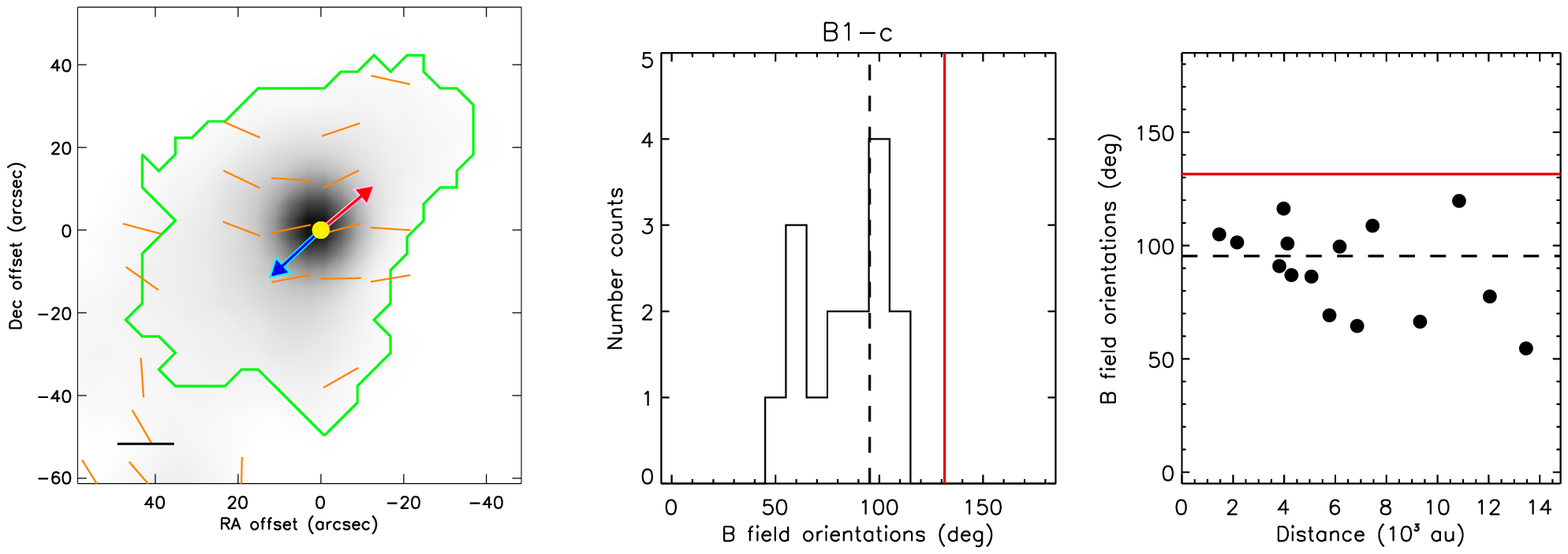}
\includegraphics[angle=90,width=0.9\textwidth]{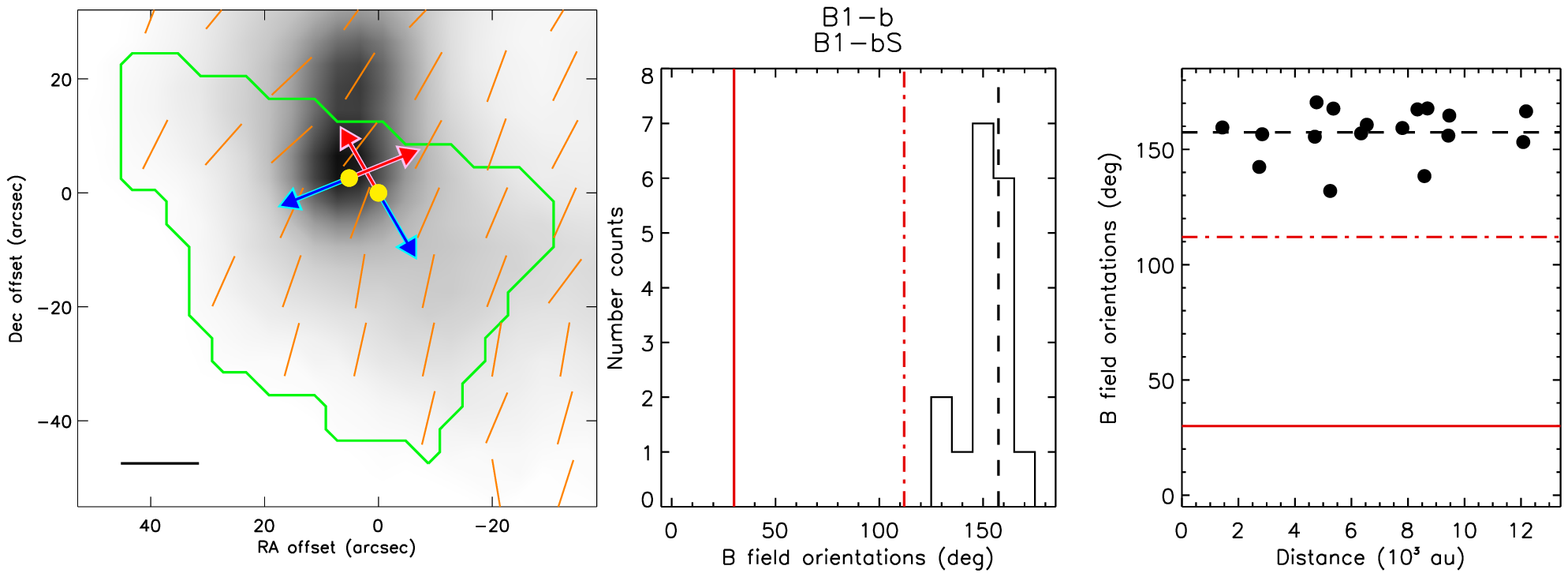}
\includegraphics[angle=90,width=0.9\textwidth]{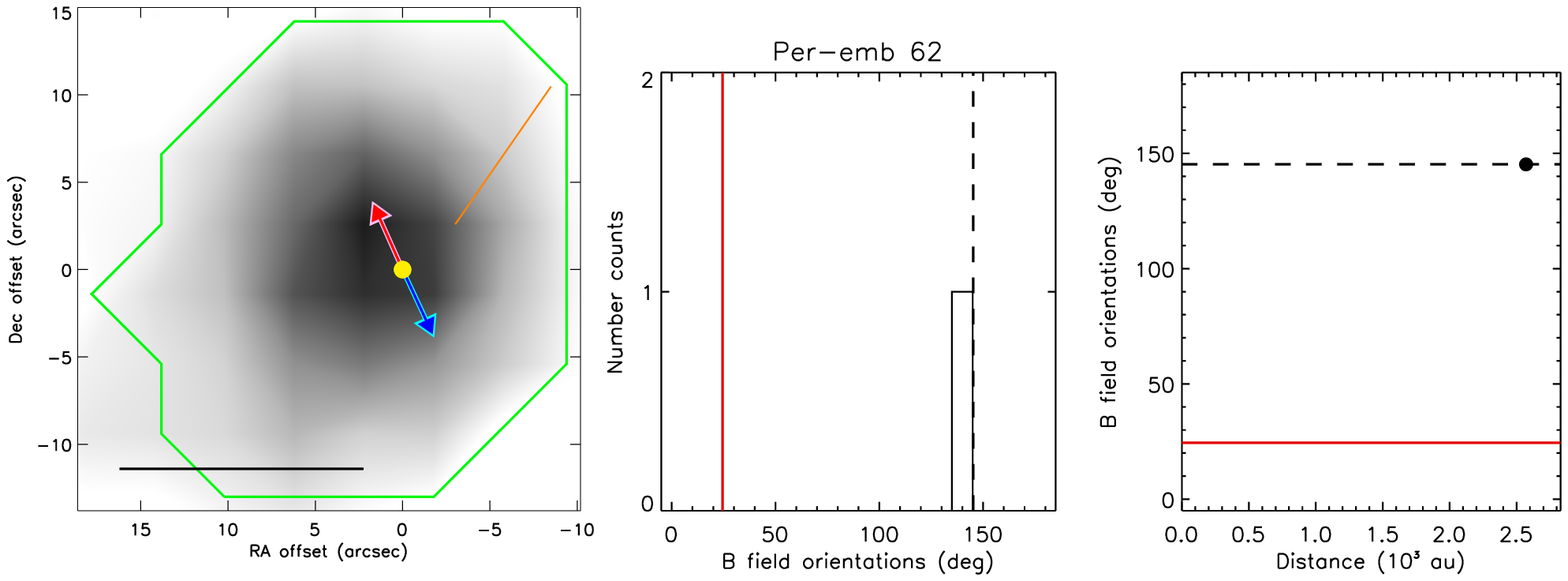}
\par
Figure \ref{subplt} --- continued.
\end{figure*}

\begin{figure*}
\centering
\includegraphics[angle=90,width=0.9\textwidth]{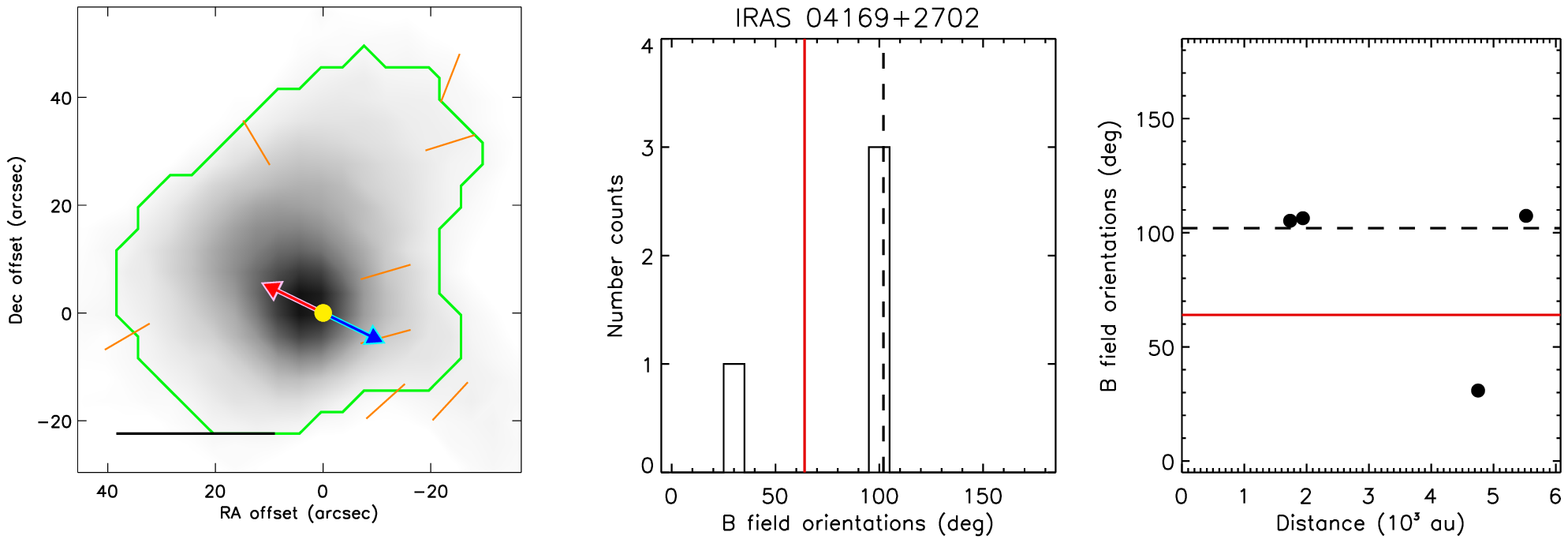}
\includegraphics[angle=90,width=0.9\textwidth]{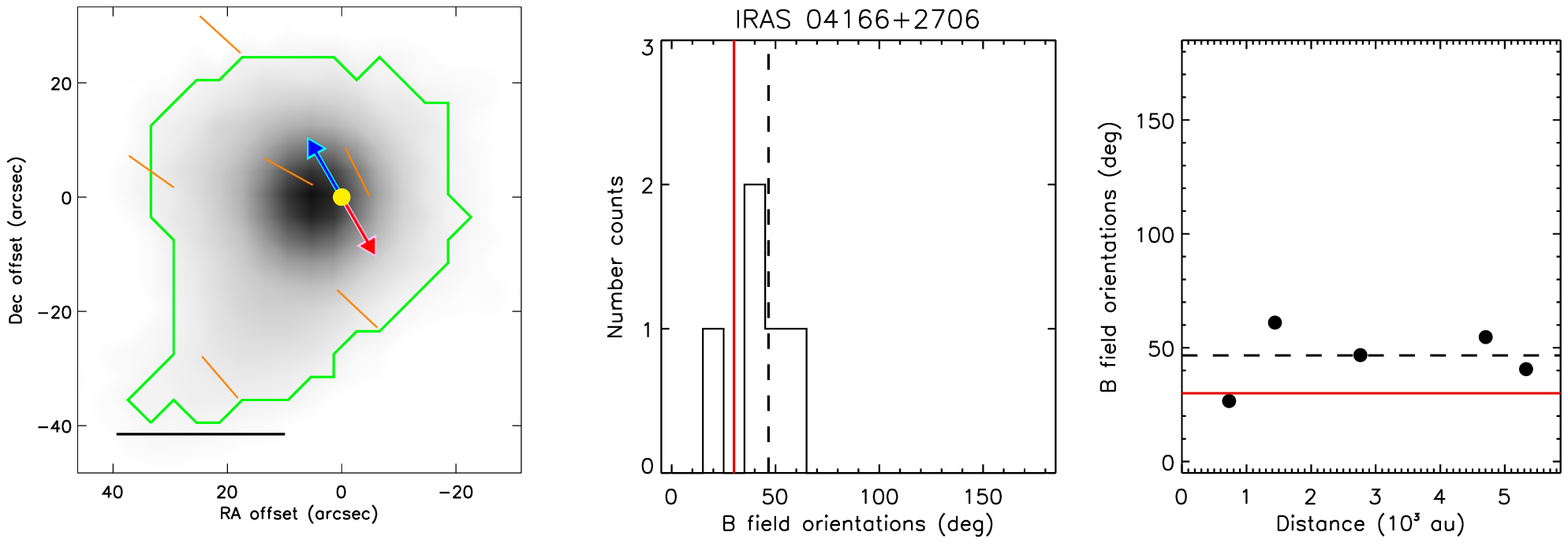}
\includegraphics[angle=90,width=0.9\textwidth]{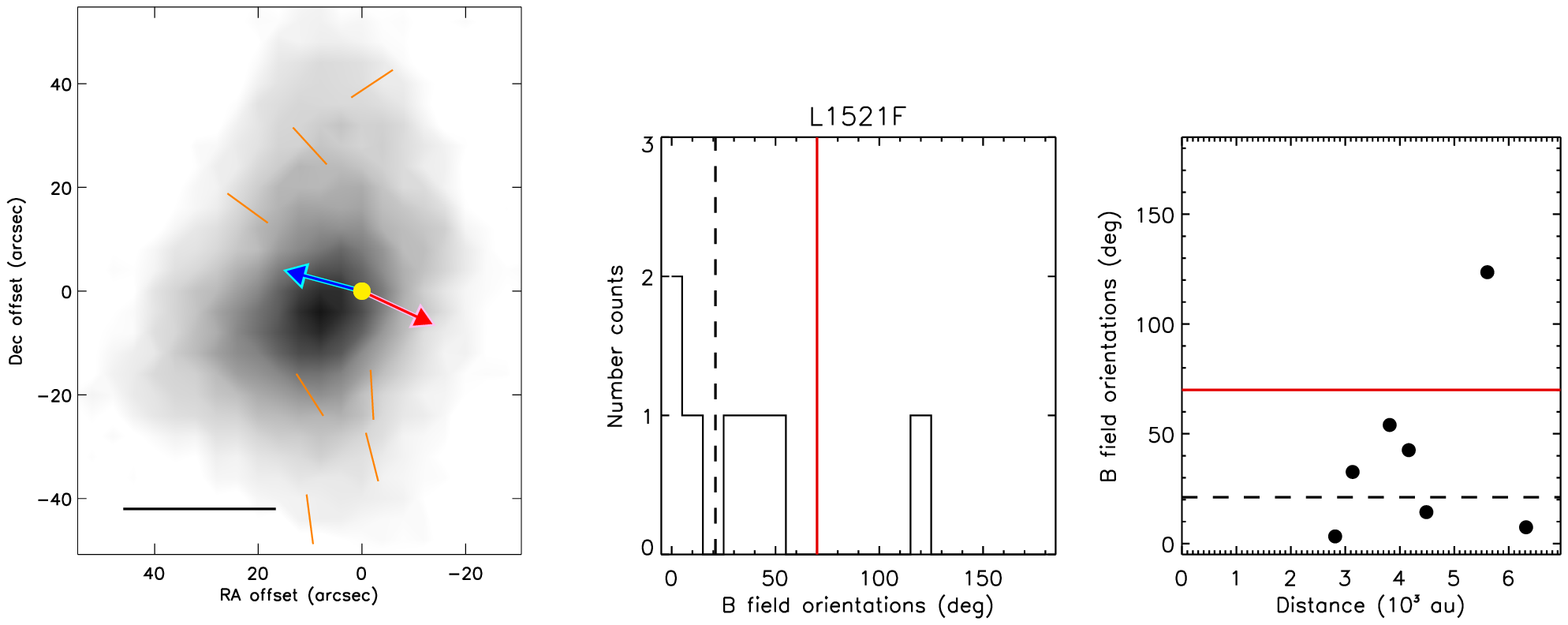}
\par
Figure \ref{subplt} --- continued.
\end{figure*}

\begin{figure*}
\centering
\includegraphics[angle=90,width=0.9\textwidth]{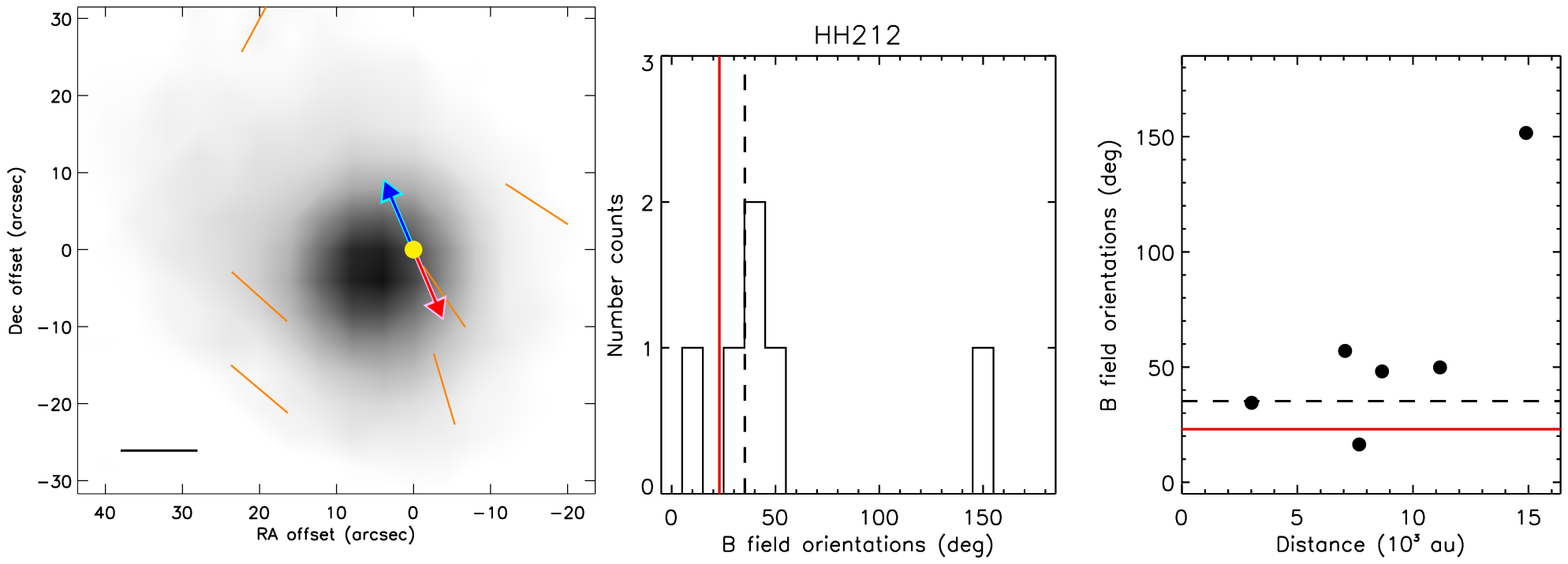}
\includegraphics[angle=90,width=0.9\textwidth]{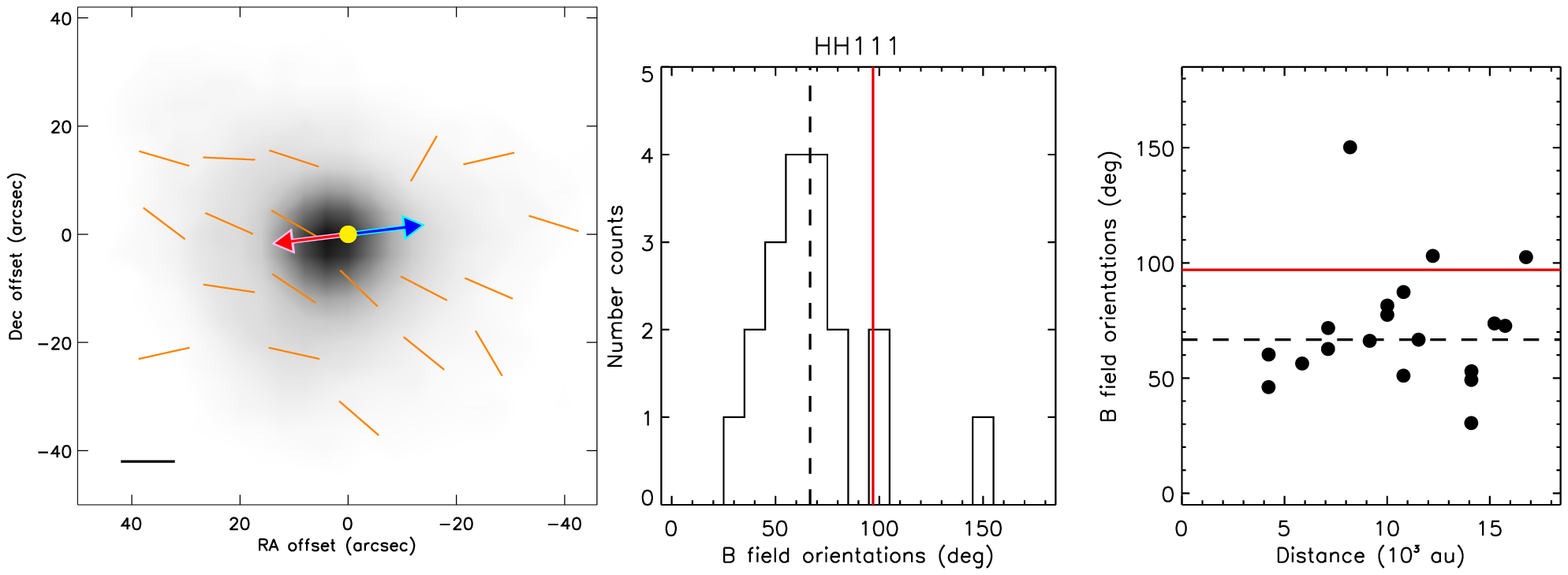}
\includegraphics[angle=90,width=0.9\textwidth]{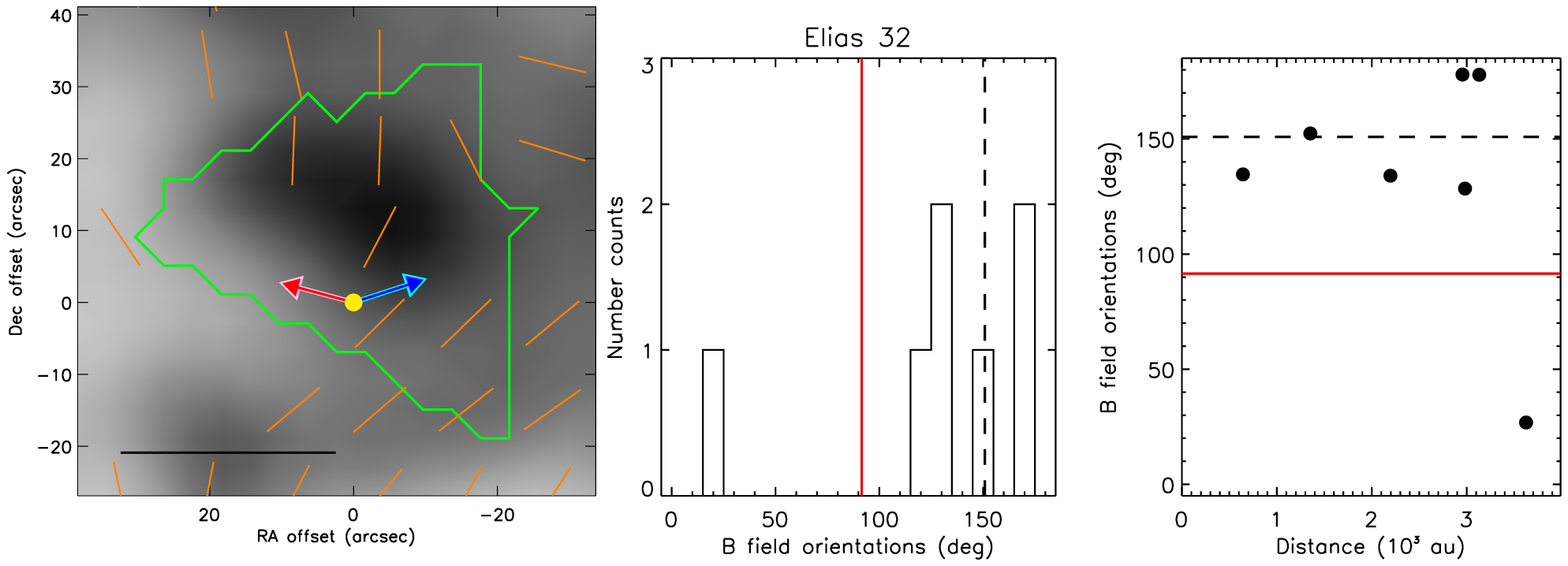}
\includegraphics[angle=90,width=0.9\textwidth]{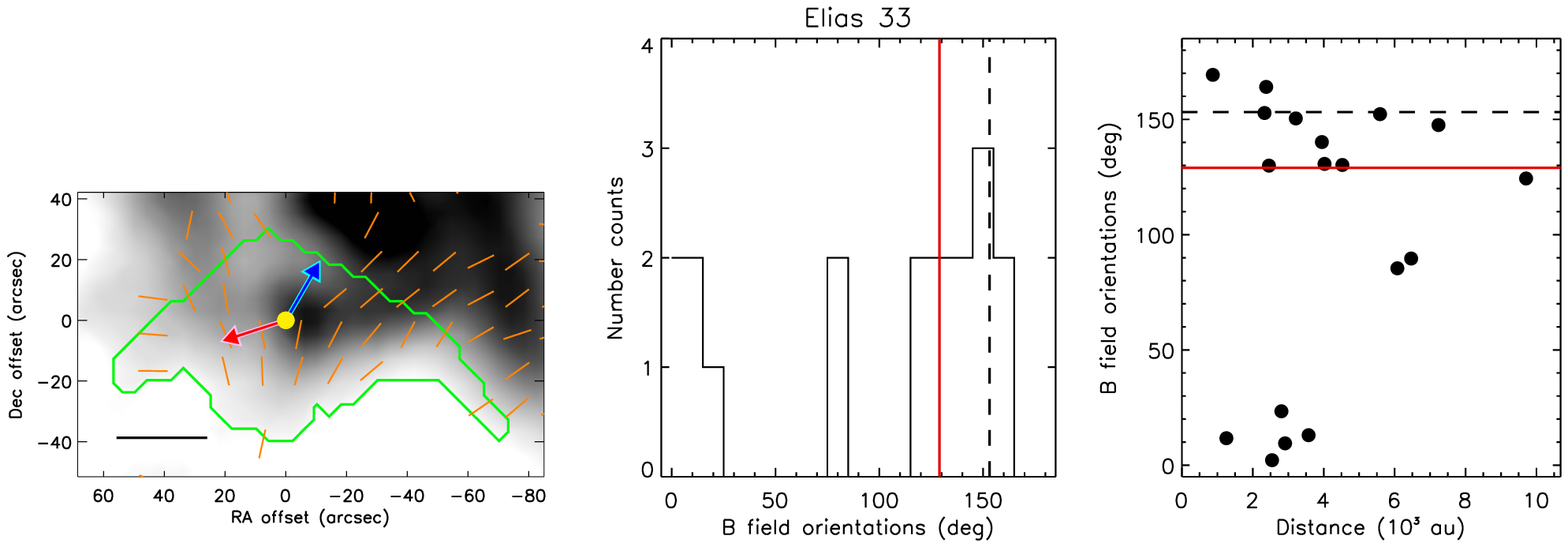}
\par
Figure \ref{subplt} --- continued.
\end{figure*}


\begin{figure*}
\centering
\includegraphics[angle=90,width=0.9\textwidth]{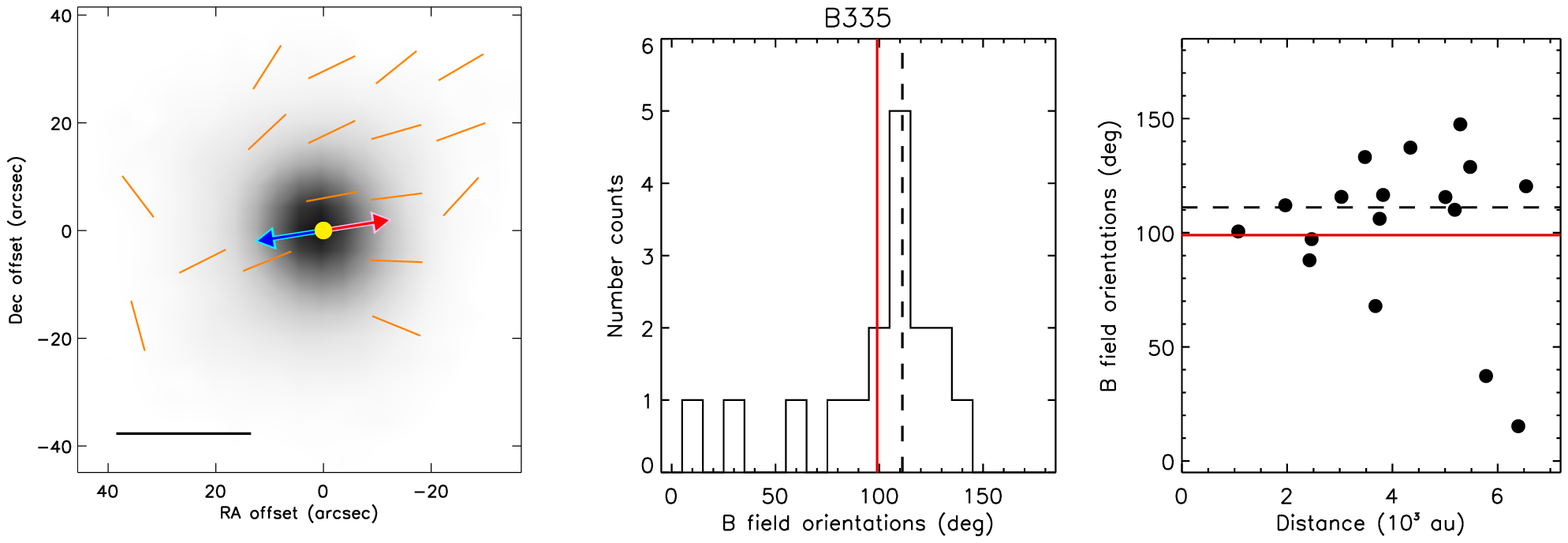}
\par
Figure \ref{subplt} --- continued.
\end{figure*}

\begin{figure*}
\centering
\includegraphics[angle=90,width=0.9\textwidth]{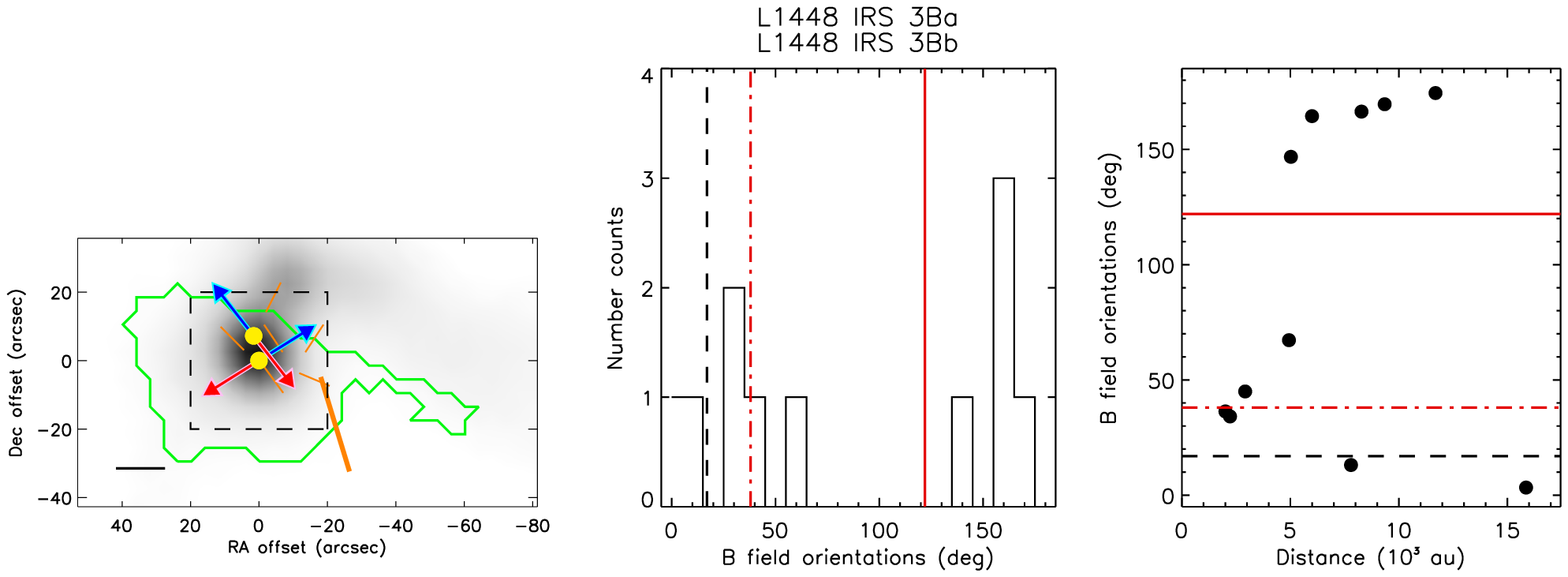}
\includegraphics[angle=90,width=0.9\textwidth]{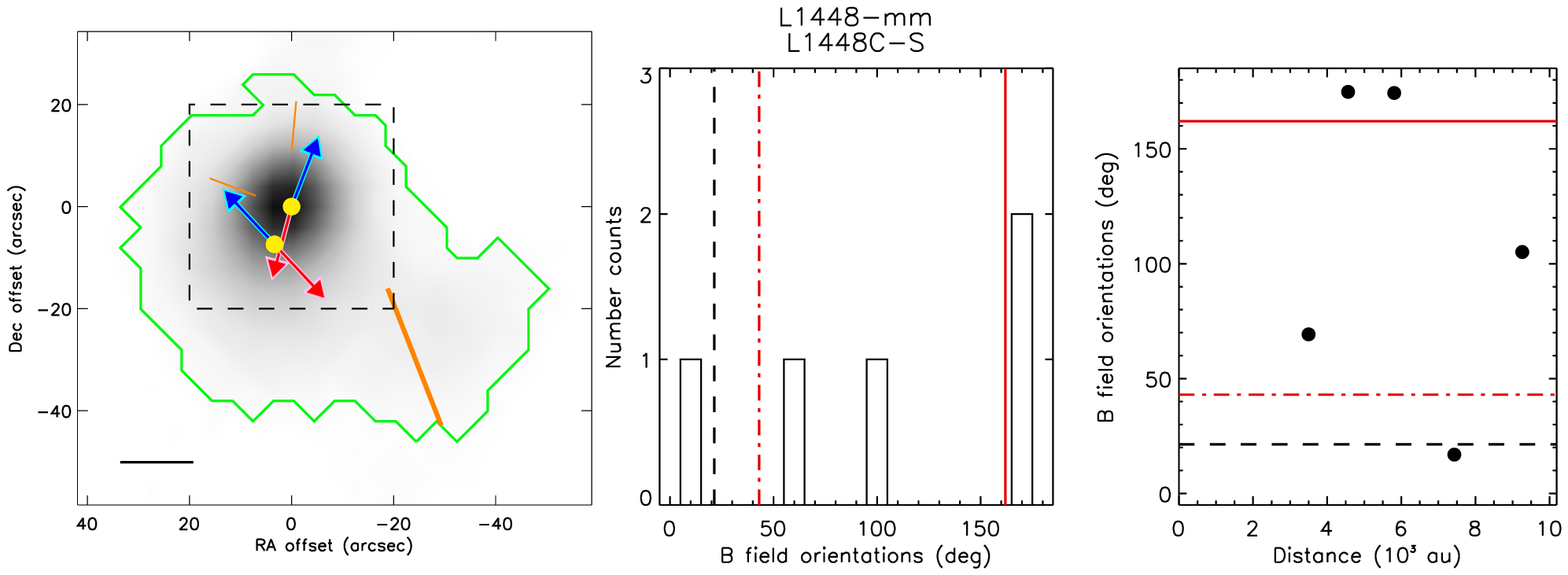}
\includegraphics[angle=90,width=0.9\textwidth]{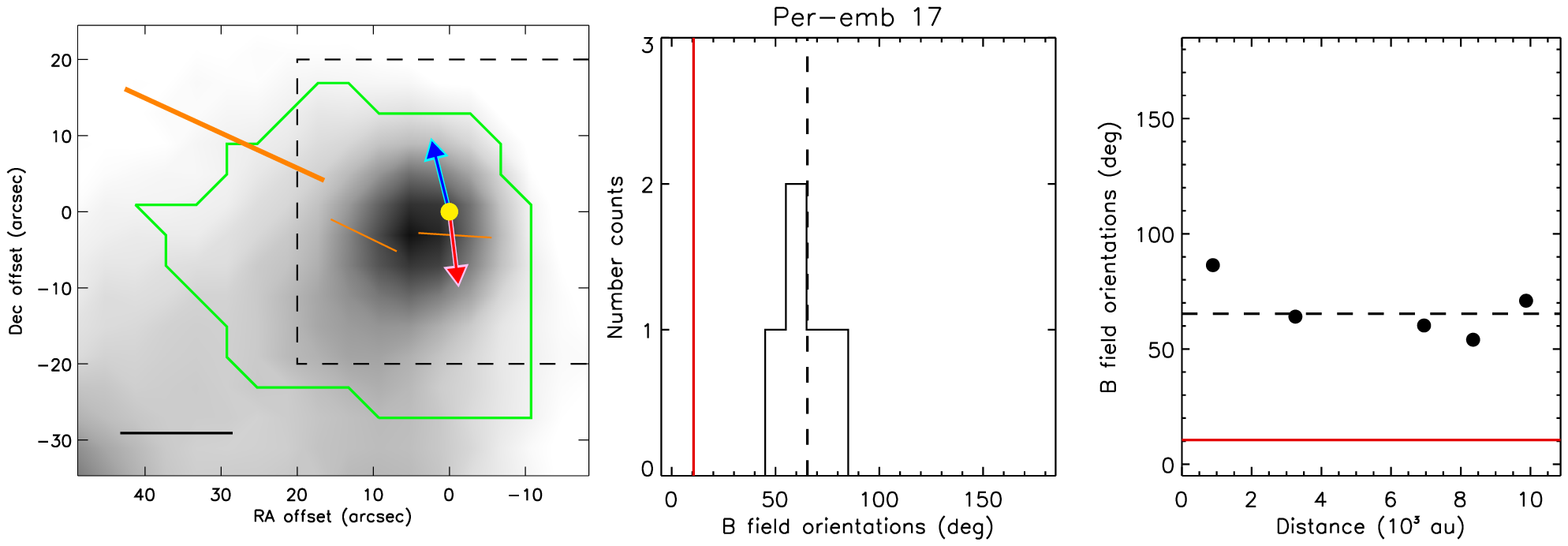}
\par
Figure \ref{subplt2} --- continued.
\end{figure*}

\begin{figure*}
\centering
\includegraphics[angle=90,width=0.9\textwidth]{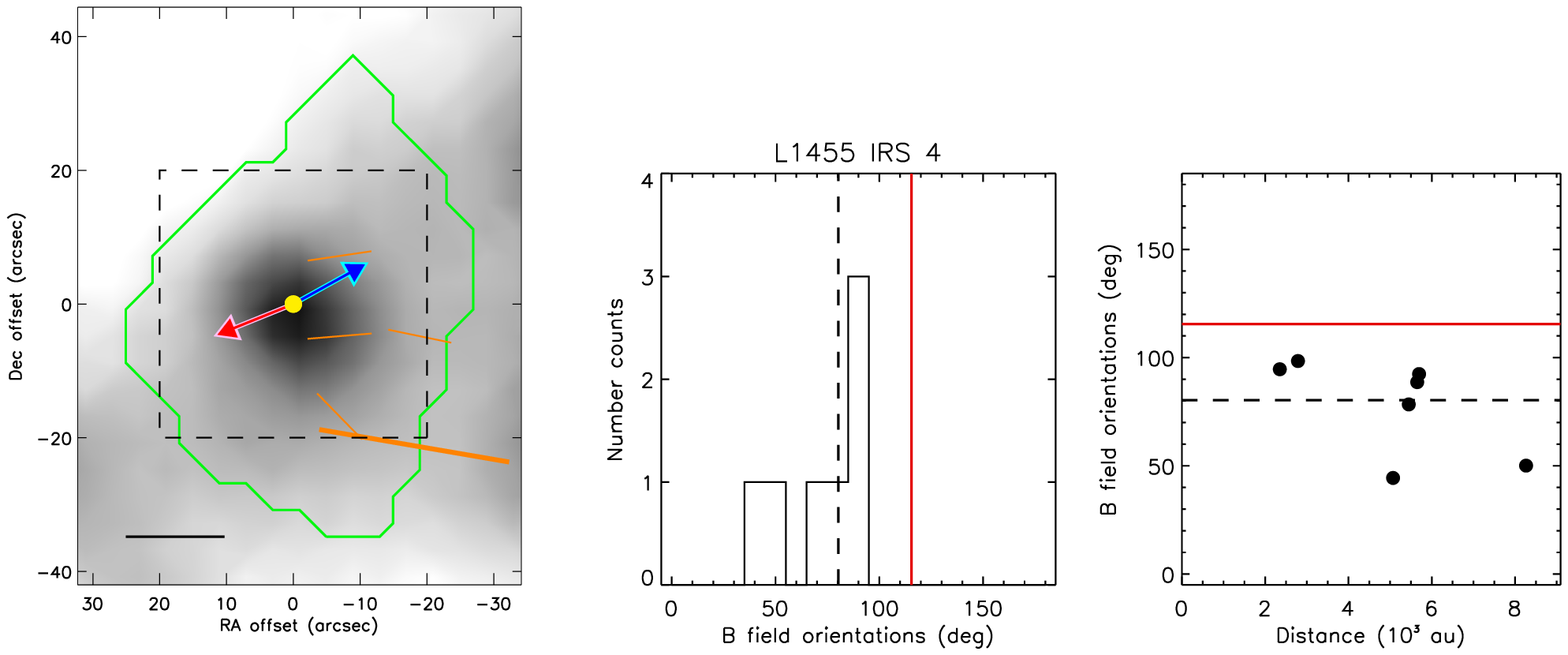}
\includegraphics[angle=90,width=0.9\textwidth]{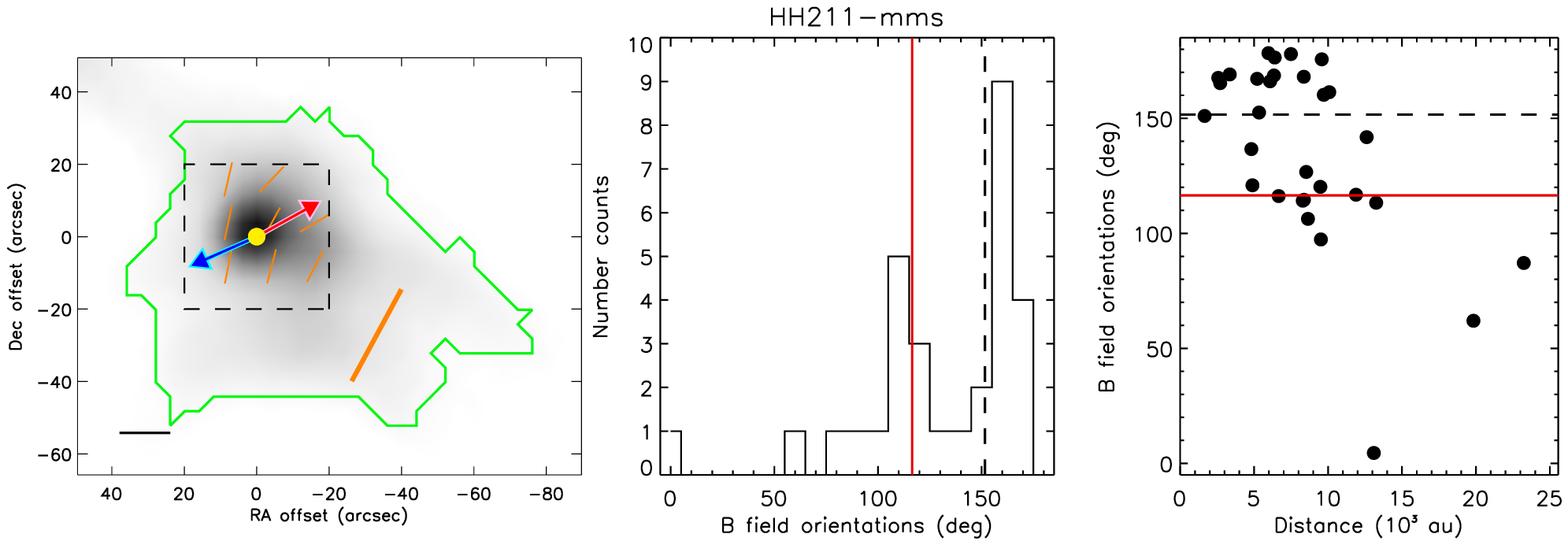}
\includegraphics[angle=90,width=0.9\textwidth]{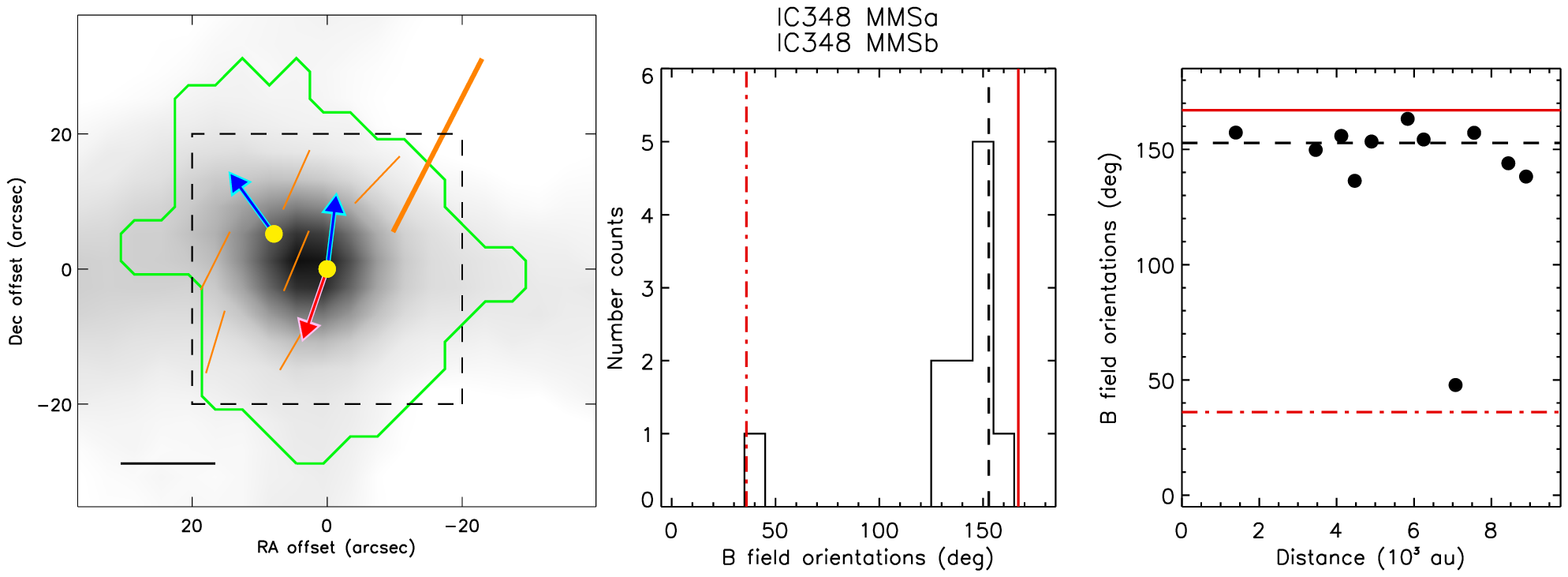}
\par
Figure \ref{subplt2} --- continued.
\end{figure*}

\begin{figure*}
\centering
\includegraphics[angle=90,width=0.9\textwidth]{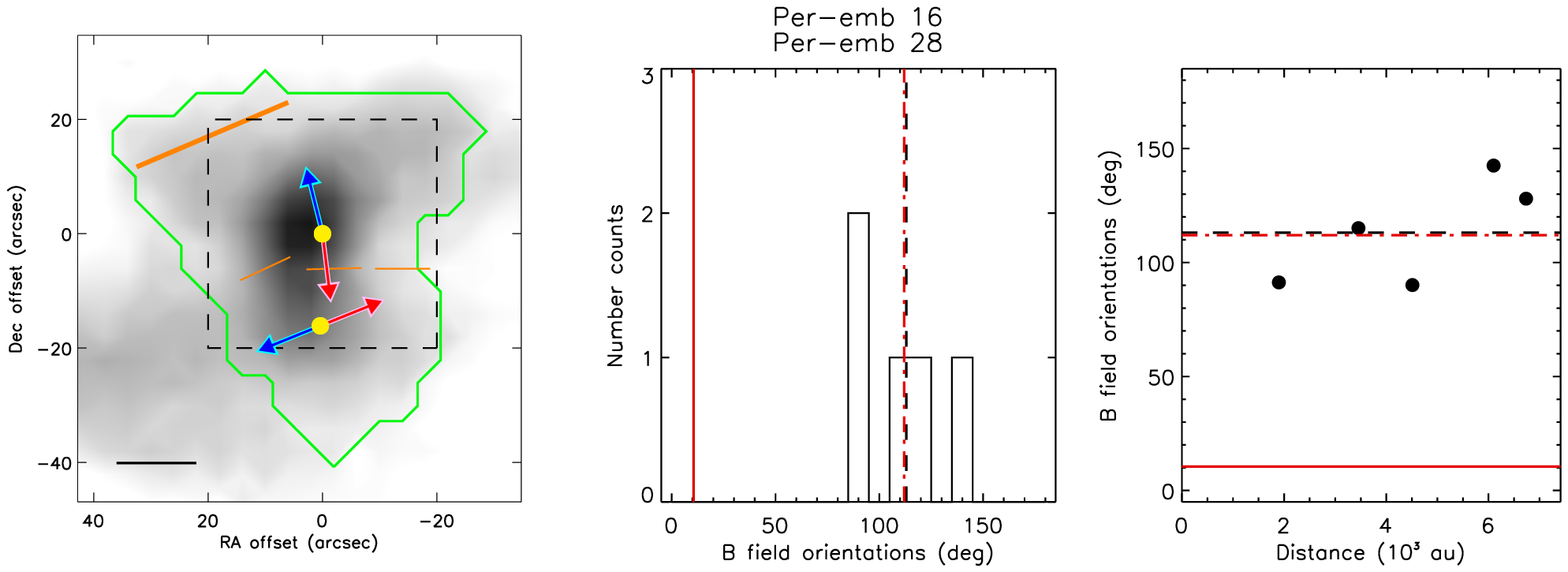}
\includegraphics[angle=90,width=0.9\textwidth]{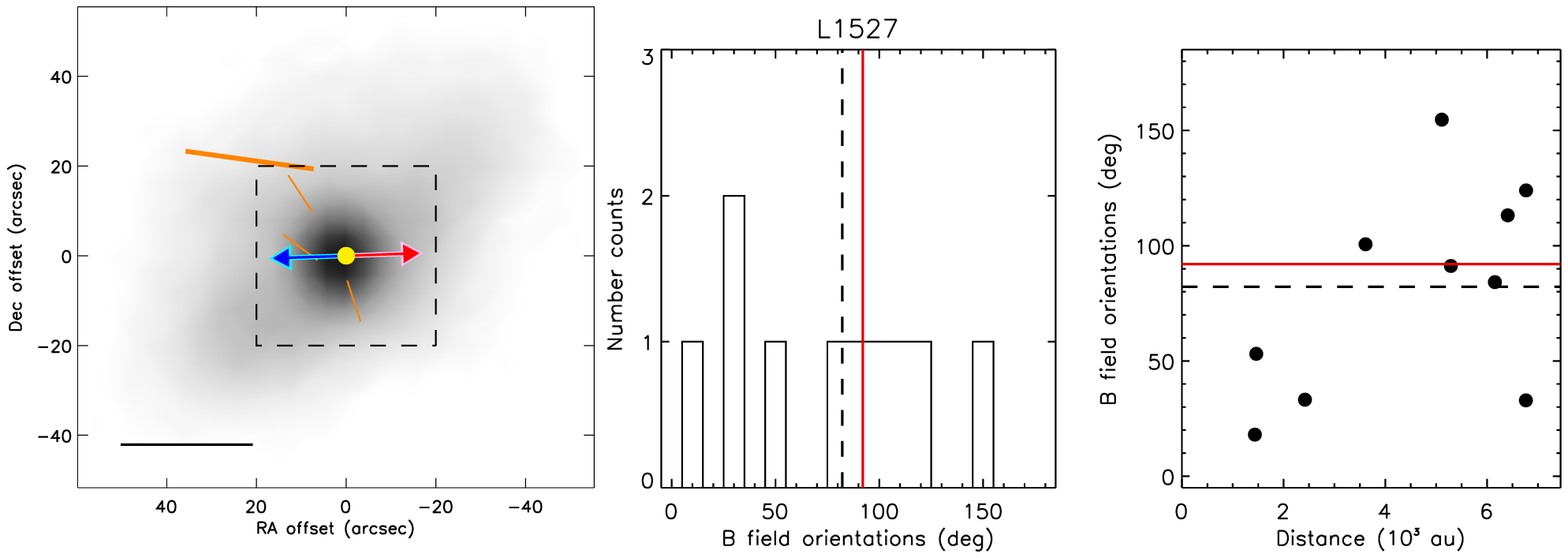}
\includegraphics[angle=90,width=0.9\textwidth]{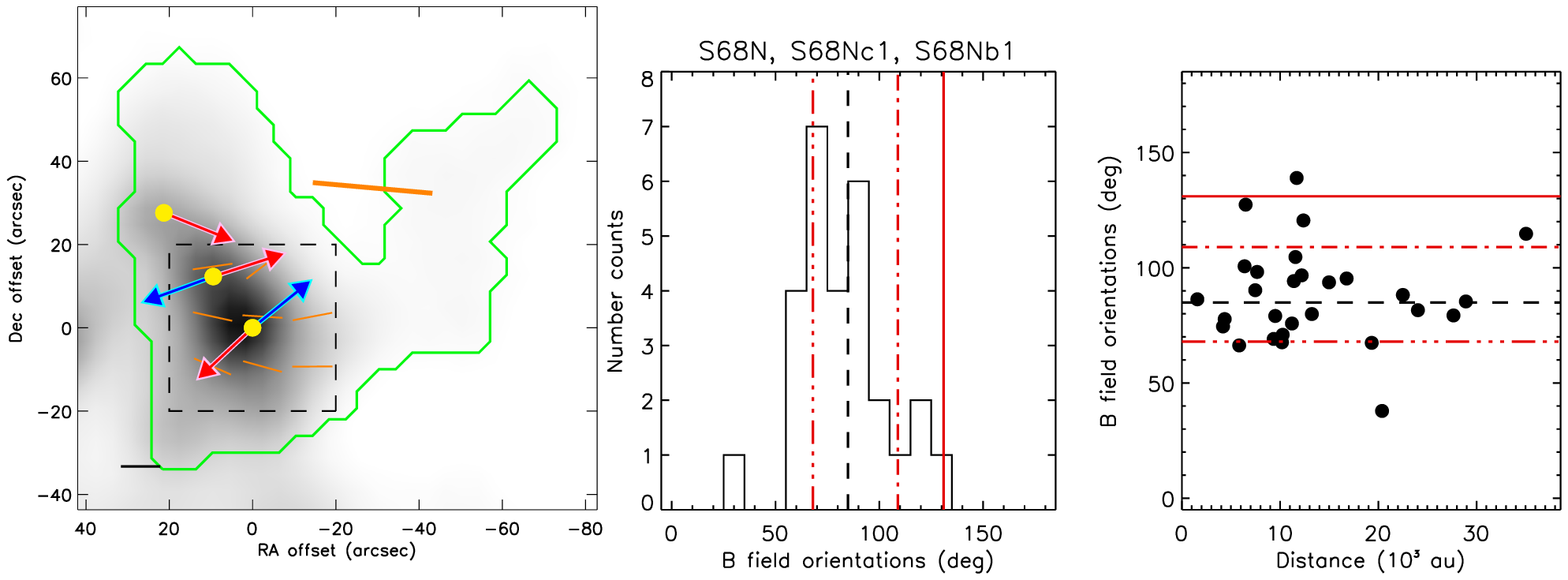}
\includegraphics[angle=90,width=0.9\textwidth]{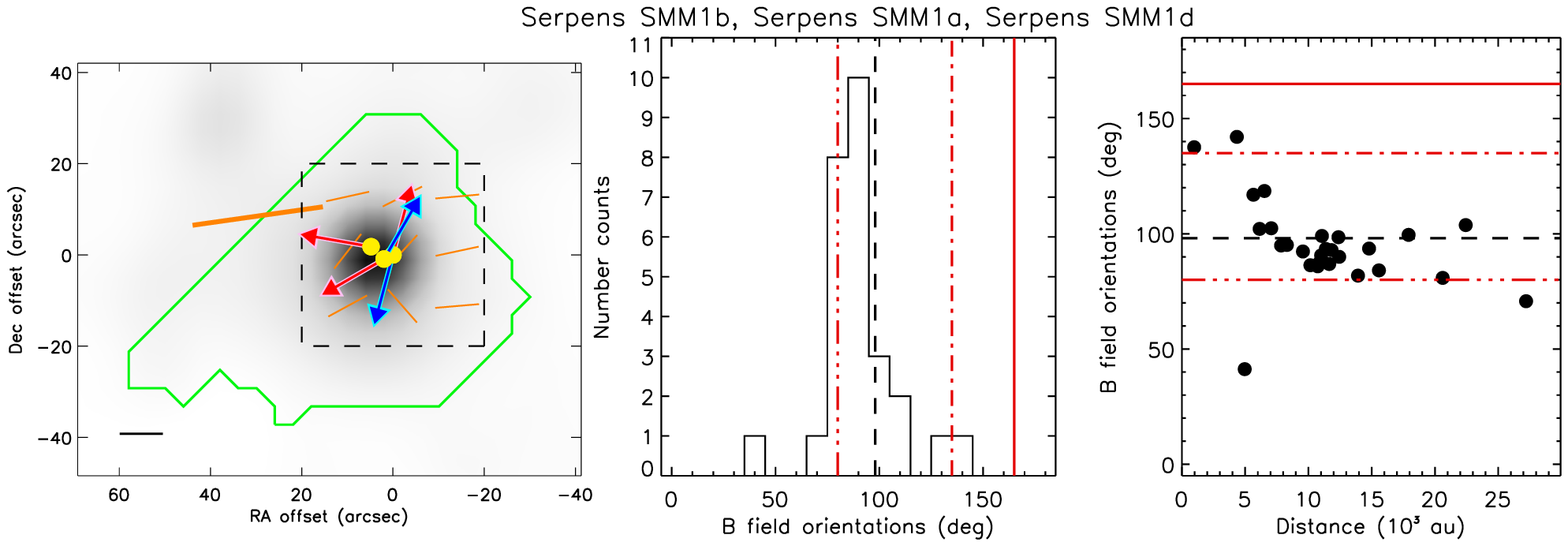}
\par
Figure \ref{subplt2} --- continued.
\end{figure*}

\begin{figure*}
\centering
\includegraphics[angle=90,width=0.9\textwidth]{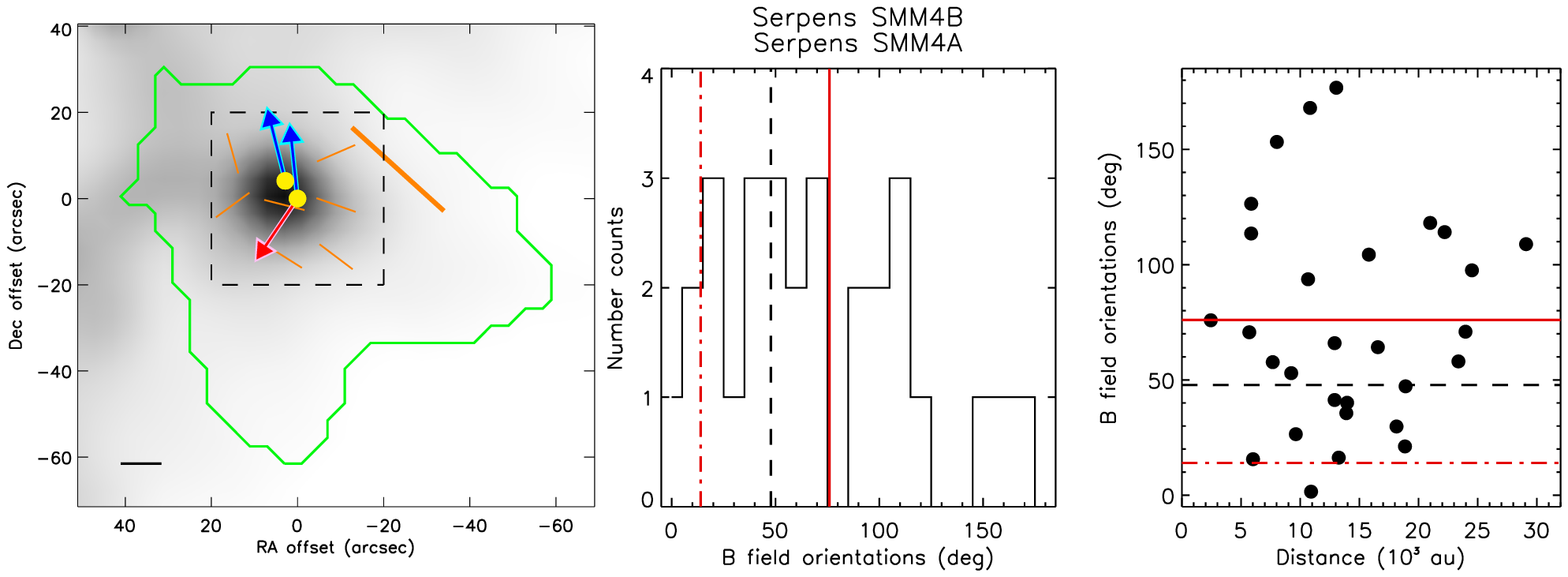}
\includegraphics[angle=90,width=0.9\textwidth]{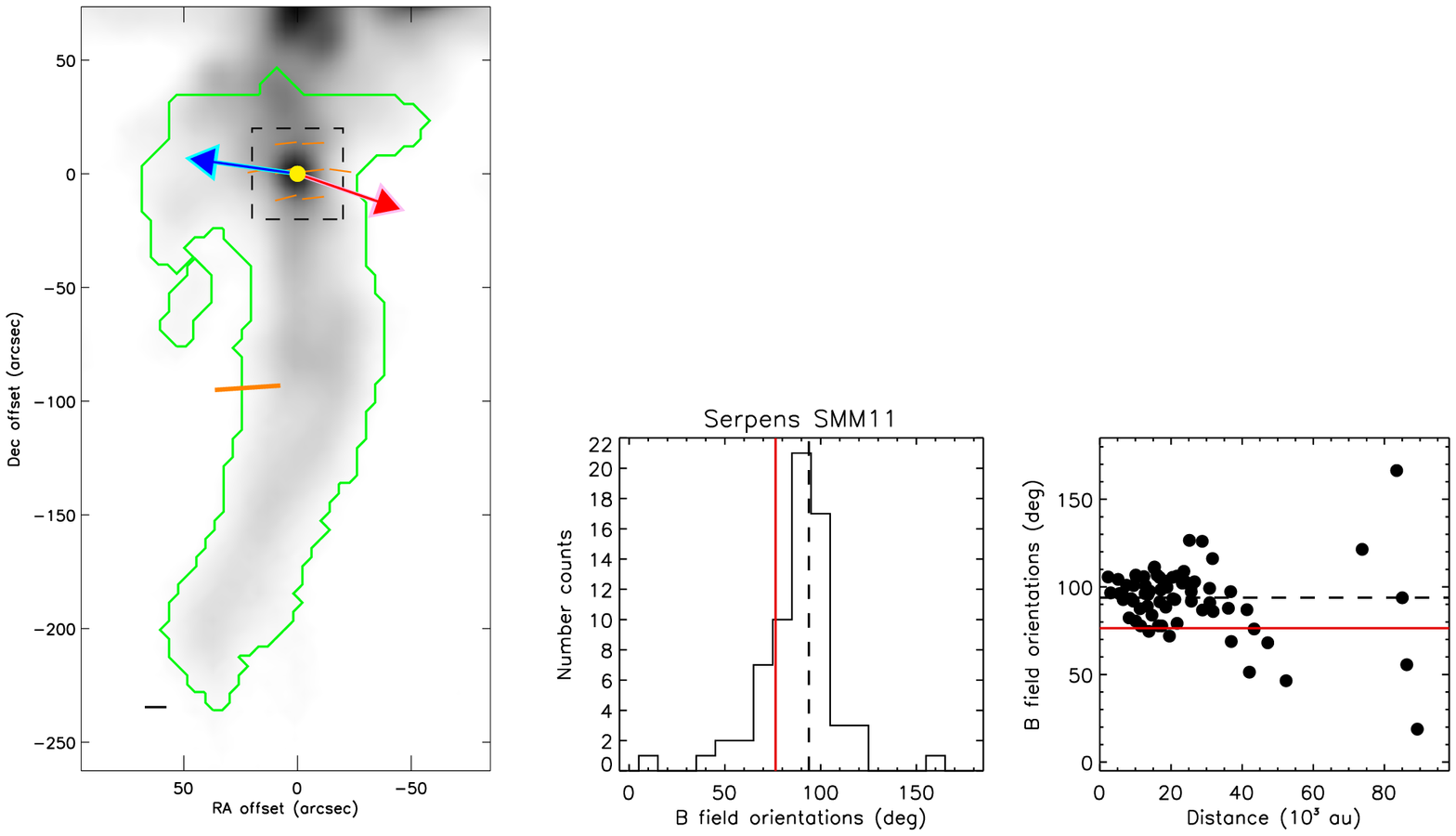}
\includegraphics[angle=90,width=0.9\textwidth]{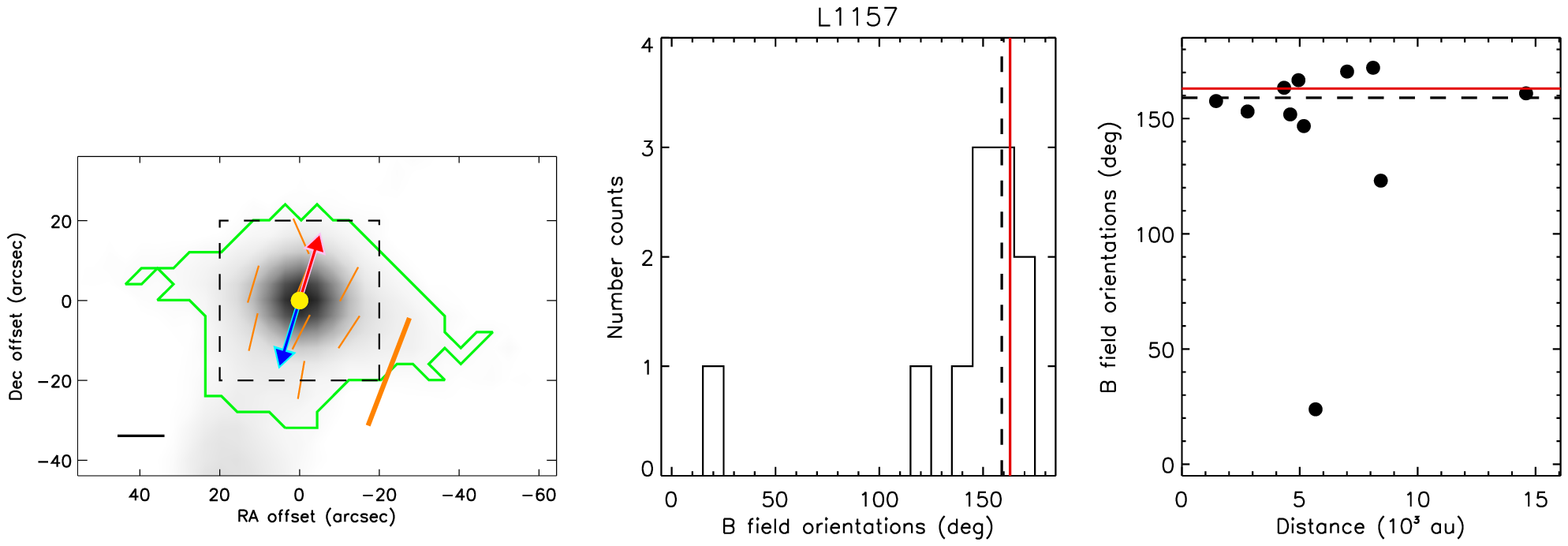}
\par
Figure \ref{subplt2} --- continued.
\end{figure*}

\section{Dependences of misalignment on source properties}\label{bias}
\begin{figure*}
\centering
\includegraphics[width=0.98\textwidth]{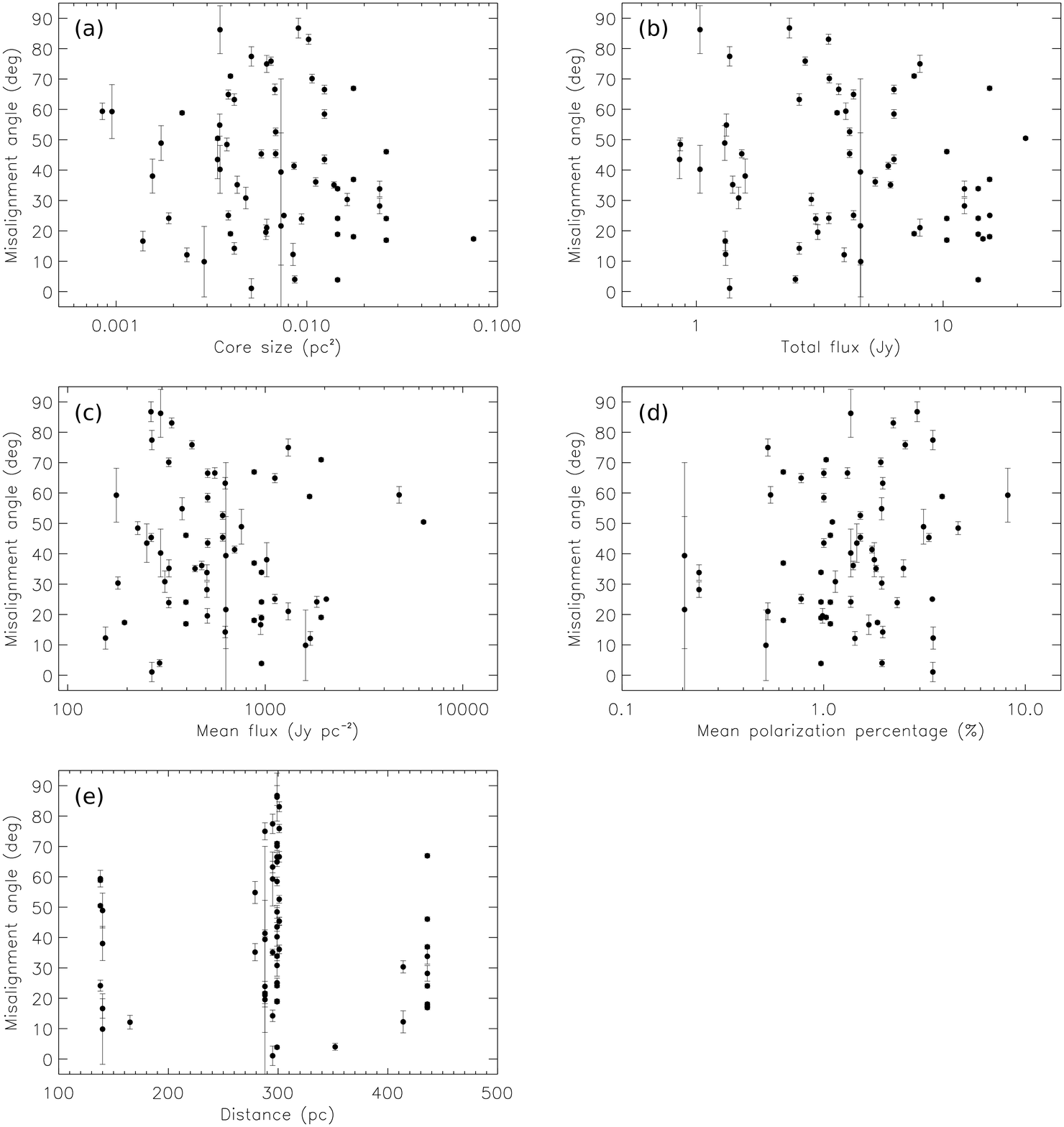}
\caption{Misalignment angles as a function of the (a) size, (b) total flux, (c) mean intensity, (d) mean polarization percentage, and (e) distance of the dense cores. Error bars present the uncertainties in the mean magnetic field orientations from the error propagation of the uncertainties of the individual polarization detections. For several sources, the error bars are smaller than the symbol size. There is additional uncertainty in the misalignment angle due to the uncertainty in the outflow orientation, which is typically 10$\arcdeg$.}\label{dep}
\end{figure*}

Figure \ref{dep} compares the measured misalignment angles with the sizes, total fluxes, mean intensities, and mean polarization percentages of the dense cores and the distances to the sources. The total flux and mean intensity are expected to be proportional to the total mass and mean column density of a dense core, respectively. These comparisons show that the measured misalignments do not depend on the mass and density of the dense cores. We note that the magnetic field structures in the nearby sources can be better resolved than in the distant sources. The comparison also shows that the measured misalignments do not depend on the distances to the sources, even though the spatial resolutions are not uniform in the sample. Therefore, there is no bias due to the properties of the dense cores or the spatial resolutions in our results.

\section{Comparison between JCMT and interferometric results}\label{int}
\begin{figure*}
\centering
\includegraphics[width=0.8\textwidth]{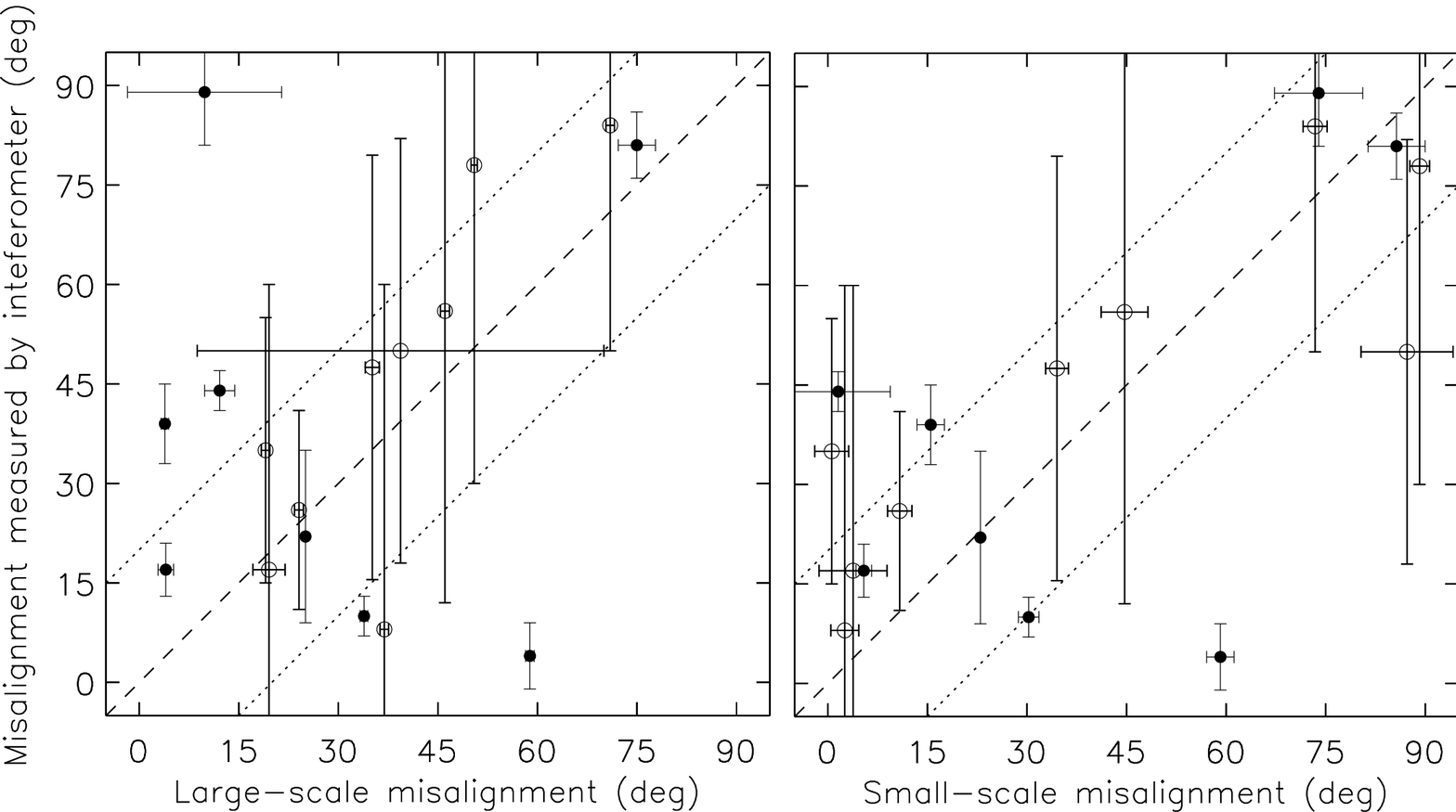}
\caption{Comparison of the large- and small-scale misalignments measured with the JCMT observations with the misalignments on a 1000 au scale measured with the interferometric observations. The horizontal axes in the left and right panels show the large- and small-scale misalignments measured with  JCMT, respectively. The vertical axes show the misalignment angles measured with the interferometer, and the results are obtained from \citet{Hull14}, \citet{Galametz18}, and \citet{Sadavoy19}. Open circles present the interferometric results from \citet{Hull14}, where the definition of the error bars is different, and their vertical error bars present the circular standard deviations of the observed magnetic field orientations. Dotted lines denote angle differences of $\pm$20$\arcdeg$.}\label{mis_int}
\end{figure*}

In 17 of our sample sources, the mean magnetic field orientations on a 1000 au scale were also measured with the CARMA, SMA, and/or ALMA observations \citep{Hull14,Galametz18,Sadavoy19}. 
We compared the large- and small-scale misalignments measured with our JCMT POL-2 data and the misalignments measured with the interferometric data (Fig.~\ref{mis_int}).
The comparison with the large-scale misalignments measured with JCMT and the interferometric results show the same trend that in most of the sources, the difference between the large- and small-scale magnetic field orientations is 10$\arcdeg$--20$\arcdeg$.
In addition, our measured small-scale misalignments are correlated with the interferometric measurements. 
Thus, the magnetic field orientations close to the stellar positions observed with JCMT can represent the overall orientations of the magnetic field structures on a 1000 au scale in the protostellar sources.

\end{appendix}

\software{Starlink \citep{Currie14}}

\begin{thebibliography}{}
\bibitem[Allen et al.(2003)]{Allen03} Allen, A., Li, Z.-Y., \& Shu, F.~H.\ 2003, \apj, 599, 363
\bibitem[Ansdell et al.(2018)]{Ansdell18} Ansdell, M., Williams, J.~P., Trapman, L., et al.\ 2018, \apj, 859, 21
\bibitem[Arce et al.(2020)]{Arce20} Arce, C., Louvet, F., Cortes, P., et al.\ 2020, arXiv:2005.12921
\bibitem[Aso et al.(2019)]{Aso19} Aso, Y., Hirano, N., Aikawa, Y., et al.\ 2019, \apj, 887, 209
\bibitem[Bachiller et al.(2001)]{Bachiller01} Bachiller, R., P{\'e}rez Guti{\'e}rrez, M., Kumar, M.~S.~N., et al.\ 2001, \aap, 372, 899
\bibitem[Basu \& Mouschovias(1994)]{Basu94} Basu, S., \& Mouschovias, T.~C.\ 1994, \apj, 432, 720
\bibitem[Blandford \& Payne(1982)]{Blandford82} Blandford, R.~D., \& Payne, D.~G.\ 1982, \mnras, 199, 883 
\bibitem[Burkert \& Bodenheimer(2000)]{Burkert00} Burkert, A., \& Bodenheimer, P.\ 2000, \apj, 543, 822
\bibitem[Chapman et al.(2013)]{Chapman13} Chapman, N.~L., Davidson, J.~A., Goldsmith, P.~F., et al.\ 2013, \apj, 770, 151
\bibitem[Chen et al.(2020)]{Chen20} Chen, C.-Y., Behrens, E.~A., Washington, J.~E., et al.\ 2020, \mnras, 494, 1971
\bibitem[Chen et al.(1995)]{Chen95} Chen, H., Myers, P.~C., Ladd, E.~F., et al.\ 1995, \apj, 445, 377
\bibitem[Chen \& Ostriker(2018)]{Chen18} Chen, C.-Y., \& Ostriker, E.~C.\ 2018, \apj, 865, 34
\bibitem[Chen \& Ostriker(2015)]{Chen15} Chen, C.-Y., \& Ostriker, E.~C.\ 2015, \apj, 810, 126
\bibitem[Chen \& Ostriker(2014)]{Chen14} Chen, C.-Y., \& Ostriker, E.~C.\ 2014, \apj, 785, 69
\bibitem[Ciardi \& Hennebelle(2010)]{Ciardi10} Ciardi, A., \& Hennebelle, P.\ 2010, \mnras, 409, L39
\bibitem[Cieza et al.(2019)]{Cieza19} Cieza, L.~A., Ru{\'\i}z-Rodr{\'\i}guez, D., Hales, A., et al.\ 2019, \mnras, 482, 698
\bibitem[Coud{\'e} et al.(2019)]{Coude19} Coud{\'e}, S., Bastien, P., Houde, M., et al.\ 2019, \apj, 877, 88
\bibitem[Cox et al.(2018)]{Cox18} Cox, E.~G., Harris, R.~J., Looney, L.~W., et al.\ 2018, \apj, 855, 92
\bibitem[Crapsi et al.(2008)]{Crapsi08} Crapsi, A., van Dishoeck, E.~F., Hogerheijde, M.~R., et al.\ 2008, \aap, 486, 245
\bibitem[Crutcher(2012)]{Crutcher12} Crutcher, R.~M.\ 2012, \araa, 50, 29
\bibitem[Crutcher et al.(2010)]{Crutcher10} Crutcher, R.~M., Wandelt, B., Heiles, C., et al.\ 2010, \apj, 725, 466
\bibitem[Currie et al.(2014)]{Currie14} Currie, M.~J., Berry, D.~S., Jenness, T., et al.\ 2014, Astronomical Data Analysis Software and Systems XXIII, 485, 391 
\bibitem[Davidson et al.(2011)]{Davidson11} Davidson, J.~A., Novak, G., Matthews, T.~G., et al.\ 2011, \apj, 732, 97
\bibitem[Dib et al.(2010)]{Dib10} Dib, S., Hennebelle, P., Pineda, J.~E., et al.\ 2010, \apj, 723, 425
\bibitem[Doi et al.(2020)]{Doi20} Doi, Y., Hasegawa, T., Furuya, R.~S., et al.\ 2020, \apj, 899, 28
\bibitem[Dunham et al.(2008)]{Dunham08} Dunham, M.~M., Crapsi, A., Evans, N.~J., et al.\ 2008, \apjs, 179, 249
\bibitem[Enoch et al.(2008)]{Enoch08} Enoch, M.~L., Evans, N.~J., Sargent, A.~I., et al.\ 2008, \apj, 684, 1240
\bibitem[Evans et al.(2009)]{Evans09} Evans, N.~J., Dunham, M.~M., J{\o}rgensen, J.~K., et al.\ 2009, \apjs, 181, 321
\bibitem[Friberg et al.(2016)]{Friberg16} Friberg, P., Bastien, P., Berry, D., et al.\ 2016, \procspie, 9914, 991403 
\bibitem[Friesen et al.(2018)]{Friesen18} Friesen, R.~K., Pon, A., Bourke, T.~L., et al.\ 2018, \apj, 869, 158
\bibitem[Galametz et al.(2018)]{Galametz18} Galametz, M., Maury, A., Girart, J.~M., et al.\ 2018, \aap, 616, A139
\bibitem[Galli \& Shu(1993)]{Galli93} Galli, D., \& Shu, F.~H.\ 1993, \apj, 417, 220
\bibitem[Gammie et al.(2003)]{Gammie03} Gammie, C.~F., Lin, Y.-T., Stone, J.~M., et al.\ 2003, \apj, 592, 203
\bibitem[Gaudel et al.(2020)]{Gaudel20} Gaudel, M., Maury, A.~J., Belloche, A., et al.\ 2020, \aap, 637, A92
\bibitem[Girart et al.(2006)]{Girart06} Girart, J.~M., Rao, R., \& Marrone, D.~P.\ 2006, Science, 313, 812
\bibitem[Gray et al.(2018)]{Gray18} Gray, W.~J., McKee, C.~F., \& Klein, R.~I.\ 2018, \mnras, 473, 2124
\bibitem[Haisch et al.(2001)]{Haisch01} Haisch, K.~E., Lada, E.~A., \& Lada, C.~J.\ 2001, \apjl, 553, L153
\bibitem[Hirano et al.(2020)]{Hirano20} Hirano, S., Tsukamoto, Y., Basu, S., et al.\ 2020, \apj, 898, 118
\bibitem[Holland et al.(2013)]{Holland13} Holland, W.~S., Bintley, D., Chapin, E.~L., et al.\ 2013, \mnras, 430, 2513
\bibitem[Hsieh et al.(2017)]{Hsieh17} Hsieh, T.-H., Lai, S.-P., \& Belloche, A.\ 2017, \aj, 153, 173
\bibitem[Hull et al.(2017a)]{Hull17a} Hull, C.~L.~H., Girart, J.~M., Tychoniec, {\L}., et al.\ 2017, \apj, 847, 92
\bibitem[Hull et al.(2017b)]{Hull17} Hull, C.~L.~H., Girart, J.~M., Tychoniec, {\L}., et al.\ 2017, \apj, 847, 92
\bibitem[Hull et al.(2020)]{Hull20} Hull, C.~L.~H., Gouellec, V.~J.~M.~L., Girart, J.~M., et al.\ 2020, \apj, 892, 152
\bibitem[Hull et al.(2013)]{Hull13} Hull, C.~L.~H., Plambeck, R.~L., Bolatto, A.~D., et al.\ 2013, \apj, 768, 159
\bibitem[Hull et al.(2014)]{Hull14} Hull, C.~L.~H., Plambeck, R.~L., Kwon, W., et al.\ 2014, \apjs, 213, 13
\bibitem[Hull \& Zhang(2019)]{Hull19} Hull, C.~L.~H. \& Zhang, Q.\ 2019, Frontiers in Astronomy and Space Sciences, 6, 3
\bibitem[Joos et al.(2012)]{Joos12} Joos, M., Hennebelle, P., \& Ciardi, A.\ 2012, \aap, 543, A128
\bibitem[Joos et al.(2013)]{Joos13} Joos, M., Hennebelle, P., Ciardi, A., et al.\ 2013, \aap, 554, A17
\bibitem[Kamazaki et al.(2019)]{Kamazaki19} Kamazaki, T., Nakamura, F., Kawabe, R., et al.\ 2019, \apj, 871, 86
\bibitem[Ko et al.(2019)]{Ko19} Ko, C.-L., Liu, H.~B., Lai, S.-P., et al.\ 2019, arXiv e-prints, arXiv:1909.09628
\bibitem[Koch et al.(2014)]{Koch14} Koch, P.~M., Tang, Y.-W., Ho, P.~T.~P., et al.\ 2014, \apj, 797, 99
\bibitem[Kristensen et al.(2012)]{Kristensen12} Kristensen, L.~E., van Dishoeck, E.~F., Bergin, E.~A., et al.\ 2012, \aap, 542, A8
\bibitem[Krumholz et al.(2013)]{Krumholz13} Krumholz, M.~R., Crutcher, R.~M., \& Hull, C.~L.~H.\ 2013, \apjl, 767, L11
\bibitem[Kuznetsova et al.(2020)]{Kuznetsova20} Kuznetsova, A., Hartmann, L., \& Heitsch, F.\ 2020, \apj, 893, 73
\bibitem[Kwon et al.(2018)]{Kwon18} Kwon, J., Doi, Y., Tamura, M., et al.\ 2018, \apj, 859, 4
\bibitem[Kwon et al.(2019)]{Kwon19} Kwon, W., Stephens, I.~W., Tobin, J.~J., et al.\ 2019, \apj, 879, 25
\bibitem[Lam et al.(2019)]{Lam19} Lam, K.~H., Li, Z.-Y., Chen, C.-Y., et al.\ 2019, \mnras, 489, 5326
\bibitem[Le Gouellec et al.(2019)]{LeGouellec19} Le Gouellec, V.~J.~M., Hull, C.~L.~H., Maury, A.~J., et al.\ 2019, \apj, 885, 106
\bibitem[Lee et al.(2017)]{LeeHull17} Lee, J.~W.~Y., Hull, C.~L.~H., \& Offner, S.~S.~R.\ 2017, \apj, 834, 201
\bibitem[Lee et al.(2017)]{Lee17} Lee, C.-F., Ho, P.~T.~P., Li, Z.-Y., et al.\ 2017, Nature Astronomy, 1, 0152
\bibitem[Lee et al.(2016)]{Lee16} Lee, C.-F., Hwang, H.-C., \& Li, Z.-Y.\ 2016, \apj, 826, 213
\bibitem[Lee et al.(2019)]{Lee19} Lee, C.-F., Kwon, W., Jhan, K.-S., et al.\ 2019, \apj, 879, 101
\bibitem[Li et al.(2014)]{Li14} Li, Z.-Y., Banerjee, R., Pudritz, R.~E., et al.\ 2014, Protostars and Planets VI, 173
\bibitem[Li et al.(2013)]{Li13} Li, Z.-Y., Krasnopolsky, R., \& Shang, H.\ 2013, \apj, 774, 82
\bibitem[Li et al.(2004)]{Li04} Li, P.~S., Norman, M.~L., Mac Low, M.-M., et al.\ 2004, \apj, 605, 800
\bibitem[Liu et al.(2019)]{Liu19} Liu, J., Qiu, K., Berry, D., et al.\ 2019, \apj, 877, 43
\bibitem[Maury et al.(2018)]{Maury18} Maury, A.~J., Girart, J.~M., Zhang, Q., et al.\ 2018, \mnras, 477, 2760
\bibitem[Masson et al.(2016)]{Masson16} Masson, J., Chabrier, G., Hennebelle, P., Vaytet, N., \& Commer{\c c}on, B.\ 2016, \aap, 587, A32 
\bibitem[Matsumoto et al.(2017)]{Matsumoto17} Matsumoto, T., Machida, M.~N., \& Inutsuka, S.\ 2017, \apj, 839, 69
\bibitem[Matthews et al.(2009)]{Matthews09} Matthews, B.~C., McPhee, C.~A., Fissel, L.~M., et al.\ 2009, \apjs, 182, 143
\bibitem[Maury et al.(2019)]{Maury19} Maury, A.~J., Andr{\'e}, P., Testi, L., et al.\ 2019, \aap, 621, A76
\bibitem[McKee \& Ostriker(2007)]{McKee07} McKee, C.~F., \& Ostriker, E.~C.\ 2007, \araa, 45, 565
\bibitem[Mellon \& Li(2008)]{Mellon08} Mellon, R.~R., \& Li, Z.-Y.\ 2008, \apj, 681, 1356
\bibitem[Menten et al.(2007)]{Menten07} Menten, K.~M., Reid, M.~J., Forbrich, J., et al.\ 2007, \aap, 474, 515
\bibitem[Motte \& Andr{\'e}(2001)]{Motte01} Motte, F. \& Andr{\'e}, P.\ 2001, \aap, 365, 440
\bibitem[Mouschovias \& Paleologou(1979)]{Mouschovias79} Mouschovias, T.~C., \& Paleologou, E.~V.\ 1979, \apj, 230, 204
\bibitem[Murillo et al.(2013)]{Murillo13} Murillo, N.~M., Lai, S.-P., Bruderer, S., et al.\ 2013, \aap, 560, A103
\bibitem[Nakano \& Nakamura(1978)]{Nakano78} Nakano, T. \& Nakamura, T.\ 1978, \pasj, 30, 671
\bibitem[Ortiz-Le{\'o}n et al.(2018a)]{Ortiz18a} Ortiz-Le{\'o}n, G.~N., Loinard, L., Dzib, S.~A., et al.\ 2018, \apjl, 869, L33
\bibitem[Ortiz-Le{\'o}n et al.(2018b)]{Ortiz18b} Ortiz-Le{\'o}n, G.~N., Loinard, L., Dzib, S.~A., et al.\ 2018, \apj, 865, 73
\bibitem[Palmeirim et al.(2013)]{Palmeirim13} Palmeirim, P., Andr{\'e}, P., Kirk, J., et al.\ 2013, \aap, 550, A38
\bibitem[Pattle \& Fissel(2019)]{Pattle19b} Pattle, K. \& Fissel, L.\ 2019, Frontiers in Astronomy and Space Sciences, 6, 15
\bibitem[Pattle et al.(2019)]{Pattle19} Pattle, K., Lai, S.-P., Hasegawa, T., et al.\ 2019, \apj, 880, 27
\bibitem[Pattle et al.(2018)]{Pattle18} Pattle, K., Ward-Thompson, D., Hasegawa, T., et al.\ 2018, \apjl, 860, L6
\bibitem[Pattle et al.(2017)]{Pattle17} Pattle, K., Ward-Thompson, D., Berry, D., et al.\ 2017, \apj, 846, 122
\bibitem[Pi{\'e}tu et al.(2007)]{Pietu07} Pi{\'e}tu, V., Dutrey, A., \& Guilloteau, S.\ 2007, \aap, 467, 163
\bibitem[Planck Collaboration et al.(2016a)]{Planck16a} Planck Collaboration, Adam, R., Ade, P.~A.~R., et al.\ 2016, \aap, 586, A135
\bibitem[Planck Collaboration et al.(2015)]{Planck15} Planck Collaboration, Ade, P.~A.~R., Aghanim, N., et al.\ 2015, \aap, 576, A104
\bibitem[Planck Collaboration et al.(2016b)]{Planck16b} Planck Collaboration, Ade, P.~A.~R., Aghanim, N., et al.\ 2016, \aap, 586, A138
\bibitem[Poidevin et al.(2013)]{Poidevin13} Poidevin, F., Falceta-Gon{\c{c}}alves, D., Kowal, G., et al.\ 2013, \apj, 777, 112
\bibitem[Pudritz \& Norman(1983)]{Pudritz83} Pudritz, R.~E. \& Norman, C.~A.\ 1983, \apj, 274, 677
\bibitem[Rao et al.(2014)]{Rao14} Rao, R., Girart, J.~M., Lai, S.-P., et al.\ 2014, \apjl, 780, L6
\bibitem[Santangelo et al.(2015)]{Santangelo15} Santangelo, G., Murillo, N.~M., Nisini, B., et al.\ 2015, \aap, 581, A91
\bibitem[Sadavoy et al.(2018)]{Sadavoy18} Sadavoy, S.~I., Myers, P.~C., Stephens, I.~W., et al.\ 2018, \apj, 859, 165
\bibitem[Sadavoy et al.(2019)]{Sadavoy19} Sadavoy, S.~I., Stephens, I.~W., Myers, P.~C., et al.\ 2019, \apjs, 245, 2
\bibitem[Simon et al.(2017)]{Simon17} Simon, M., Guilloteau, S., Di Folco, E., et al.\ 2017, \apj, 844, 158
\bibitem[Soam et al.(2019)]{Soam19} Soam, A., Lee, C.~W., Andersson, B.-G., et al.\ 2019, \apj, 883, 9
\bibitem[Soam et al.(2018)]{Soam18} Soam, A., Pattle, K., Ward-Thompson, D., et al.\ 2018, \apj, 861, 65
\bibitem[Soler et al.(2013)]{Soler13} Soler, J.~D., Hennebelle, P., Martin, P.~G., et al.\ 2013, \apj, 774, 128
\bibitem[Stephens et al.(2017)]{Stephens17} Stephens, I.~W., Dunham, M.~M., Myers, P.~C., et al.\ 2017, \apj, 846, 16
\bibitem[Stephens et al.(2013)]{Stephens13} Stephens, I.~W., Looney, L.~W., Kwon, W., et al.\ 2013, \apjl, 769, L15
\bibitem[Tafalla et al.(2015)]{Tafalla15} Tafalla, M., Bachiller, R., Lefloch, B., et al.\ 2015, \aap, 573, L2
\bibitem[Tafalla et al.(2013)]{Tafalla13} Tafalla, M., Liseau, R., Nisini, B., et al.\ 2013, \aap, 551, A116
\bibitem[Takahashi et al.(2013)]{Takahashi13} Takahashi, S., Ohashi, N., \& Bourke, T.~L.\ 2013, \apj, 774, 20
\bibitem[Takakuwa et al.(2018)]{Takakuwa18} Takakuwa, S., Tsukamoto, Y., Saigo, K., et al.\ 2018, \apj, 865, 51
\bibitem[Tobin et al.(2020)]{Tobin20} Tobin, J.~J., Sheehan, P., Megeath, S.~T., et al.\ 2020, arXiv e-prints, arXiv:2001.04468
\bibitem[Torres et al.(2009)]{Torres09} Torres, R.~M., Loinard, L., Mioduszewski, A.~J., et al.\ 2009, \apj, 698, 242
\bibitem[Tobin et al.(2016)]{Tobin16} Tobin, J.~J., Looney, L.~W., Li, Z.-Y., et al.\ 2016, \apj, 818, 73
\bibitem[Tsukamoto et al.(2017)]{Tsukamoto17} Tsukamoto, Y., Okuzumi, S., Iwasaki, K., et al.\ 2017, \pasj, 69, 95
\bibitem[Tychoniec et al.(2019)]{Tychoniec19} Tychoniec, {\L}., Hull, C.~L.~H., Kristensen, L.~E., et al.\ 2019, \aap, 632, A101
\bibitem[van Kempen et al.(2009)]{Kempen09} van Kempen, T.~A., van Dishoeck, E.~F., Salter, D.~M., et al.\ 2009, \aap, 498, 167
\bibitem[Verliat et al.(2020)]{Verliat20} Verliat, A., Hennebelle, P., Maury, A.~J., et al.\ 2020, \aap, 635, A130
\bibitem[Wang, J.-W.~et al.(2019)]{WangJ19} Wang, J.-W., Lai, S.-P., Eswaraiah, C., et al.\ 2019, \apj, 876, 42
\bibitem[Wang, L.-Y.~et al.(2019)]{WangL19} Wang, L.-Y., Shang, H., \& Chiang, T.-Y.\ 2019, \apj, 874, 31
\bibitem[Ward-Thompson et al.(2017)]{Ward17} Ward-Thompson, D., Pattle, K., Bastien, P., et al.\ 2017, \apj, 842, 66
\bibitem[Watson(2020)]{Watson20} Watson, D.~M.\ 2020, Research Notes of the American Astronomical Society, 4, 88
\bibitem[Williams \& Cieza(2011)]{Williams11} Williams, J.~P., \& Cieza, L.~A.\ 2011, \araa, 49, 67
\bibitem[Williams et al.(2019)]{Williams19} Williams, J.~P., Cieza, L., Hales, A., et al.\ 2019, \apjl, 875, L9
\bibitem[Williams et al.(1994)]{Williams94} Williams, J.~P., de Geus, E.~J., \& Blitz, L.\ 1994, \apj, 428, 693
\bibitem[Wolf et al.(2003)]{Wolf03} Wolf, S., Launhardt, R., \& Henning, T.\ 2003, \apj, 592, 233
\bibitem[Wurster, \& Bate(2019)]{Wurster19b} Wurster, J., \& Bate, M.~R.\ 2019, \mnras, 486, 2587
\bibitem[Wurster et al.(2019)]{Wurster19a} Wurster, J., Bate, M.~R., \& Price, D.~J.\ 2019, \mnras, 489, 1719
\bibitem[Yen et al.(2018)]{Yen18} Yen, H.-W., Koch, P.~M., Manara, C.~F., et al.\ 2018, \aap, 616, A100
\bibitem[Yen et al.(2015)]{Yen15} Yen, H.-W., Koch, P.~M., Takakuwa, S., et al.\ 2015, \apj, 799, 193
\bibitem[Yen et al.(2019)]{Yen19} Yen, H.-W., Zhao, B., Hsieh, I.-T., et al.\ 2019, \apj, 871, 243
\bibitem[Yen et al.(2020)]{Yen20} Yen, H.-W., Zhao, B., Koch, P., et al.\ 2020, \apj, 893, 54
\bibitem[Young \& Evans(2005)]{Young05} Young, C.~H. \& Evans, N.~J.\ 2005, \apj, 627, 293
\bibitem[Young et al.(2003)]{Young03} Young, C.~H., Shirley, Y.~L., Evans, N.~J., et al.\ 2003, \apjs, 145, 111
\bibitem[Zhang et al.(2018)]{Zhang18} Zhang, S., Hartmann, L., Zamora-Avil{\'e}s, M., et al.\ 2018, \mnras, 480, 5495
\bibitem[Zhang et al.(2014)]{Zhang14} Zhang, Q., Qiu, K., Girart, J.~M., et al.\ 2014, \apj, 792, 116
\bibitem[Zhao et al.(2016)]{Zhao16} Zhao, B., Caselli, P., Li, Z.-Y., et al.\ 2016, \mnras, 460, 2050 
\bibitem[Zhao et al.(2018)]{Zhao18} Zhao, B., Caselli, P., Li, Z.-Y., \& Krasnopolsky, R.\ 2018, \mnras, 473, 4868 
\bibitem[Zucker et al.(2019)]{Zucker19} Zucker, C., Speagle, J.~S., Schlafly, E.~F., et al.\ 2019, \apj, 879, 125
\end{thebibliography}
\end{document}